\documentclass[12pt]{article}

  \usepackage{subeqn}
  \usepackage{amsthm}
  \usepackage{amssymb}
  \usepackage{cite}

  \def\be{\begin{equation}}
  \def\ee{\end{equation}}
  \def\ba{\begin{array}}
  \def\ea{\end{array}}
  \def\bea{\begin{eqnarray}}
  \def\eea{\end{eqnarray}}
  \def\bean{\begin{eqnarray*}}
  \def\eean{\end{eqnarray*}}
  \def\hf{\frac{1}{2}}

  \def\e{{\rm e}}
  \def\z{{\zeta}}
\def\ii{\mathrm{i}} \def\d{{\rm d} \hspace{0.0mm}}
  \def\a{\alpha}

  \newtheorem{theorem}{Theorem}[section]
  \newtheorem{proposition}[theorem]{Proposition}

  \newcommand{\sca}[2]{\langle #1, #2 \rangle}
  \newcommand{\vt}[1]{\mbox{\boldmath$#1$}}

  \newcommand{\CMP}{{\em Commun.\ Math.\ Phys.} } 
  \newcommand{\CPAM}{{\em Commun.\ Pure Appl.\ Math.} } 
  \newcommand{\FAA}{{\em Func.\ Anal.\ Appl.} } 
  \newcommand{\IP}{{\em Inver.\ Probl.} }

  \newcommand{\JDE}{{\em J.\ Diff.\ Eq.} }
  
  \newcommand{\JPA}{{\em J.\ Phys.\ A: Math.\ Gen.} }
  
  \newcommand{\JNMP}{{\em J.\ Nonlin.\ Math.\ Phys.} }
  \newcommand{\JMP}{{\em J.\ Math.\ Phys.} }

  \newcommand{\JPSJ}{{\em J.\ Phys.\ Soc.\ Jpn.} }
  \newcommand{\JSC}{{\em J.\ Symbolic Comput.} }
  \newcommand{\LMP}{{\em Lett.\ Math.\ Phys.} }

  \newcommand{\NC}{{\em Nuovo Cimento\/} }

  \newcommand{\PL}{{\em Phys.\ Lett.} }
  
  \newcommand{\PRL}{{\em Phys.\ Rev.\ Lett.} }

  \newcommand{\PTP}{{\em Prog.\ Theor.\ Phys.} }

  \newcommand{\SAM}{{\em Stud.\ Appl.\ Math.} }
  
  \newcommand{\TMP}{{\em Theor.\ Math.\ Phys.} }

  \title{\bf Classification of polynomial integrable systems
     of mixed scalar and vector evolution equations.\ I}
  \author{Takayuki Tsuchida\\
 Department of Physics, 
 Kwansei Gakuin University, \\
 2-1 Gakuen, Sanda 
 669-1337, Japan \\
  E-mail: tsuchida@ms.u-tokyo.ac.jp\\ \\
	  and\\ \\
  Thomas Wolf\\ Department of Mathematics,
  Brock University\\ 
  St.Catharines, Ontario, Canada L2S 3A1\\
  E-mail: twolf@brocku.ca}

\begin{document}
  \maketitle

  \newpage

 \begin{abstract}
 We 
 perform a 
 classification 
 of integrable systems of mixed scalar and vector 
 evolution equations with respect to higher symmetries. 
 We consider 
 polynomial 
 systems 
 that are homogeneous 
 under a suitable weighting 
 of 
 variables. 
  This paper deals with 
 the KdV weighting, the Burgers (or potential KdV or modified KdV) 
  weighting, the Ibragimov--Shabat weighting and two unfamiliar weightings. 
 The case of other weightings 
  will be studied
  in a subsequent paper. 
 Making an ansatz for undetermined coefficients and 
 using a computer package 
 for solving 
 bilinear algebraic systems, 
 we give the complete lists of 
 $2^{\mbox{\scriptsize nd }}$order systems with a 
 $3^{\mbox{\scriptsize rd }}$order or a 
 $4^{\mbox{\scriptsize th }}$order symmetry 
 and $3^{\mbox{\scriptsize rd }}$order 
 systems with a $5^{\mbox{\scriptsize th }}$order symmetry. 
  For all but a few systems in the lists, 
  we show 
  that the system (or, at least a subsystem of it) 
  admits either a Lax representation or a linearizing transformation. 
  A thorough comparison with 
 recent 
  work of Foursov and Olver is made. 
  \end{abstract}


  \newpage
  \noindent 
 \tableofcontents

  \newpage
  \noindent
  \section{Introduction}
  \setcounter{equation}{0}

 The symmetry approach has been proven 
 to be the most efficient integrability test 
for $(1+1)$-dimensional nonlinear evolution 
equations \cite{IbSha,Fokas2,SoSh,MS1,MS2,Fokas,MikShYam,MSY,FujiWata,MiShSok} 
(see also a recent review \cite{ASY}). 
It is useful in classifying 
 both 
 scalar evolution equations 
 and 
 evolutionary systems of 
 equations (see, {\it e.g.\ }\cite{MiShSok}). 
 A mile stone 
 in this direction 
 is 
 the work of 
 Mikhailov, Shabat and Yamilov \cite{MS1,MS2,MikShYam,MSY} 
 on the classification of $2^{\mbox{\scriptsize nd }}$order systems 
 with two components. 
 Their work dealt with a large 
 class 
 of systems that are non-polynomial in general 
 and have a nondegenerate leading part with respect to 
 $x$-derivatives. 
 They 
 obtained 
 a complete 
 list of 
 systems possessing 
 higher conservation laws, 
 up to some (almost) invertible 
 transformations \cite{MikShYam,MSY}. 
 Systems with both higher conservation laws and higher symmetries 
 are believed to be 
 integrable 
 by the {\it inverse Scattering method}, 
 for short 
 ``S-integrable'' 
 in the terminology of Calogero \cite{Calo1,Calo2}. 
 The aim of this paper is to extend 
 the classification 
 of Mikhailov {\it et al.}\ 
 and to make it easily accessible. 
 To be 
 specific, 
 we pursue the following 
 goals with 
 this paper:
 \begin{itemize}
 \item
 To provide a ``user-friendly'' complete list of systems 
 without any freedom of nontrivial transformations. 
 By that the user does not have to find transformations to locate a given 
 system in our list. Trivial
 scaling parameters are removed.  
 Naturally, this is possible only for a much more 
 restricted class of systems than 
 that considered by Mikhailov {\it et al.} 
 \item
 To 
 include systems 
 without higher conservation laws\footnote{Here we mean 
 conservation laws that do not depend 
 on $x$ and $t$ explicitly.}, 
 but with higher symmetries, in the classification. 
 Systems of this sort are believed to be 
 linearizable by an appropriate 
 {\it Change of variables} \/and, if so, 
 said to be ``C-integrable'' 
in the terminology of Calogero \cite{Calo1,Calo2}. 
  \item
  To allow systems to have a 
 degenerate 
 leading part. 
 This means that 
 the coefficient matrix of leading terms may have a 
 zero eigenvalue. 
  \item
 To classify systems of 
 higher order ($3^{\mbox{\scriptsize rd }}$order, ...). 
%$\ldots$). 
  \item
  To classify 
 systems with more than two components. 
  \end{itemize}
 Here we mention 
 earlier studies 
 devoted to 
 these extensions, although 
 we do not know any work dealing with all these extensions 
 simultaneously. 
 A rather user-friendly list of integrable 
 systems of $2^{\mbox{\scriptsize nd }}$order 
 with two components was presented in \cite{Sanders} 
 (see also a similar list in \cite{SWo0}). 
 Some classifications of ``C-integrable'' systems 
 including 
 coupled Burgers-type equations 
 have been reported in \cite{SWo0,Bakirov,Svi2,Foursov1}\footnote{The list 
given in \cite{Bakirov} seems to be 
 incomplete, because we 
 cannot identify an integrable system of the Burgers type 
 (cf.\ (\ref{sol25'}) in this paper) with any system in the list.}. 
 Classification of integrable 
 coupled KdV-type equations has been performed in 
 \cite{Zharkov1,Zharkov2,Meshkov,Foursov0} 
 using the symmetry approach 
 and in \cite{Karasu,Sakov} using the Painlev\'{e} PDE test. 
 Coupled potential KdV (coupled pKdV) equations 
 and coupled modified KdV (coupled mKdV) equations 
 with higher symmetries were listed 
 in \cite{Foursov2} 
 (see also \cite{Foursov3}). 
 Classification of 
 coupled KdV equations and coupled mKdV equations 
 was studied 
 in connection with Jordan algebras 
 in \cite{Svi0,Svi}, where the coefficient matrix of leading terms 
 is restricted to the identity. 
 The Painlev\'{e} PDE test was 
 applied to 
 coupled higher-order nonlinear 
 Schr\"{o}dinger equations 
 in \cite{Sak}, 
 where 
 integrable 
 coupled mKdV equations and 
 coupled derivative nonlinear Schr\"{o}dinger (coupled DNLS) equations 
 were obtained. 
 Classification of non-commutative 
 generalizations of integrable systems on an associative algebra 
 was addressed in \cite{Olver1} 
 (see also \cite{Linden1,Olver2,TW3} for DNLS-type systems), while 
 vector generalizations of 
 integrable systems 
 were discussed in \cite{SWo}. 

 In this paper, we investigate evolutionary systems 
 for one scalar unknown 
 $u(x,t)$ and one vector unknown 
 $U(x,t) \equiv (U_1, U_2, \ldots, U_N )$ 
 using the symmetry approach. In particular, 
 we classify 
 $2^{\mbox{\scriptsize nd }}$order 
 and $3^{\mbox{\scriptsize rd }}$order systems 
 that are polynomial in $u$, $U$ and their 
 derivatives. 
 This work was initiated by 
 Vladimir 
 Sokolov and the second author (T.W.) in \cite{SWo}. 
 Here, $N$ is an arbitrary positive integer and 
 the product between two vectors is defined by 
 the scalar product: 
 $$\sca{\partial_x^m U}{\partial_x^n U} \equiv 
\sum_{j=1}^N (\partial_x^m U_j) (\partial_x^n U_j), \hspace{5mm}m, n \ge 0.$$
 We do not 
 consider constant vectors 
$C_j$ 
or matrices 
$C_{jk}$ 
as in 
$\sum_{j=1}^N C_{j} (\partial_x^m U_j)$ or, for example, 
$\sum_{j,k=1}^N C_{jk} (\partial_x^m U_j) (\partial_x^n U_k)$. 
 Moreover, we 
 require that the scalar and vector evolution equations are
{\it truly} \/coupled, that is, $U$ occurs in $u_t=\ldots$ and $u$ occurs 
 in $U_t=\ldots \; $. 
 Classifications described in this paper are
 restricted to 
 $(\lambda_1, \lambda_2)$-{\em homogeneous} \/systems of {\em weight} $\mu$.
 These are systems that admit the one-parameter group of scaling symmetries
$$(x, \hspace{1pt}\, t, \hspace{1pt}\, u, \hspace{1pt}\, U_j) 
\longrightarrow
(a^{-1}x, \hspace{1pt}\, a^{-\mu} t, \hspace{1pt}\, 
 a^{\lambda_1} u, \hspace{1pt}\, a^{\lambda_2} U_j), 
\hspace{4mm} a \neq 0.$$
We consider only systems with 
$\lambda_1, \lambda_2 > 0$ and 
a differential order equal to $\mu$. 
For systems with $\lambda_1 = \lambda_2$, 
this would imply 
the existence of a linear leading part (dispersion), 
but not in the case of mixed systems with 
$\lambda_1 \neq \lambda_2$. 
For example, 
for $\mu=2$ and $\lambda_1=2 \lambda_2$, 
the two terms $u_{xx}$ and $\langle U,U_{xx}\rangle$ 
have the same weight and a differential order equal to $\mu$. 
In 
either case, 
systems having a degenerate leading part are 
also 
included in our classification. 

 For the scalar case, it was proven 
 in \cite{Wang} 
 that a $\lambda$-homogeneous
 polynomial evolution equation with $\lambda>0$ and a dispersion term 
 may possess a 
 polynomial higher symmetry only if
 \begin{description}
 \item[\hspace{10mm}] $\lambda=2$    \ \ \  
	 (KdV weighting), or
 \item[\hspace{10mm}] $\lambda=1$    \ \ \  (Burgers/pKdV/mKdV weighting), or
 \item[\hspace{10mm}] $\lambda={1\over 2}$ \ \ \ (Ibragimov--Shabat 
 weighting \cite{Ibragimov}).
 \end{description}
 It was also proven in \cite{Wang} that 
 any symmetry-integrable\footnote{The symmetry-integrable 
 equations are such equations that possess an infinite set 
 of (commuting) higher symmetries.} equation of 
 $2^{\mbox{\scriptsize nd }}$($3^{\mbox{\scriptsize rd}}$) order 
 in the considered classes 
 {\it does} \/possess 
 a symmetry of $3^{\mbox{\scriptsize rd }}$($5^{\mbox{\scriptsize th}}$) 
 order, respectively. 
 Similar results on 
 $(\lambda_1, \lambda_2)$-homogeneous 
 polynomial systems 
 of weight $2$ 
 with two components 
 were 
 obtained in \cite{Sanders}. 
 Under the conditions of 
 $\lambda_1, \lambda_2 > 0$, 
 $| \lambda_1 -\lambda_2 | \notin {\mathbb N}_{>0}
$, a nondegeneracy of the linear part\footnote{For 
full details, 
see \cite{Sanders}.} and the order of 
the nonlinear part less than $2$, 
such a system may possess 
polynomial higher symmetries 
only if 
 $\lambda_1 = \lambda_2 = 2, \, 1, \, {1\over 2}$ or 
 \begin{description}
 \item[\hspace{10mm}] $\lambda_1 = {1 \over 3}, \; \lambda_2 = {2 \over 3}$; 
 \item[\hspace{10mm}] $\lambda_1 = {2 \over 3}, \; \lambda_2 = {1 \over 3}$.
 \end{description}
 In these classes, 
 any symmetry-integrable 
 system of $2^{\mbox{\scriptsize nd }}$order 
 {\it does} \/possess a symmetry of $3^{\mbox{\scriptsize rd }}$order 
 or $4^{\mbox{\scriptsize th }}$order. 
 Since we 
 study $(1+N)$-component systems of 
 $2^{\mbox{\scriptsize nd }}$and $3^{\mbox{\scriptsize rd }}$order 
 that may have a 
 degenerate leading part, 
 we cannot 
 entirely rely on 
 these results. They 
 neither give 
 all possible 
 pairs of $(\lambda_1, \lambda_2)$ for 
 integrable cases 
 nor indicate the order of a higher symmetry to 
 exist. 
 Nevertheless, in this paper, 
 we concentrate our attention 
 on systems that are 
 homogeneous 
 in 
 $(\lambda_1, \lambda_2)
 = (2,2), (1,1), 
 \bigl({1\over 2},{1\over 2}\bigr), \bigl({1\over 3},{2\over 3}\bigr)$ 
 or $\bigl({2\over 3},{1\over 3}\bigr)$ 
 and $\mu = 2$ or $3$. 
 Indeed, 
as we will see 
below, 
there exist 
 a lot of interesting 
 integrable systems in these classes. 
 Classifications for other pairs of 
 $(\lambda_1, \lambda_2)$ will be reported 
 in a subsequent paper. 

 The search for integrable systems in this paper is based on 
 the simplest version of the symmetry approach \cite{IbSha,Fokas2}, i.e.\ the 
 existence of one higher symmetry. 
 It is considered 
 as a necessary, but in general not sufficient 
 condition for integrability\footnote{Some systems 
of the 
 Bakirov type 
 are known to possess only 
 a finite number of higher symmetries \cite{Beukers,Beukers2,Kamp2}. 
 However, 
 all such examples 
 are pathological and less interesting, 
 because they are 
%both triangular and already in their given form almost linear.
already in their given form triangular linear. 
In this paper, we 
 encounter systems 
 with a higher symmetry 
 that are 
 reducible to 
 a triangular 
 form 
 by a {\it nonlinear} \/and {\it non-ultralocal} \/change 
of 
 variables. 
 It is an open question whether 
 such systems are 
 symmetry-integrable in general. 
 }. Both ``S-integrable'' and ``C-integrable'' systems can be 
 detected by the existence of one higher symmetry. 
 To do concrete computations using the computer algebra program {\sc Crack} 
 \cite{Wolf}
 we 
 assume 
 the existence of 
 a $3^{\mbox{\scriptsize rd }}$order 
 or a $4^{\mbox{\scriptsize th }}$order symmetry 
 for a $2^{\mbox{\scriptsize nd }}$order system 
 and a $5^{\mbox{\scriptsize th }}$order symmetry for a 
 $3^{\mbox{\scriptsize rd }}$order system. 
 Although 
 the existence of symmetries of a specific order
 may be 
 too restrictive 
 and not necessary 
 for 
 integrability\footnote{As a result, 
 we may 
 miss 
 some integrable cases.
 }, 
 it allows us to perform exhaustive searches and 
 obtain complete lists for these cases.
 An overview of the performed computations 
 is given in the next section. 
 For the 
 lists generated by 
 computer, we first remove unessential parameters 
 by scaling independent and dependent variables. 
 We note that 
 a linear change of dependent variables mixing the scalar $u$ 
 with components of the vector $U$ 
 would give systems with more than one scalar unknown 
 and is therefore not considered. 
 Next, we 
prove integrability 
for 
nearly all listed systems 
 by constructing either 
 a Lax representation 
 or a linearizing transformation\footnote{In 
 this paper, we are not going to pursue the symmetry integrability.}. 
 Some systems in the lists can be reduced to 
 triangular systems by a nonlinear 
 transformation 
 of dependent variables. 
 If that is possible then 
 systems contain a 
 closed subsystem 
 in a nontrivial manner. 
 In that case, 
 we first prove 
 the integrability of the subsystem 
 and then 
 discuss how to solve the remaining equations. 
 We can reduce 
 the task of proving the integrability 
 through establishing 
 relationships 
 among the listed systems. 
 We construct 
 a rich set of Miura-type transformations, 
 including 
 Miura maps plus potentiation, 
 that connect 
 different systems in the lists. 
 Thus we 
 have only to 
 investigate 
 one representative 
 for each group of connected 
 systems. 

 This paper is organized as follows. In section 2, we 
 explain briefly 
 how 
 the lists of systems 
 in this paper 
 are 
 generated by 
 computer. 
 In section 3, we 
 perform a classification of 
 $2^{\mbox{\scriptsize nd }}$order and 
 $3^{\mbox{\scriptsize rd }}$order systems in the 
 $\lambda_1 = \lambda_2 =2$ (KdV weighting) 
 case. 
 The list of $2^{\mbox{\scriptsize nd }}$order 
 systems with a $3^{\mbox{\scriptsize rd }}$order 
 or a $4^{\mbox{\scriptsize th }}$order 
 symmetry is empty, 
 while 
 that of 
 $3^{\mbox{\scriptsize rd }}$order systems 
 with 
 a $5^{\mbox{\scriptsize th }}$order symmetry 
 consists of four members. 
 The list itself is already known \cite{SWo} 
 but we prove their integrability in section 3.

 Section 4 
 forms the {\it main part} \/of 
this paper. 
 We classify 
 $2^{\mbox{\scriptsize nd }}$order and 
 $3^{\mbox{\scriptsize rd }}$order systems in the 
 $\lambda_1 = \lambda_2 =1$ 
 (Burgers/pKdV/mKdV weighting) case. 
 The list of $2^{\mbox{\scriptsize nd }}$order 
 systems consists of 
 three members 
 that generalize the Burgers equation. 
 All 
 these 
 systems possess both 
 a $3^{\mbox{\scriptsize rd }}$order 
 and a $4^{\mbox{\scriptsize th }}$order symmetry. 
 We can (triangular) 
%(almost) 
linearize 
 two of them 
 through an extension 
 of the Hopf--Cole 
 transformation, 
 while 
 integrability\footnote{We 
 mean the existence of either a Lax representation or 
 a linearizing transformation.} of the other 
 system, (\ref{sol12'}), 
 {\it remains unproven}. 
 We 
 discuss travelling-wave solutions 
 of (a subsystem of) this system 
 to 
 indicate its 
 nontrivial nature. 
 The interested reader 
 is referred to 
 section~\ref{Solution-12} 
 for 
 the details. 
 The list of $3^{\mbox{\scriptsize rd }}$order 
 systems consists of 
 25 members, 
 three of which 
 are 
 symmetries of the $2^{\mbox{\scriptsize nd }}$order systems 
 from the previous list. 
 Consequently, 
 we can (triangular) 
%(almost) 
linearize two of the three 
 $3^{\mbox{\scriptsize rd }}$order 
 systems, 
 while 
 integrability of the other 
 $3^{\mbox{\scriptsize rd }}$order 
 system 
 remains to be seen. 
 Another $3^{\mbox{\scriptsize rd }}$order 
 system in the list (system (\ref{sol11})) 
is very close to the latter 
 system and 
 we 
 do not know how to 
 integrate 
 it 
 either. 
 For 
 the other $21\hspace{2pt}(=25-3-1)$ 
 systems of $3^{\mbox{\scriptsize rd }}$order, 
 we 
 prove 
 that 
 they are integrable 
 or, at least, 
 they contain 
 an integrable 
 closed subsystem. 
 We point out 
 that 
 one of the 21 systems is the 
 $3^{\mbox{\scriptsize rd }}$order symmetry of 
 a nontrivial $1^{\mbox{\scriptsize st }}$order system. 
 Miura-type transformations 
 that connect 
 $3^{\mbox{\scriptsize rd }}$order 
 systems 
 in the 
 lists 
 of 
 $\lambda_1 = \lambda_2 = 1$ 
 and $\lambda_1 = \lambda_2 = 2$ 
 are presented. 

 In section 5, we classify 
 $2^{\mbox{\scriptsize nd }}$order and $3^{\mbox{\scriptsize rd }}$order 
 systems in the $\lambda_1 = \lambda_2 = {1 \over 2}$ 
 (Ibragimov--Shabat weighting) case. 
 The list of $2^{\mbox{\scriptsize nd }}$order 
 systems 
 is empty, 
 while 
 that of $3^{\mbox{\scriptsize rd }}$order systems 
 consists of two members. 
 We can linearize both of them 
 through a generalization of the linearizing transformation 
 for the 
 Ibragimov--Shabat equation. 
 In section 6, we 
 obtain 
 negative 
 results regarding a classification 
 in 
 the case of $\lambda_1 = {1 \over 3}, \; \lambda_2 = {2 \over 3}$. 
 In section 7, we 
 perform a classification 
 of $2^{\mbox{\scriptsize nd }}$order and $3^{\mbox{\scriptsize rd }}$order 
 systems 
 in the case of 
 $\lambda_1 = {2 \over 3}, \; \lambda_2 = {1 \over 3}$. 
 We obtain one $2^{\mbox{\scriptsize nd }}$order system 
 with a $3^{\mbox{\scriptsize rd }}$order symmetry, 
 two $2^{\mbox{\scriptsize nd }}$order systems 
 with a $4^{\mbox{\scriptsize th }}$order symmetry 
 and two $3^{\mbox{\scriptsize rd }}$order systems 
 with a $5^{\mbox{\scriptsize th }}$order symmetry. 
 Thereby we 
 have 
 one $2^{\mbox{\scriptsize nd }}$order system 
 without a $3^{\mbox{\scriptsize rd }}$order symmetry, 
 but with a $4^{\mbox{\scriptsize th }}$order symmetry. 
 All the 
 listed 
 systems 
 can be linearized through an ultralocal change of 
 dependent variables. 
 Section 8 is devoted to concluding remarks. 

 Finally, we 
 would like to 
 mention that 
 our results 
 in section 4 
 refine and generalize 
 the recent work of Foursov 
 and Olver \cite{Foursov1,Foursov2,Foursov3}. 
 Their work 
 focused on 
 polynomial 
 systems of two symmetrically 
 coupled 
 nonlinear 
 evolution equations, i.e.\ 
 symmetric systems for two scalar unknowns. 
 They 
 obtained 
 the complete lists of 
 $\lambda_1 = \lambda_2=1$ homogeneous 
 systems of $2^{\mbox{\scriptsize nd }}$and 
 $3^{\mbox{\scriptsize rd }}$order 
 with 
 two 
 higher 
 symmetries 
 of specific 
 orders. 
 Most of 
 the 
 $(1+N)$-component systems 
 listed in section 4 
 generalize 
 two-component 
 systems of Foursov--Olver, 
 up to a linear change of dependent variables. 
 To see this, we remark that, because of our assumption 
 on 
 the admissible 
 multiplications, 
 the evolution equation for the scalar 
 $u$ 
 is even in the vector 
 $U$, while 
 the equation for 
 $U$ is odd in $U$. 
 Therefore, 
 we can symmetrize 
 our systems 
 in the special case $N=1$ 
 through the linear change of variables: 
 $u= a (q+r)$, $U=b (q-r)$ 
 where $a$ and $b$ are nonzero constants. 

 On the basis of this re-formulation, 
 we 
 compare 
 in section 4 
 our lists 
 with 
 those of
 Foursov--Olver.
 A 
 brief 
 summary of 
 the comparison results 
 is as follows: 
 \begin{itemize}
 \item 
 Any system in Foursov--Olver's lists\footnote{We 
mean the lists of 
 systems 
 that 
 are not reducible to 
 a 
 triangular form 
 by a linear change of dependent variables.} 
 corresponds to 
 the $N=1$ case of 
 one 
 or two 
 systems 
 in our lists. 
 This means that, 
 after 
 the linear change of dependent variables 
 mentioned above, 
 their two-component systems 
 always 
 admit 
 $(1+N)$-component 
 generalization(s) 
 preserving 
 the 
 integrability. 
 This result 
 is 
 quite interesting, but 
 unlikely to hold true 
 in general for other classes 
 of 
 two-component systems. 
 \item 
 Some systems in our lists 
 do not 
 have 
 any counterpart 
 in Foursov--Olver's lists. 
 They 
 are systems that 
 become the trivial equation $u_t=0$ under the reduction $U=\vt{0}$. 
 Such systems were excluded 
 from consideration in the work of 
 Foursov--Olver 
 by their assumption on 
 strong 
 nondegeneracy of 
 the linear 
 part (see, {\it e.g.\ }\/section 2 of \cite{Foursov1}). 
 In this respect, 
 our lists are 
 richer than Foursov--Olver's lists 
 even 
 in the $N=1$ case. 
 \end{itemize}
 Besides 
 extending Foursov--Olver's lists \cite{Foursov1,Foursov2,Foursov3}, 
 we prove the integrability of 
 many 
 systems in their lists 
 for the first time. 
 We also correct 
 errors in \cite{Foursov1,Foursov2,Foursov3} and 
 point out 
 overlooked 
 references 
 in which 
 some 
 systems 
 in 
 their lists 
 were 
 studied earlier. 

 \newpage
 \noindent
 \section{Computational aspects}
 \label{Comp_as} 
 Before describing the 
 classification results 
 in 
sections~\ref{lambda2}--\ref{lambda2-3}, 
 in this section, 
 we would like to make 
some 
comments on the computations performed.

 As the first step a homogeneous ansatz for a 
 system consisting of a scalar equation $u_t=\ldots$ and a vector equation 
 $U_t=\ldots$ is generated together with 
 a system of higher symmetry equations 
 $u_\tau=\ldots,\ U_\tau=\ldots \; $. 
 Each term has a different 
 undetermined coefficient. 
We assume that these 
coefficients do not depend on 
$N$
(the number of components of $U$), 
and that 
 $N$ is not fixed at 
 any specific value\footnote{The 
 arbitrariness of $N$ is crucial 
 for 
 functional independence of the scalar products 
 $\sca{\partial_x^m U}{\partial_x^n U}$ $(0 \le m \le n)$ \cite{SWo}.}. 

 Computationally more expensive is the formulation of the symmetry
 conditions $u_{[t,\tau]}=0$, $ U_{[t,\tau]}=\vt{0}$.
 For low values of $\lambda_1, \lambda_2$ and high 
 differential order the right-hand sides of the system and the
 symmetry involve many terms, 
 and in addition each of the terms
 has an increasing number of factors. Higher-order $x$-derivatives of
 such terms cause a large expression swell, too large to compute the 
 commutators in one step. We therefore perform the
 computation of $u_{[t,\tau]}$ and $U_{[t,\tau]}$ in stages.
 Because right-hand sides of the system and the 
symmetry do not involve
 $\partial_t, \partial_\tau$, substitutions of $u_t, U_t, u_\tau, U_\tau$
 in the commutators are done only once. Consequently, commutators 
 are linear in the coefficients of the system and coefficients 
 of the symmetry.
 To exploit this linearity we partition
 \begin{equation}
 u_t=\sum_i F_i,\ \ U_t=\sum_i G_i,\ \ u_\tau=\sum_i H_i,\ \ 
 U_\tau=\sum_i K_i,\ \ u_{[t,\tau]} =\sum_i P_i,\ \ U_{[t,\tau]}=\sum_i Q_i,
 \end{equation}
where the expressions $F_i, G_i, H_i, K_i, P_i, Q_i$ contain only terms with 
a total degree $i$ of all scalar vector products of $U$ and 
$x$-derivatives of $U$ (for example, $\langle U,U_x \rangle U$ having 
degree $1$). 
By using the observation that the number of scalar 
vector products in a term 
 does not change when a term is differentiated we can compute each $P_i$ 
 independently through
 \[P_i = \sum_{j=0}^i u_{[t,\tau]}
         \left|_{u_t=F_j,\ U_t=G_j,\ u_\tau=H_{i-j},\ U_\tau=K_{i-j}} \right.,
\]
 and similarly for each $Q_i$. Because they are the only terms that have
 $i$-th degree powers of scalar 
 vector products, all $P_i, Q_i$ must vanish identically.
 After one single $P_i$ or $Q_i$ is computed, it can be 
 split\footnote{By {\em splitting} \/we mean the extraction and setting to zero
 of all coefficients of all products of all powers of all scalar and vector
 functions and their derivatives.} and some of the
 consequences, like the vanishing of some coefficients, can be used to simplify
 $F_i,G_i,H_i,K_i$ before computing the next $P_j$ and $Q_j$.

 For large problems (low $\lambda$ and high differential order) the
 computation of $Q_i$ was still too memory intensive\footnote{The
 computation was too memory demanding to
 be performed on the computers available to one of the authors in 2000.} so
 that another partitioning of the computation was implemented. 
In this level of partitioning, first those terms in $F_i, G_i, H_i,
K_i$ which can contribute to the highest derivative vectorial factor
$\partial_x^j U$ in $Q_i$ were considered. Let us call the partial
commutator that comes out of this computation 
$\hat{C}_{i,j}$. 
From $\hat{C}_{i,j}$ 
all the terms with vectorial factor $\partial_x^j U$ are extracted, split, 
and the remaining terms from $\hat{C}_{i,j}$ 
(with a vectorial factor of 
\mbox{$\textrm{order} < j$}) 
are carried over for the computation of the next terms with 
derivative vectorial factor $\partial_x^{j-1} U$ in $Q_i$.\footnote{More 
details can be obtained at request.}

 In this way the computation of single
 large expressions $u_{[t,\tau]}, U_{[t,\tau]}$ is avoided and replaced
 by the computation of many partial commutators 
 resulting in bilinear algebraic equations for the undetermined 
 coefficients.

 To each list of conditions is attached a list of inequalities which
 have to be fulfilled by any solution.  A first inequality results from
 the requirement that at least one of both equations involves at least
 one $x$-derivative of the required order ($2$ or $3$).  Similarly the
 symmetry equations have to involve at least one $x$-derivative of the
 required order and the right-hand sides of both symmetry equations
 must be 
non-zero. Two further conditions prevent the generation of
 triangular integrable systems by requiring that $U$ occurs in
 $u_t=\ldots$ and $u$ occurs in $U_t=\ldots \; $. 

 The solution of the overdetermined bilinear algebraic systems was
 accomplished with the computer program {\sc Crack} written for the
 solution of overdetermined algebraic but also differential
 systems. One technique that proved to be quite useful in general,
 especially for the solution of larger systems with 
 $\lambda_1=\lambda_2=1$ and 
 $\lambda_1=\lambda_2={1\over 2}$, is an equation shortening 
 method described in \cite{TWshort}.

 The following tables 
 give an overview on the complexity of computations. These
 have been performed on a 1.7GHz Pentium 4 PC running the
 computer algebra system REDUCE 3.7 in a 120 MB session under Linux.
 Quoted execution times are sensitive to settings of computing
 parameters and should be taken only as rough 
indicators.
 {\small
 \begin{table}[p]
 \begin{center}
 \begin{tabular}{|l|r|r|r|r|r|} \hline 
 $\lambda_1, \, \lambda_2$                       & 2,$\,$2 &  1,$\,$1 & 
	 $\frac{1^{\vphantom A}}{2_{\vphantom A}}, \, \frac{1}{2}$ 
   &   	$\frac{1^{\vphantom A}}{3_{\vphantom A}}, \, \frac{2}{3}$ 
   &   	$\frac{2^{\vphantom A}}{3_{\vphantom A}}, \, \frac{1}{3}$ 
	 \\  \hline
 \# of unknowns in the system    &    5 &    10 &    15 
	 & 10 & 13  \\  \hline
 \# of unknowns in the symmetry  &    6 &   21 &   36 
	 & 24 & 22  \\  \hline
 total \# of  unknowns           &    11 &   31 &  51 
	 & 34 & 35 \\  \hline
 \# of conditions                &    13 &   66 & 149 
	 & 102 & 114 \\  \hline
 total \# of terms in all conditions   
	 &  34 &  341 & 1093 & 529 & 694 \\  \hline
 average \# of terms in a condition    
	 &   2.6 &   5.2 &   7.3 & 5.2 & 6.1 \\  \hline
 time to formulate alg.\ conditions 
	 &   0.5s & 1.8s & 8s & 3.2s & 6.3s \\ \hline
 time to solve conditions           
	 & 0.5s & 29s & 29s & 45s & 22s  \\  \hline
 \# of solutions   &     0 &    3 &   0 & 0 & 1 \\  \hline
 \end{tabular} 
 \caption{Computations in 
the orders $2+3$ problem for the 5 weightings
 }
 \end{center}
 \end{table}
 \begin{table}[p]
 \begin{center}
 \begin{tabular}{|l|r|r|r|r|r|} \hline 
 $\lambda_1, \, \lambda_2$                       & 2,$\,$2 &  1,$\,$1 & 
	 $\frac{1^{\vphantom A}}{2_{\vphantom A}}, \, \frac{1}{2}$ 
   &   	$\frac{1^{\vphantom A}}{3_{\vphantom A}}, \, \frac{2}{3}$ 
   &   	$\frac{2^{\vphantom A}}{3_{\vphantom A}}, \, \frac{1}{3}$ 
	 \\  \hline
 \# of unknowns in the system    &    5 &    10 &    15 
	 & 10 & 13  \\  \hline
 \# of unknowns in the symmetry  &    12 &   39 &   79 
	 & 54 & 66  \\  \hline
 total \# of  unknowns           &    17 &   49 &   94 
	 & 64 & 79 \\  \hline
 \# of conditions                &    26 &  123 & 313 
	 & 215 & 276 \\  \hline
 total \# of terms in all conditions   
	 &  77 &  770 & 3096 & 1462 & 2435 \\  \hline
 average \# of terms in a condition    
	 &   3.0 &    6.3 &   9.9 & 6.8 & 8.8 \\  \hline
 time to formulate alg.\ conditions 
	 &   1s & 5s & 48s & 13s & 48s \\ \hline
 time to solve conditions           
	 & 0.4s & 1m58s & 3m44s & 1m23s & 3m40s  \\  \hline
 \# of solutions   &     0 &    3\hspace{2pt}\footnotemark 
	 &   0 & 0 & 2 \\  \hline
 \end{tabular} 
 \caption{Computations 
in the orders $2+4$ problem for the 5 weightings
 }
 \end{center}
 \end{table}
 \footnotetext{Although the program {\sc Crack} originally 
 produced $4$ solutions, 
 we could easily recognize that one solution is a special case of another.}
 \begin{table}[p]
 \begin{center}
 \begin{tabular}{|l|r|r|r|r|r|} \hline 
 $\lambda_1, \, \lambda_2$                       & 2,$\,$2 &  1,$\,$1 & 
	 $\frac{1^{\vphantom A}}{2_{\vphantom A}}, \, \frac{1}{2}$ 
   &   	$\frac{1^{\vphantom A}}{3_{\vphantom A}}, \, \frac{2}{3}$ 
   &   	$\frac{2^{\vphantom A}}{3_{\vphantom A}}, \, \frac{1}{3}$ 
	 \\  \hline
 \# of unknowns in the system    &     6 &    21 &    36 
	 & 24 & 22  \\  \hline
 \# of unknowns in the symmetry  &    17 &    74 &   164 
	 & 115 & 126  \\  \hline
 total \# of  unknowns           &    23 &    95 &   200 
	 &139 & 148 \\  \hline
 \# of conditions                &    50 &   386 &  1154 
	 & 798 & 955 \\  \hline
 total \# of terms in all conditions   
	 &   218 &  5000 & 27695 & 12694 & 17385 \\  \hline
 average \# of terms in a condition    
	 &   4.4 &    13 &    24 & 16 & 18 \\  \hline
 time to formulate alg.\ conditions 
	 &   5s & 2m52s   
	 & 2h7m & 23m45s & 41m18s \\ \hline
 time to solve conditions           
	 & 6.5s & 5h47m 
	 & 1day\hspace{1pt}\footnotemark & 1h20m & 1h7m  \\  \hline
 \# of solutions   &     4 &    25 &   2 & 0 & 2 \\  \hline
 \end{tabular} 
 \caption{Computations 
in the orders $3+5$ problem for the 5 weightings
 }
 \end{center}
 \end{table}
 \footnotetext{The computation involved one manual interference.}
}

For $\lambda_1=\lambda_2=1$ 
and orders $3+5$ the large number of
 solutions has the consequence that the system of algebraic conditions
does not simplify 
so readily and is more complicated to solve.
 Hence the program has 
 more often to impose case distinctions where an unknown is assumed to
 be at first zero and then non-zero.  As a result solutions may be found which
 can be unified into a single solution.  This is the case if, for example,
 one solution $S_1$ includes the condition $a_{17}=0$ while the other
 solution $S_2$ requires $a_{17} \neq 0$ for some undetermined coefficient
 $a_{17}$, and if setting $a_{17}=0$ in $S_2$ makes both
 solutions equivalent, in the system and in the symmetry. 
Sometimes a substitution like $a_{17}=0$ in $S_2$ may cause a 
division by zero which can be avoided by re-parametrizing $S_2$. 
An algorithm and its implementation in a 
computer program analyzing such situations have
 recently been developed and applied.

 On the web page 
 \\
 \hspace*{22mm}
 {\tt http://lie.math.brocku.ca/twolf/htdocs/sv/over.html}
 \\
one can inspect the original systems of conditions and the solutions 
as well as download them in machine readable form.  In addition to 
investigating systems of differential orders $2+3$ 
(for ${\rm system}+{\rm symmetry}$), 
$2+4$ and 
$3+5$, 
we also investigated orders $1+2$ and $1+3$. 
The main purpose was to recognize 
whether a $2^{\mbox{\scriptsize nd }}$order or a 
$3^{\mbox{\scriptsize rd }}$order system is 
actually the symmetry of a nontrivial 
$1^{\mbox{\scriptsize st }}$order system. 
Details can also be found on the above-mentioned web page. 
The package {\sc Crack} can be obtained from
 \\
 \hspace*{22mm}
 {\tt http://lie.math.brocku.ca/twolf/crack/}~. 

 After all solutions have been determined the task of proving
 integrability follows.  In the process of identifying and classifying
 some of the constant coefficient systems, the 
 {\it Mathematica} \/package
 ``InvariantsSymmetries.m'' \cite{Hereman} has been used to compute
 conservation laws and higher symmetries.

 \newpage
 \noindent
 \section{The case 
 $\lambda_1 = \lambda_2 =2$ 
     $\;\,$-- coupled KdV 
 equations --} 
 \label{lambda2}
 \setcounter{equation}{0}

 In this section, we 
 classify $2^{\mbox{\scriptsize nd }}$order 
 and $3^{\mbox{\scriptsize rd }}$order 
 systems in the $\lambda_1=\lambda_2 =2$ (KdV weighting) case. 
 In the first part (section~\ref{com-lis1}), we present 
 a complete list of 
 such systems with 
 a specific order symmetry 
 (the list 
 is already known, see \cite{SWo}). 
 In the second part (section~\ref{int-all1}), we 
 prove the integrability of the listed systems. 

 \subsection{List of systems with a higher symmetry}
 \label{com-lis1}
 The general ansatz for a 
 $\lambda_1 = \lambda_2 =2$ homogeneous evolutionary 
 system of $2^{\mbox{\scriptsize nd }}$order 
 for a scalar function 
 $u$ and a vector function $U$ takes the form
 \bea
 \label{2nd-order}
 \cases{
 u_{t_2} = a_1 u_{xx} + a_2 u^2 + a_3 \sca{U}{U}
     \vspace{1.5mm}, \cr
 U_{t_2} = a_{4} U_{xx} + a_{5} u U.
 \cr}
 \eea
 The following constraints guarantee the order to be 2 and
 the system not to be triangular:
 \[
 (a_1, a_4) \neq (0,0), \hspace{5mm} a_3 \neq 0, \hspace{5mm} 
 a_5 \neq 0.
 \]
 Similarly, the general ansatz for a 
 $3^{\mbox{\scriptsize rd }}$order system 
 takes the form
 \bea
 \label{KdV}
 \cases{
 u_{t_3} = b_1 u_{xxx} + b_2 uu_{x} + b_3 \sca{U}{U_x}
     \vspace{1.5mm}, \cr
 U_{t_3} = b_{4} U_{xxx} + b_{5} u_x U + b_{6} u U_x,
 \cr}
 \eea
 for which 
 the following constraints guarantee the order to be 3 and 
 the system not to be triangular:
 \[
 (b_1, b_4) \neq (0,0), \hspace{5mm} b_3 \neq 0, \hspace{5mm} 
 (b_5, b_6) \neq (0,0).
 \]
 However, 
 we 
 relax these constraints 
 as 
 \[
 (b_1, b_4) \neq (0,0), \hspace{5mm} 
 (b_1, b_2, b_3) \neq (0,0,0), 
 \hspace{5mm} 
 (b_4, b_5, b_6) \neq (0,0,0), 
 \]
 when 
 we consider 
 a $3^{\mbox{\scriptsize rd }}$order symmetry 
 for a $2^{\mbox{\scriptsize nd }}$order system, 
 as stated 
 in section~\ref{Comp_as}. 
 We omit 
 the general ansatz for a $4^{\mbox{\scriptsize th }}$order or a 
 $5^{\mbox{\scriptsize th }}$order 
 system here (in the 
 $\lambda_1 = \lambda_2 =2$ case) and 
 hereafter (in the 
 other weightings) 
 because of its increased length. 
 However, it is 
 available on the above-mentioned 
internet 
%Internet 
site. 
 \vspace{6pt}
 \begin{proposition}
 {\it
 No $2^{\mbox{\scriptsize nd }}$order system 
 of the form 
 $(\ref{2nd-order})$ 
 with a $3^{\mbox{\scriptsize rd }}$order symmetry 
 of the form $(\ref{KdV})$ 
 or a $4^{\mbox{\scriptsize th }}$order symmetry 
 exists. 
 }
 \end{proposition}
 \vspace{6pt}
 \begin{theorem}
 {\it
 Any $3^{\mbox{\scriptsize rd }}$order 
 system of the form $(\ref{KdV})$ with a 
 $5^{\mbox{\scriptsize th }}$order symmetry has to coincide
 with one of the following 
 four 
 systems up to a scaling of 
 $t_3,x,u,U \,($we omit the subscript of $t_3)$:
 \bea
 \label{cDS}
 \cases{
 u_t = \sca{U}{U_x}
     \vspace{1.5mm}, \cr
 U_t = U_{xxx} + u_x U + 2u U_x,
 \cr}
 \eea
 \vspace{-2.5mm}
 \bea
 \label{cKdV}
 \cases{
 u_t = u_{xxx} + 6 uu_x - 6 \sca{U}{U_x}
     \vspace{1.5mm}, \cr
 U_t = U_{xxx} + 6 u_x U + 6 u U_x,
 \cr}
 \eea
 \vspace{-2.5mm}
 \bea
 \label{cIto}
 \cases{
 u_t = u_{xxx} + 3 uu_x +3 \sca{U}{U_x}
     \vspace{1.5mm}, \cr
 U_t = u_x U + u U_x,
 \cr}
 \eea
 \vspace{-2.5mm}
 \bea
 \label{cHS}
 \cases{
 u_t = u_{xxx} + 6 uu_x -12 \sca{U}{U_x}
     \vspace{1.5mm}, \cr
 U_t = -2 U_{xxx} - 6u U_x.
 \cr}
 \eea
 }
 \end{theorem}
 \vspace{6pt}
 \noindent
 All 
 systems (\ref{cDS})--(\ref{cHS}) 
 admit the reduction $U =\vt{0}$. From this viewpoint, 
 (\ref{cDS}) is 
 a generalization of the trivial equation $u_t=0$, while 
 (\ref{cKdV})--(\ref{cHS}) are generalizations of the KdV equation. 

 \subsection{Integrability of systems (\ref{cDS})--(\ref{cHS})}
 \label{int-all1}

 \subsubsection{System (\ref{cDS})}
 \label{sub-cDS}

 System (\ref{cDS}) is a multi-component generalization of one 
 of the Drinfel'd--Sokolov systems \cite{Wilson,Drinfeld}. 
 The integrability of this system 
 has been 
 established in the literature \cite{Melnikov,Strampp1,Strampp2}. 

 \subsubsection{System (\ref{cKdV})}
 \label{sub-cKdV}

 System (\ref{cKdV}) 
 is known as a Jordan KdV system \cite{SWo,Svi0,Svi,Ad}. 
 Let us briefly summarize its integrability. 
 It is well-known 
 that the matrix KdV equation, 
 \begin{equation}
 Q_t = Q_{xxx} + 3 ( Q^2 )_x,
 \label{matrix-KdV}
 \end{equation}
 admits a Lax representation \cite{Lax,Kamijo,CaloDega,Zakh}. 
 Then, 
 system (\ref{cKdV}) is also integrable, because it is 
 obtained from (\ref{matrix-KdV}) 
 through the following reduction: 
 \[
 Q = u {\mathbf 1} + \sum_{j=1}^{N} U_j {\mathbf e}_j. 
 \]
Here ${\mathbf 1}$ is the identity matrix and 
$\{ {\mathbf e}_1, \ldots, {\mathbf e}_N \}$ are mutually 
anti-commuting matrices that satisfy the 
condition, 
 \[
 \{ {\mathbf e}_i, {\mathbf e}_j \}_{+} 
 \equiv {\mathbf e}_i {\mathbf e}_j + {\mathbf e}_j {\mathbf e}_i 
 = -2 \delta_{i j} {\mathbf 1}.
 \]

 \subsubsection{System (\ref{cIto})}
 \label{sub-cIto}

System (\ref{cIto}) is a multi-component 
generalization of the 
Zakharov--Ito 
system \cite{Zakh,Ito} 
and 
corresponds to a special case of the coupled KdV 
equations considered by Kupershmidt \cite{Kuper}. 
Introducing 
a new variable $w$ 
by 
\mbox{$w \equiv \sqrt{\sca{U}{U}}$}, 
we find 
that (\ref{cIto}) contains the original 
Zakharov--Ito system, 
 \bea
 \label{Itoeq}
 \cases{
 u_t = u_{xxx} + 3uu_x + 3 w w_x
 \vspace{1.5mm} ,\cr
 w_t = (u w)_x.
 \cr}
 \eea
Therefore, 
system (\ref{cIto}) is a triangular system that consists 
of the Zakharov--Ito system 
and the linear equation for $U$ with 
Zakharov--Ito-system-dependent 
coefficients. 

To demonstrate the 
integrability of the whole system (\ref{cIto}), 
we first summarize a 
Lax representation for the Zakharov--Ito 
system \cite{Zakh,Boiti1,Bogo}. 
We consider a 
pair of linear equations for a scalar function $\psi$,
 \bea
 \label{ItoLax}
 \cases{
 \psi_{xx} = (\z + q + \z^{-1} r) \psi
 \vspace{1.5mm} ,\cr
 \psi_t = (4\z - 2q)\psi_x +q_x \psi,
 \cr}
 \eea
where $\z$ is the spectral parameter. Then the compatibility condition 
$\psi_{xxt} = \psi_{txx}$ for (\ref{ItoLax}) 
implies 
the following 
system: 
 \[
 \cases{
 q_t = q_{xxx} -6qq_x + 4r_x
 \vspace{1.5mm} ,\cr
 r_t = -4q_x r - 2qr_x.
 \cr}
 \]
This system 
coincides with the Zakharov--Ito 
system (\ref{Itoeq}) 
through 
the 
change 
of dependent variables, 
$q=-u/2$, $r= -3w^2/16$. 
It is noteworthy that the quantity 
$1/\psi^2$ in the limit 
$\z \to 0$ obeys the same 
evolution equation as 
that for $U$, namely, $U_t = (uU)_x$. 

 Next, we fix a solution of 
 subsystem (\ref{Itoeq}) and 
 discuss solutions of the linear equation for $U$. 
For the sake of simplicity, 
we assume that $w(x,t)$ 
in the fixed solution is not a trivial function. 
Then, 
noting the relation $w_t = (uw)_x$, 
we 
obtain 
the following solution to 
the equation for $U$:
 \[
  U_j = w \cdot f_j \Bigl( \int^x w \, \d x' \Bigr), \hspace{5mm} 
  j=1, 2, \ldots, N.
 \]
Here $f_1(z), \ldots, f_N(z)$ are arbitrary functions of $z$, 
except that they must 
satisfy 
one constraint, 
$\sum_{j=1}^N \left[ f_j(z) \right]^2 =1$, 
due to the relation 
$\sca{U}{U} = w^2$. 
For the case in which 
$w(x,t)$ is identically zero, 
we mention some references in 
section~\ref{Solution-10}.

 \subsubsection{System (\ref{cHS})}
 \label{sub-cHS}

System (\ref{cHS}) is a multi-component generalization\footnote{This 
multi-component generalization 
was proposed in \cite{HiroOh}, 
but the integrability was not discussed in that paper.} 
of the two-component KdV system [(\ref{cHS}) with $N=1$] 
proposed by Hirota and Satsuma \cite{HiSa}. 
Actually, the 
Hirota--Satsuma system 
is also understood 
as an 
example of the Kac--Moody KdV systems 
studied independently 
by Drinfel'd and Sokolov \cite{Wilson,Drinfeld}. 
A Lax representation for the Hirota--Satsuma system was 
constructed 
in \cite{Dodd}\footnote{Vladimir Sokolov commented that 
the Lax representation was reported 
earlier in the Russian paper \cite{DS2}, which is 
not accessible 
to the authors.}. 
Recently, it was generalized to 
the three-component case [(\ref{cHS}) with 
$N=2$] by Wu {\it et al}.\ \cite{Wu}. 
Let us demonstrate 
that (\ref{cHS}) admits a Lax representation 
in the general 
case of $N$-component vector $U$. We consider a set of 
linear equations for two 
column-vector functions $\vt{\psi}$ and $\vt{\phi}$,
 \bea
 \nonumber 
 \cases{
 \vt{\psi}_{xx} + P \vt{\psi} + Q \vt{\phi} = \z \vt{\psi}
 \vspace{1.5mm} ,\cr
 \vt{\phi}_{xx} + P \vt{\phi} + R \vt{\psi} = -\z \vt{\phi}
 \vspace{1.5mm} ,\cr
 \vt{\psi}_{t} = 4 \z \vt{\psi}_x + 2P \vt{\psi}_x -4Q \vt{\phi}_x 
	 - P_x \vt{\psi} + 2Q_x \vt{\phi}
 \vspace{1.5mm} ,\cr
 \vt{\phi}_{t} = -4 \z \vt{\phi}_x + 2P \vt{\phi}_x -4R \vt{\psi}_x 
	 - P_x \vt{\phi} + 2R_x \vt{\psi}.
 }
 \eea
Here, $\z$ is the spectral parameter and 
$P$, $Q$ and $R$ are square matrices 
with the same dimension. 
The compatibility 
conditions $\vt{\psi}_{xxt} = \vt{\psi}_{txx}$, 
 $\vt{\phi}_{xxt} = \vt{\phi}_{txx}$ 
imply 
a system of three matrix equations,
 \bea
 \label{matrixHS}
 \cases{
 P_t = P_{xxx} + 3(P^2)_x -6(QR)_x
 \vspace{1.5mm} ,\cr
 Q_t = -2 Q_{xxx} -6Q_x P + 3[P_x,Q]
 \vspace{1.5mm} ,\cr
 R_t = -2 R_{xxx} -6R_x P + 3[P_x,R],
 \cr}
 \eea
together with three constraints,
 \[
 [P,Q] =O, \hspace{5mm} [P,R]=O, \hspace{5mm} 
 [Q,R]_x =O.
 \]
If we consider the reduction,
 \[
 P = u {\mathbf 1}, \hspace{2.5mm}
 Q = U_1 {\mathbf 1} +
	  \sum_{j=1}^{N-1} U_{j+1} {\mathbf e}_j, 
 \hspace{2.5mm}
 R = U_1 {\mathbf 1} -\sum_{j=1}^{N-1} U_{j+1} {\mathbf e}_j, 
 \hspace{2.5mm} 
 \{ {\mathbf e}_i, {\mathbf e}_j \}_{+} = -2 \delta_{i j} {\mathbf 1}, 
 \]
the three constraints are 
automatically 
satisfied 
and system (\ref{matrixHS}) is 
reduced to the multi-component 
Hirota--Satsuma system (\ref{cHS}). 
 \newpage
 \noindent
 \section{The case $\lambda_1 = \lambda_2=1$ $\;\,$-- coupled 
 Burgers, 
 pKdV, 
 mKdV equations --} 
 \label{lambda1}
 \setcounter{equation}{0}

 In this section, we 
 classify $2^{\mbox{\scriptsize nd }}$order 
 and $3^{\mbox{\scriptsize rd }}$order systems 
 in the $\lambda_1=\lambda_2 =1$ (Burgers/pKdV/mKdV weighting) case. 
 In the first part (section~\ref{com-lis2}), we present 
 complete lists of such systems 
 with a specific order symmetry. 
 In the second part (section~\ref{int-all2}), we 
 discuss the integrability 
 of the listed systems 
 and compare 
 them with 
 Foursov--Olver's 
 two-component 
 systems \cite{Foursov1,Foursov2,Foursov3} 
 through symmetrization as stated in the introduction. 

 \subsection{Lists of systems with a higher symmetry}
 \label{com-lis2}

 The general ansatz for a 
 $\lambda_1=\lambda_2 =1$ homogeneous evolutionary 
 system of $2^{\mbox{\scriptsize nd }}$order for a scalar function 
 $u$ and a vector function $U$ takes the form
 \bea
 \label{general'}
 \cases{
 u_{t_2} = a_1 u_{xx} + a_2 uu_{x} + a_3 u^3 
	 + a_4 u \sca{U}{U} +a_5 \sca{U}{U_x} 
     \vspace{1.5mm}, \cr
 U_{t_2} = a_{6} U_{xx} + a_{7} u_{x} U 
	 + a_{8} u U_x + a_{9} u^2 U + a_{10} 
	 \sca{U}{U}U.
 \cr}
 \eea
 The following constraints guarantee the order to be 2 and
 the system not to be triangular:
 \[
 (a_1, a_6) \neq (0,0), \hspace{5mm} (a_4,a_5) \neq (0,0), \hspace{5mm} 
 (a_7, a_8, a_9) \neq (0,0,0).
 \]
 Similarly, the general ansatz for a 
 $3^{\mbox{\scriptsize rd }}$order system 
 takes the form
 \bea
 \label{general}
 \cases{
 u_{t_3} = b_1 u_{xxx} + b_2 uu_{xx} + b_3 u_x^2 +b_4 u^2 u_x +b_5 u^4 
	 + b_6 u_x \sca{U}{U} 
 \cr \qquad \; \mbox{} + b_7 u \sca{U}{U_x} 
	 + b_8 \sca{U}{U_{xx}} + b_9 \sca{U_x}{U_x} 
	 + b_{10} u^2 \sca{U}{U} 
 \cr \qquad \; \mbox{} +b_{11} \sca{U}{U}^2
     \vspace{1.5mm}, \cr
 U_{t_3} = b_{12} U_{xxx} + b_{13} u_{xx} U + b_{14} u_x U_x + b_{15} u U_{xx} 
	+ b_{16} uu_x U 
 \cr \qquad \; \mbox{} 
	 + b_{17} u^2 U_x  + b_{18} \sca{U}{U}U_x + b_{19} \sca{U}{U_x}U 
	 + b_{20} u^3 U 
 \cr \qquad \; \mbox{} + b_{21} u \sca{U}{U} U,
 \cr}
 \eea
 for which 
 the following constraints guarantee the order to be 3 and
 the system not to be triangular: $(b_1, b_{12}) \neq (0,0)$ and
 at least one of $b_6, \ldots, b_{11}$ 
 and one of $b_{13}, \ldots, b_{17}, b_{20}, b_{21}$
 must not vanish. 
 However, 
 when 
 we consider 
 a $3^{\mbox{\scriptsize rd }}$order symmetry 
 for a $2^{\mbox{\scriptsize nd }}$order system, 
 we 
 relax these constraints as follows (cf.\ section~\ref{Comp_as}): 
 $(b_1, b_{12}) \neq (0,0)$ and
 at least one of $b_1, \ldots, b_{11}$ 
 and one of $b_{12}, \ldots, b_{21}$
 must not vanish. 
 \vspace{6pt}
 \begin{theorem}
 {\it
 Any $2^{\mbox{\scriptsize nd }}$order 
 system of the form $(\ref{general'})$ with a 
 $3^{\mbox{\scriptsize rd }}$order symmetry 
 of the form $(\ref{general})$ 
 has to coincide
 with one of the following three 
 systems up to a scaling of 
 $t_2,x,u,U \,($we omit the subscript of $t_2)$:
 \bea
 \label{sol25'}
 \cases{
 u_{t} = \frac{1}{3} (1+2a) (u_{xx} + 2uu_x) + \frac{4}{3} \sca{U}{U_x}
 \vspace{1.5mm} ,\cr
 U_{t} = U_{xx} + \frac{1}{3} (1-a) u_x U + uU_x + \frac{1}{12} (1-4a) u^2 U 
 \cr \qquad \mbox{} - \frac{1}{3} \sca{U}{U}U,
 \hspace{30mm} a :  {\rm arbitrary},
 \cr}
 \eea
 \vspace{-2.5mm}
 \bea
 \label{sol13'}
 \cases{
 u_{t} = u_{xx} + 2uu_x +2 \sca{U}{U_x}
 \vspace{1.5mm} ,\cr
 U_{t} = - \hf u_x U - \hf u^2 U - \hf \sca{U}{U}U,
 \cr}
 \eea
 \vspace{-2.5mm}
 \bea
 \label{sol12'}
 \cases{
 u_{t} = u_{xx} + 2uu_x + \sca{U}{U_x}
 \vspace{1.5mm} ,\cr
 U_{t} = \hf u_x U + u U_x.
 \cr}
 \eea
 }
 \end{theorem}
 \vspace{6pt}
 \begin{proposition}
 {\it
 Any $2^{\mbox{\scriptsize nd }}$order 
 system of the form $(\ref{general'})$ with a 
 $4^{\mbox{\scriptsize th }}$order symmetry has to coincide with one of 
 the 
 three systems 
 $(\ref{sol25'})$--$(\ref{sol12'})$ up to a scaling of $t_2,x,u,U$.
 }
 \end{proposition}
 \vspace{6pt}
 \noindent
 All 
 systems (\ref{sol25'})--(\ref{sol12'}) 
 admit the reduction $U ={\mathbf 0}$. From this viewpoint, 
 (\ref{sol25'})--(\ref{sol12'}) are considered as generalizations of 
 the Burgers equation. 
 \vspace{6pt}
 \begin{theorem}
 {\it
 Any $3^{\mbox{\scriptsize rd }}$order 
 system of the form $(\ref{general})$ with a 
 $5^{\mbox{\scriptsize th }}$order symmetry has to coincide
 with one of the following $25$ systems up to a scaling of 
 $t_3,x,u,U \,($we omit the subscript of $t_3)$:
 \bea
 \label{sol25}
 \cases{
 u_t = a (u_{xxx} + 3 uu_{xx} + 3 u_x^2 +3 u^2 u_x) + u_x \sca{U}{U} 
    +2u \sca{U}{U_x} 
 \cr \qquad \mbox{}
	 + 2\sca{U}{U_{xx}} + 2\sca{U_x}{U_x} 
     \vspace{1.5mm}, \cr
 U_t = U_{xxx} + \hf (1-a) u_{xx} U + \frac{3}{2} u_x U_x 
	+ \frac{3}{2}u U_{xx} +\frac{3}{4} (1-2a) uu_x U 
 \cr \qquad \mbox{} + \frac{3}{4} u^2 U_x 
	  - \sca{U}{U_x}U +\frac{1}{8}(1-4a) u^3 U - \hf u \sca{U}{U} U,
 \cr \hspace{87mm} a :  {\rm arbitrary},
 \cr}
 \eea
 \vspace{-2.5mm}
 \bea
 \label{sol13}
 \cases{
 u_t = u_{xxx} + 3uu_{xx} + 3u_x^2 +3 u^2 u_x + u_x \sca{U}{U} 
    +2u \sca{U}{U_x} 
 \cr \qquad \mbox{}
	 + 2\sca{U}{U_{xx}} + 2\sca{U_x}{U_x} 
     \vspace{1.5mm}, \cr
 U_t = -\hf u_{xx} U -\frac{3}{2} uu_x U - \sca{U}{U_x}U -\hf u^3 U 
       - \hf u \sca{U}{U} U,
 \cr}
 \eea
 \vspace{-2.5mm}
 \bea
 \label{sol12}
 \cases{
 u_t = u_{xxx} + 3uu_{xx} + 3u_x^2 +3 u^2 u_x + u_x \sca{U}{U} 
    +2u \sca{U}{U_x} 
 \cr \qquad \mbox{}
	 + \sca{U}{U_{xx}} + \sca{U_x}{U_x} 
     \vspace{1.5mm}, \cr
 U_t = \hf u_{xx} U + u_x U_x + uu_x U +u^2 U_x + \hf \sca{U}{U} U_x 
	 + \hf \sca{U}{U_x}U,
 \cr}
 \eea
 \vspace{-2.5mm}
 \bea
 \label{sol11}
 \cases{
 u_t = u_{xxx} + 3uu_{xx} + 3u_x^2 +3 u^2 u_x + u_x \sca{U}{U} 
    +2u \sca{U}{U_x} 
 \cr \qquad \mbox{}
	 + \sca{U}{U_{xx}} + \sca{U_x}{U_x} 
     \vspace{1.5mm}, \cr
 U_t = \hf u_{xx} U + u_x U_x + uu_x U +u^2 U_x + \sca{U}{U} U_x,
 \cr}
 \eea
 \vspace{-2.5mm}
 \bea
 \label{sol1}
 \cases{
 u_t = 3u_x \sca{U}{U} +3 \sca{U}{U_{xx}} 
       -3 \sca{U}{U}^2  \vspace{1.5mm}, \cr
 U_t = U_{xxx} + u_{xx} U + u_x U_x -3\sca{U}{U_x}U,
 \cr}
 \eea
 \vspace{-2.5mm}
 \bea
 \label{sol2}
 \cases{
 u_t = 2u_x \sca{U}{U} +2 \sca{U}{U_{xx}}- \sca{U_x}{U_x}
       -2 \sca{U}{U}^2  \vspace{1.5mm}, \cr
 U_t = U_{xxx} + u_{xx} U + 2u_x U_x -2\sca{U}{U}U_x-2\sca{U}{U_x}U,
 \cr}
 \eea
 \vspace{-2.5mm}
 \bea
 \label{sol3}
 \cases{
 u_t = u_x \sca{U}{U} +2u \sca{U}{U_x} 
       + \sca{U}{U_{xx}} +  \sca{U_x}{U_x}  \vspace{1.5mm}, \cr
 U_t = U_{xxx} + u_{xx} U + u_x U_x -2 uu_x U -u^2 U_x 
       + \sca{U}{U}U_x - \sca{U}{U_x}U,
 \cr}
 \eea
 \vspace{-2.5mm}
 \bea
 \label{sol9}
 \cases{
 u_t = u_{xxx} + \frac{3}{2} u_x^2 + \frac{3}{2}\sca{U_x}{U_x} 
     \vspace{1.5mm}, \cr
 U_t = u_x U_x,
 \cr}
 \eea
 \vspace{-2.5mm}
 \bea
 \label{sol10}
 \cases{
 u_t = u_{xxx} + 3 u_x^2 + 2a u_x \sca{U}{U} 
    + a \sca{U}{U_{xx}}+a \sca{U_x}{U_x} +b \sca{U}{U}^2  
     \vspace{1.5mm}, \cr
 U_t = u_{xx} U + 2 u_x U_x + a \sca{U}{U}U_x + a \sca{U}{U_x} U, 
 \quad (a,b) \neq (0,0),
 \cr}
 \eea
 \vspace{-2.5mm}
 \bea
 \label{sol14}
 \cases{
 u_t = u_{xxx} + 3 u_x^2 - 3\sca{U_x}{U_x} 
     \vspace{1.5mm}, \cr
 U_t = U_{xxx} + 6 u_x U_x,
 \cr}
 \eea
 \vspace{-2.5mm}
 \bea
 \label{sol19}
 \cases{
 u_t = u_{xxx} + 3 u_x^2 + u_x \sca{U}{U} + \sca{U}{U_{xx}}
     \vspace{1.5mm}, \cr
 U_t = U_{xxx} + 3 u_{xx} U + 3 u_x U_x + \sca{U}{U_x}U,
 \cr}
 \eea
 \vspace{-2.5mm}
 \bea
 \label{sol18}
 \cases{
 u_t = u_{xxx} + 3 u_x^2 + 2 u_x \sca{U}{U} + \sca{U}{U_{xx}}
 + \hf \sca{U_x}{U_x} 
     \vspace{1.5mm}, \cr
 U_t = U_{xxx} + 6 u_{xx} U + 6u_x U_x + 2 \sca{U}{U_x}U,
 \cr}
 \eea
 \vspace{-2.5mm}
 \bea
 \label{sol20}
 \cases{
 u_t = u_{xxx} +3 u_x^2 + 4 u_x \sca{U}{U} 
       + 2 \sca{U}{U_{xx}} + \sca{U_x}{U_x} + \frac{2}{3} \sca{U}{U} ^2
     \vspace{1.5mm}, \cr
 U_t = -2 U_{xxx} - 6 u_{xx} U - 6u_x U_x -4 \sca{U}{U_x} U,
 \cr}
 \eea
 \vspace{-2.5mm}
 \bea
 \label{sol21}
 \cases{
 u_t = u_{xxx} + u_x^2 
       -12 \sca{U}{U_{xx}} + 12 \sca{U_x}{U_x} -4 \sca{U}{U} ^2
     \vspace{1.5mm}, \cr
 U_t = 4 U_{xxx} + u_{xx} U +2 u_x U_x +4 \sca{U}{U} U_x +4 \sca{U}{U_x}U,
 \cr}
 \eea
 \vspace{-2.5mm}
 \bea
 \label{sol4}
 \cases{
 u_t = u_{xxx} - \frac{3}{2} u^2 u_x + \frac{3}{2} u_x \sca{U}{U} 
    +u \sca{U}{U_x} + \sca{U}{U_{xx}}  + \sca{U_x}{U_x} 
 \vspace{1.5mm}, \cr
 U_t = -u_x U_x - \hf u^2 U_x + \frac{3}{2} \sca{U}{U}U_x,
 \cr}
 \eea
 \vspace{-2.5mm}
 \bea
 \label{sol5}
 \cases{
 u_t = u_{xxx} - \frac{3}{2} u^2 u_x + \frac{3}{2} u_x \sca{U}{U} 
    +u \sca{U}{U_x} + \sca{U}{U_{xx}}  + \sca{U_x}{U_x} \vspace{1.5mm}, \cr
 U_t = -u_x U_x - \hf u^2 U_x + \hf \sca{U}{U}U_x + \sca{U}{U_x}U,
 \cr}
 \eea
 \vspace{-2.5mm}
 \bea
 \label{sol6}
 \cases{
 u_t = u_{xxx} - \frac{3}{2} u^2 u_x + \frac{1}{2} u_x \sca{U}{U} 
    +u \sca{U}{U_x} + \sca{U}{U_{xx}}  + \sca{U_x}{U_x} 
 \vspace{1.5mm}, \cr
 U_t = u_{xx} U + u_x U_x -u u_x U - \hf u^2 U_x 
       + \hf \sca{U}{U}U_x + \sca{U}{U_x} U,
 \cr}
 \eea
 \vspace{-2.5mm}
 \bea
 \label{sol7}
 \cases{
 u_t = u_{xxx} - \frac{3}{2} u^2 u_x + \frac{3}{2} u_x \sca{U}{U} 
    +u \sca{U}{U_x} + \sca{U}{U_{xx}} + \sca{U_x}{U_x} 
 \cr \qquad \mbox{}+ \hf \sca{U}{U}^2 
     \vspace{1.5mm}, \cr
 U_t = - u_x U_x - \hf u^2 U_x 
       - \hf \sca{U}{U}U_x + \hf u \sca{U}{U} U,
 \cr}
 \eea
 \vspace{-2.5mm}
 \bea
 \label{sol8}
 \cases{
 u_t = u_{xxx} - \frac{3}{2} u^2 u_x + u_x \sca{U}{U} 
    +u \sca{U}{U_x} + \sca{U}{U_{xx}} + \sca{U_x}{U_x} 
 \cr \qquad \mbox{}-\frac{1}{4} u^2 \sca{U}{U} + \frac{1}{4} \sca{U}{U}^2 
     \vspace{1.5mm}, \cr
 U_t = \hf u_{xx} U + \hf \sca{U}{U_x}U -\frac{1}{4} u^3 U 
       + \frac{1}{4}u \sca{U}{U} U,
 \cr}
 \eea
 \vspace{-2.5mm}
 \bea
 \label{sol15}
 \cases{
 u_t = u_{xxx} + u^2 u_x + u_x \sca{U}{U}
     \vspace{1.5mm}, \cr
 U_t = U_{xxx} + u^2 U_x + \sca{U}{U} U_x,
 \cr}
 \eea
 \vspace{-2.5mm}
 \bea
 \label{sol17}
 \cases{
 u_t = u_{xxx} + 2 u^2 u_x + u_x \sca{U}{U} + u \sca{U}{U_x}
     \vspace{1.5mm}, \cr
 U_t = U_{xxx} + uu_x U + u^2 U_x + \sca{U}{U} U_x + \sca{U}{U_x}U,
 \cr}
 \eea
 \vspace{-2.5mm}
 \bea
 \label{sol16}
 \cases{
 u_t = u_{xxx} -6 u^2 u_x + 6u_x \sca{U}{U} +12u \sca{U}{U_x} 
     \vspace{1.5mm}, \cr
 U_t = U_{xxx} - 12 uu_x U - 6u^2 U_x + 6 \sca{U}{U}U_x,
 \cr}
 \eea
 \vspace{-2.5mm}
 \bea
 \label{sol22}
 \cases{
 u_t = u_{xxx} -6 u^2 u_x + u_x \sca{U}{U} + 2u \sca{U}{U_x} 
	 + \sca{U}{U_{xx}}+ \sca{U_x}{U_x} 
     \vspace{1.5mm}, \cr
 U_t = U_{xxx} + 3 u_{xx} U + 3 u_x U_x - 6 uu_x U -3 u^2 U_x 
	 + \sca{U}{U} U_x 
 \cr \qquad \mbox{} + 3 \sca{U}{U_x}U,
 \cr}
 \eea
 \vspace{-2.5mm}
 \bea
 \label{sol23}
 \cases{
 u_t = u_{xxx} -6 u^2 u_x + u_x \sca{U}{U} + 2u \sca{U}{U_x} 
	 + \sca{U}{U_{xx}}+ \sca{U_x}{U_x} 
     \vspace{1.5mm}, \cr
 U_t = U_{xxx} + 6 u_{xx} U + 6u_x U_x -12 uu_x U -6 u^2 U_x 
	 + \sca{U}{U} U_x 
 \cr \qquad \mbox{} + 4 \sca{U}{U_x}U,
 \cr}
 \eea
 \vspace{-2.5mm}
 \bea
 \label{sol24}
 \cases{
 u_t = u_{xxx} -6 u^2 u_x + u_x \sca{U}{U} + 2u \sca{U}{U_x} 
       + \sca{U}{U_{xx}} + \sca{U_x}{U_x} 
     \vspace{1.5mm}, \cr
 U_t = -2 U_{xxx} - 6 u_{xx} U - 6u_x U_x + 12uu_x U + 6u^2 U_x 
	 + \sca{U}{U} U_x 
 \cr \qquad \mbox{} -2 \sca{U}{U_x}U.
 \cr}
 \eea
 }
 \end{theorem}
 \vspace{6pt}
 \noindent
 All 
 systems (\ref{sol25})--(\ref{sol24}) 
 admit the reduction $U ={\mathbf 0}$. From this viewpoint, 
 (\ref{sol25})--(\ref{sol11}), 
 (\ref{sol1})--(\ref{sol3}), (\ref{sol9})--(\ref{sol21}) 
 and (\ref{sol4})--(\ref{sol24}) 
 are 
 generalizations of 
 the $3^{\mbox{\scriptsize rd }}$order 
 Burgers equation, 
 the trivial equation 
 $u_t=0$, the 
 pKdV equation 
 and the mKdV equation, respectively. 
 Actually, 
 (\ref{sol25})--(\ref{sol12}) are 
 the $3^{\mbox{\scriptsize rd }}$order 
 symmetries of the 
 $2^{\mbox{\scriptsize nd }}$order systems 
 (\ref{sol25'})--(\ref{sol12'}), respectively. 

 \subsection{Integrability of systems (\ref{sol25'})--(\ref{sol24})}
 \label{int-all2}

 \subsubsection{Systems (\ref{sol25'}) and (\ref{sol25})}
 \label{Solution-25}

 We investigate 
 the $2^{\mbox{\scriptsize nd }}$order system (\ref{sol25'}) 
 and its $3^{\mbox{\scriptsize rd }}$order symmetry 
 (\ref{sol25}) together. 
We note that the 
linear 
term $u_{xx}$ in (\ref{sol25'}) 
 vanishes iff 
$a=-{1 \over 2}$, 
while the term $u_{xxx}$ in (\ref{sol25}) 
 vanishes iff $a=0$. 
 This indicates 
 that the case distinctions of 
$a \neq -{1 \over 2}$ or $a = -{1 \over 2}$ 
and $a \neq 0$ or $a = 0$ are not very essential for the whole 
 hierarchy of systems 
 starting from 
 (\ref{sol25'}). 
 As an extension of the Hopf--Cole transformation, we consider 
 the change of variables, 
  \bea
 \nonumber \cases{
  w= \e^{\int^x u\hspace{1pt}\d x'}\vspace{1.5mm}\hspace{-1pt}, \cr
  W= U \e^{\hf \int^x u\hspace{1pt}\d x'}\hspace{-1pt}. 
 \cr
}
  \eea
Then we can triangular 
%almost 
linearize 
%(cf.\ \cite{Bakirov}) 
 (\ref{sol25'}) and (\ref{sol25}) 
 simultaneously as
  \bea
  \label{lin1}
  \cases{
  w_t = \frac{1}{3} (1+2a) w_{xx} + \frac{2}{3} \sca{W}{W}
  \vspace{1.5mm}, \cr
  W_t = W_{xx},
  \cr }
  \eea
 and
  \bea
  \label{lin2}
  \cases{
  w_t = a w_{xxx} + \sca{W}{W}_x
  \vspace{1.5mm}, \cr
  W_t = W_{xxx}.
  \cr }
  \eea
For some values of $a$, 
we can solve these systems easily. 
When $a=1$, 
we can 
fully linearize 
 systems (\ref{lin1}) and (\ref{lin2}) 
 through defining new variables $V$ and $v$ by $W= V_x$, 
 $w+ \frac{1}{3} \sca{V}{V}=v$ (see \cite{Svi2}). 
When $a=-{1 \over 2}$, 
 we integrate the equation for $w$ in (\ref{lin1}) to obtain 
 \[
 w(x,t) = w(x,0) +  \frac{2}{3} 
	 \int^t_0 \sca{W(x,t')}{W(x,t')} \, \d t'.
 \]
Similarly, 
when $a=0$, we can integrate the equation for $w$ in (\ref{lin2}). 
We mention that 
Beukers, Sanders and Wang \cite{Beukers,Beukers2} 
studied higher symmetries 
of the triangular 
%almost 
linear systems 
(\ref{lin1}) and (\ref{lin2}) 
in the case of scalar $W$. 
\vspace{12pt} 
\\
{\it Symmetrization}.  
We discuss symmetrization for the 
$2^{\mbox{\scriptsize nd }}$order system (\ref{sol25'}), 
since it is more fundamental than its 
$3^{\mbox{\scriptsize rd }}$order symmetry 
(\ref{sol25}). To 
identify (\ref{sol25'}) as a multi-component generalization of 
a system in \cite{Foursov1,Foursov3}, we assume the condition 
$a \neq -{1 \over 2}$ 
and rescale variables as 
 \[
 \partial_t = \frac{1}{3}(1+2a) \partial_s, \hspace{5mm} 
 u = 4u', \hspace{5mm} U= \sqrt{6} U'. 
 \]
In addition, 
we introduce a new parameter $\a$ by 
the relation 
 \[
 \frac{3}{1+2a}  = 1 - 2\a, 
 \]
where $\a \neq {1 \over 2}$. 
Then 
(\ref{sol25'}) is rewritten as 
 \bea
 \cases{
 u'_s = u'_{xx} + 8 u' u'_x + (2-4\a) \sca{U'}{U'_x}
 \vspace{1.5mm} ,\cr
 U'_s = (1-2\a) U'_{xx} -4 \a u'_x U' + (4-8\a) u' U'_x - 
 (4+8\a) u'^{\, 2} U' 
 \cr \qquad \mbox{} 
 - (2-4\a) \sca{U'}{U'}U'. 
 \cr}
 \nonumber
 \eea
 In the case where $U'$ is scalar, this system is 
 identical to (3.7) in \cite{Foursov1}. 
In that case, considering the linear change of variables 
 \[
 u' = q+r, \quad U' = q-r,
 \]
we obtain a system of two 
symmetrically coupled Burgers equations, which coincides with 
(3.6) in \cite{Foursov1} or (3.7) in \cite{Foursov3}. 
We note that system (\ref{sol25'}) with 
$a=-{1 \over 2}$ 
does not 
have any counterpart in \cite{Foursov1,Foursov3}, 
because of its degeneracy of the linear part 
(cf.\ the introduction). 

 \subsubsection{Systems (\ref{sol13'}) and (\ref{sol13})}
 \label{Solution-13}

We investigate 
the $2^{\mbox{\scriptsize nd }}$order system (\ref{sol13'}) 
and its $3^{\mbox{\scriptsize rd }}$order symmetry 
(\ref{sol13}) together. 
Here, we note that (\ref{sol13'}) and (\ref{sol13}) 
are obtained from (\ref{sol25'}) and (\ref{sol25}), respectively, by 
rescaling $t$, $U$ appropriately 
and taking the limit $a \to \infty$. 
Then, through the same 
change of variables as in section~\ref{Solution-25}, 
\bea
 \nonumber \cases{
  w= \e^{\int^x u\hspace{1pt}\d x'}
\vspace{1.5mm}\hspace{-1pt}, \cr
  W= U \e^{\hf \int^x u\hspace{1pt}\d x'}\hspace{-1pt},
  \cr}
  \eea
we can transform systems (\ref{sol13'}) and (\ref{sol13}) to 
  \bea
  \label{lin1'}
  \cases{
  w_t = w_{xx} + \sca{W}{W}
  \vspace{1.5mm}, \cr
  W_t = \vt{0},
  \cr }
  \eea
 and
  \bea
  \label{lin2'}
  \cases{
  w_t = w_{xxx} + \sca{W}{W}_x
  \vspace{1.5mm}, \cr
  W_t = \vt{0}.
  \cr }
  \eea
Moreover, introducing 
a function $g(x)$ 
such that 
$g''(x) = \sca{W}{W}$, 
we can linearize 
the equations for $w$ in (\ref{lin1'}) and (\ref{lin2'}) 
with respect to 
the variable $w+ g(x)$. 
 \vspace{12pt}
 \\
 {\it Symmetrization}. 
 We discuss symmetrization for the 
 $2^{\mbox{\scriptsize nd }}$order system (\ref{sol13'}). 
To identify (\ref{sol13'}) as a multi-component generalization 
of a system 
in \cite{Foursov1,Foursov3}, we rescale 
the dependent 
variables as 
 \[
 u = 4u', \hspace{5mm} U= 2 \sqrt{5} U'. 
 \]
Then 
(\ref{sol13'}) is rewritten as 
 \bea
 \nonumber
 \cases{
 u'_t = u'_{xx} + 8 u' u'_x + 10 \sca{U'}{U'_x}
 \vspace{1.5mm} ,\cr
 U'_t = -2 u'_x U' -8 u'^{\, 2} U' - 10 \sca{U'}{U'}U'. 
 \cr}
 \eea
In the case where $U'$ is scalar, this system 
should coincide with 
(3.10) in \cite{Foursov1}, 
if it were written correctly. 
Unfortunately, in \cite{Foursov1}, 
Foursov made a mistake in deriving 
the equation for $z$ in (3.10) 
from (3.9). 
It should be corrected as 
$z_t = -2w_x z -8w^2 z -10z^3$. 
If we consider the linear change of variables
 \[
 u' = q+r, \quad U' = q-r,
 \]
we obtain a system of two symmetrically coupled Burgers equations, 
which coincides with 
(3.9) in \cite{Foursov1} or (3.6) in \cite{Foursov3}. 

 \subsubsection{Systems (\ref{sol12'}) and (\ref{sol12})}
 \label{Solution-12}

 We concentrate our attention on 
 the $2^{\mbox{\scriptsize nd }}$order system (\ref{sol12'}), 
 and do not study 
 its $3^{\mbox{\scriptsize rd }}$order 
 symmetry 
 (\ref{sol12}). 
 Defining a new variable $w$ by $w \equiv \hf \sca{U}{U}$, we 
 find that (\ref{sol12'}) contains a two-component Burgers system, 
  \bea
  \label{cBurgers}
  \cases{
  u_{t} = u_{xx} + 2uu_x + w_x
  \vspace{1.5mm} ,\cr
  w_{t} = (uw)_x.
  \cr}
  \eea
 Therefore, 
 system (\ref{sol12'}) is a triangular system that consists 
 of the 
 Burgers system (\ref{cBurgers}) 
 and the linear 
 equation for $U$ with 
 Burgers-system-dependent coefficients. 
 We mention that in the long-wave limit 
 (disappearance of 
 $u_{xx}$), (\ref{cBurgers}) 
 reduces to the Leroux system 
 and 
 that (\ref{cBurgers}) can be rewritten as a 
 non-evolutionary scalar equation in (at least) two different ways. 
 The symmetry 
 integrability of (\ref{cBurgers}) 
 as well as the existence of a recursion operator 
 has already been demonstrated \cite{Foursov1,Ma}. 
 Nevertheless, we could find 
 neither a linearizing transformation nor a
{\it true} \/Lax representation for (\ref{cBurgers}). 
 In what follows, we discuss travelling-wave solutions of 
 (\ref{cBurgers}), 
 which are expected to give useful information 
 on its properties. 

 Substituting the travelling-wave form
 \[
 u(x,t) = f(z)-a, \hspace{5mm} w(x,t)= g(z),
 \hspace{5mm} z = x -a t,
 \]
 into 
 (\ref{cBurgers}), we get 
 a system of two ordinary differential equations. 
 Integrating it once, we obtain 
  \bea
  \label{ODEs}
  \cases{
  f' + f^2 -af + g + b =0
  \vspace{1.5mm} ,\cr
  fg + c=0.
  \cr}
  \eea
 Here, $b$ and $c$ are integration constants that 
 are determined 
 from the boundary 
 conditions for 
 $u$ and $w$. 
 Plunging 
$g= -c/f$ 
into the first equation 
 in 
 (\ref{ODEs}), we obtain the 
 ordinary differential equation for $f$, 
 \begin{equation}
 \frac{\d f}{\d z} 
 = - \frac{f^3-af^2 + bf -c}{f}.
 \label{ODE}
 \end{equation}
 For the sake of 
 simplicity, we assume that $f^3-af^2 + bf -c$ can be 
 factorized into the product 
 $(f-\a_1) (f-\a_2) (f-\a_3)$ 
 with three distinct real roots 
 $\a_1, \a_2, \a_3$. Thus we have 
  \[
  a = \a_1 + \a_2 + \a_3, \hspace{5mm}
  b = \a_1 \a_2 + \a_2 \a_3 + \a_3 \a_1, \hspace{5mm}
  c = \a_1 \a_2 \a_3.
  \]
 Furthermore, we assume the conditions $\a_j \neq 0 \; (j=1,2,3)$ 
 to obtain nontrivial solutions of 
 (\ref{cBurgers}). 
 Indeed, if $\a_j=0$, then 
 $c=0$ and we obtain from (\ref{ODEs}) either 
 a trivial solution or 
 a solution of the scalar Burgers equation. 
 Noting the identity
 \bea
 \frac{f}{(f-\a_1)(f-\a_2)(f-\a_3)} 
 & \hspace{-1.5mm} = \hspace{-1.5mm} & \frac{\a_1}{(\a_1-\a_2)(\a_1-\a_3)} 
  \left( \frac{1}{f-\a_1} - \frac{1}{f-\a_3} \right)
 \nonumber \\
 && 
 \mbox{} + \frac{\a_2}{(\a_2-\a_1)(\a_2-\a_3)}
  \left( \frac{1}{f-\a_2} - \frac{1}{f-\a_3} \right),
 \nonumber
 \eea
 we can integrate (\ref{ODE}) to 
 obtain
 \bea
 \label{f_eq}
 \left( 1+ \frac{\a_3-\a_1}{f-\a_3} 
 \right)^{\frac{\a_1}{(\a_1-\a_2)(\a_1-\a_3)}}
 \left( 1+ \frac{\a_3-\a_2}{f-\a_3} 
 \right)^{\frac{\a_2}{(\a_2-\a_1)(\a_2-\a_3)}} 
 = d \e^{-z},
 \eea
 where $d$ is a constant. 
 Now, it is clear that the functional form of $1/(f-\a_3)$ 
 depends on the ratio 
 of two powers on the left-hand side, 
 $(\a_3^{-1} - \a_2^{-1})/(\a_1^{-1} - \a_3^{-1})$. 
 Let us consider the simplest case 
 in which this ratio is unity, i.e., 
 \[
 \frac{1}{\a_3} = \frac{1}{2} \left( \frac{1}{\a_1} + \frac{1}{\a_2} \right).
 \]
 In this case, we have $\a_1+\a_2 \neq 0$ 
 for the existence of $\a_3$. 
 Then we can solve (\ref{f_eq}) explicitly 
 for $1/(f-\a_3)$: 
 \bea
 \frac{1}{f-\a_3} 
 = - \frac{
	 \a_1 + \a_2 + \frac{(\a_1 + \a_2 )^2}{\a_1-\a_2}
 \sqrt{1+ \exp \left[ 
 -\frac{(\a_1 -\a_2)^2}{\a_1 + \a_2}(z-z_0) \right] }}{2 \a_1 \a_2}.
 \label{f_form}
 \eea
 Here, $z_0$ is the constant given 
 by 
 \[
 \e^{z_0}
 \equiv  d \left[\frac{-4\a_1 \a_2}{(\a_1+\a_2)^2} 
	 \right]^{\frac{\a_1+\a_2}{(\a_1-\a_2)^2}}.
 \]
 Using the relation $\a_3 = 2 \a_1 \a_2 / (\a_1 + \a_2)$, 
 we can rewrite (\ref{f_form}) as 
 \begin{equation}
 f(z) =  \frac{2 \a_1 \a_2}{\a_1 + \a_2 + \frac{\a_1-\a_2}{
 \sqrt{1+ \exp \left[ -\frac{(\a_1 -\a_2)^2}{\a_1 + \a_2}(z-z_0) \right]} } }.
 \label{travel}
 \end{equation}
 In order 
 for the function $f(z)$ to be non-singular, we should 
 assume the condition $\a_1 (\a_1 + \a_2) >0$. 
 To summarize, we have obtained in the simplest case 
 a travelling-wave solution of 
 (\ref{cBurgers}) given by 
 \[
 u(x,t) = f(x-at)-a, \hspace{5mm} w(x,t)
	 = -\frac{c}{f(x-at)},
 \]
 with (\ref{travel}), 
 $a = \a_1 + \a_2 + 2 \a_1 \a_2/(\a_1+\a_2)$ 
 and $c = 2 (\a_1 \a_2)^2 /(\a_1+\a_2)$. 

 When a nontrivial solution, like the above, of 
 subsystem (\ref{cBurgers}) is given, we can 
 solve the remaining equation for $U$ 
 in the original system 
 (\ref{sol12'}), $\bigl( U_j^2 \bigr)_t = \bigl( u U_j^2 \bigr)_x$, in the 
 same way as in section~\ref{sub-cIto}. 
 \vspace{12pt}
 \\
 {\it Symmetrization}. 
 We discuss symmetrization for the $2^{\mbox{\scriptsize nd }}$order 
 system (\ref{sol12'}). 
 With the rescaling of dependent variables
 \[
 u = 2u', \hspace{5mm} U= \sqrt{6} U',
 \]
 (\ref{sol12'}) is rewritten as 
  \bea
  \nonumber
  \cases{
  u'_{t} = u'_{xx} + 4u' u'_x + 3\sca{U'}{U'_x}
  \vspace{1.5mm} ,\cr
  U'_{t} = u'_x U' + 2u' U'_x.
  \cr}
  \eea
 In the case where $U'$ is scalar, this system 
 is identical to (3.4) in \cite{Foursov1}. If we 
 consider the linear change of variables 
 \[
 u' = q+r, \quad U' = q-r,
 \]
 we obtain a system of 
 two symmetrically coupled Burgers equations, 
 which coincides with 
 (3.3) in \cite{Foursov1} or (3.5) in \cite{Foursov3}. 

 \subsubsection{System (\ref{sol11})}
 \label{Solution-11}

 In the case where $U$ is scalar, 
 system (\ref{sol11}) coincides with system (\ref{sol12}). 
 However, unlike (\ref{sol12}), 
 (\ref{sol11}) in the general $N$ case 
 does not possess a $2^{\mbox{\scriptsize nd }}$order symmetry 
 of the form (\ref{general'}). 
 System (\ref{sol11}) contains the 
 $3^{\mbox{\scriptsize rd }}$order symmetry of the 
 two-component Burgers system 
 (\ref{cBurgers}), where $w$ is again given by 
 $w = \hf \sca{U}{U}$. 
 Therefore, in order to demonstrate the integrability of (\ref{sol11}), 
 we first of all need to know either a linearizing transformation 
 or a Lax representation for (\ref{cBurgers}). 
 This remains as an open question. 
 \vspace{12pt}
 \\
 {\it Symmetrization}. 
 Through symmetrization of (\ref{sol11}) in the case of 
 scalar $U$, 
 we just 
 obtain 
 the $3^{\mbox{\scriptsize rd }}$order symmetry of the two 
 symmetrically coupled Burgers 
 equations in section~\ref{Solution-12}. 

 \subsubsection{System (\ref{sol1})}
 \label{Solution-1}
 System (\ref{sol1}) is 
 connected with system (\ref{sol3}) 
 through a Miura-type transformation. 
 We 
 discuss the integrability of these two systems 
 in section~\ref{Solution-3}. 
 \vspace{12pt}
 \\
 {\it Symmetrization}. 
 In the case where $U$ is 
 scalar, 
 we consider 
 the linear change of variables
 \[
 u = q+r, \quad U = 
	 \sqrt{\a} (q-r).
 \]
 Here, $\a$ is a nonzero constant. 
 Then we can rewrite
 (\ref{sol1}) as a system of two 
 symmetrically coupled 
 equations, 
 \bea
 \nonumber 
 \cases{
 q_t = \hf q_{xxx} -\hf r_{xxx} + \hf (1+ 3\a)(q-r) q_{xx} 
	 + \hf (1-3\a)(q-r) r_{xx}
 \cr \qquad \mbox{}
	 + \hf q_x^2 - \hf r_x^2 +3\a (q-r)^2 r_x 
	 - \frac{3}{2} \a^2 (q-r)^4
     \vspace{1.5mm}, \cr
 r_t = -\hf q_{xxx} +\hf r_{xxx} - \hf (1- 3\a)(q-r) q_{xx} 
	 - \hf (1+3\a)(q-r) r_{xx}
 \cr \qquad \mbox{}
	 - \hf q_x^2 + \hf r_x^2 +3\a (q-r)^2 q_x 
	 - \frac{3}{2} \a^2 (q-r)^4.
 \cr}
 \eea
 This system does not belong to the class 
 of 
 systems studied in \cite{Foursov2,Foursov3}, because 
 of its degeneracy of the linear part. 

 \subsubsection{System (\ref{sol2})}
 \label{Solution-2}

 For 
 system (\ref{sol2}), we 
 have the relation 
 $(u_x - \sca{U}{U})_t =0$. 
 Thus, we can set 
 \begin{equation}
 u_x - \sca{U}{U} \equiv \phi (x),
 \label{constr}
 \end{equation}
 where the function $\phi(x)$ 
 does not depend on 
 $t$. 
 Then the equation for $U$ is rewritten in terms of 
 $\phi(x)$ as 
 \begin{equation}
 U_t = U_{xxx} + 2\phi U_x + \phi_x U.
 \label{U-line}
 \end{equation}
 The 
 solutions of 
 (\ref{U-line}) are given by 
 \[
 U(x,t) = \int \d \lambda \, 
 \e^{\lambda t} \Psi (x; \lambda), 
 \]
 where $\Psi(x;\lambda)$ is a solution of the ordinary differential equation
 \begin{equation}
 \Psi_{xxx} + 2 \phi \Psi_x + \phi_x \Psi = \lambda \Psi.
 \label{KKscat}
 \end{equation}
 Once we obtain $\phi(x)$ and $U(x,t)$, 
 we can determine 
 $u(x,t)$ 
 by using (\ref{constr}). 
 The vector equation (\ref{KKscat}) 
 is of the same form as the 
 scattering problem 
associated with 
the Kaup--Kupershmidt equation \cite{KK,Gibbons}. 
 We can 
 see 
 that this is not a coincidence 
 through investigation of the $5^{\mbox{\scriptsize th }}$order symmetry 
 of system (\ref{sol2}). 
 Indeed, 
 the $5^{\mbox{\scriptsize th }}$order symmetry 
is 
rewritten 
 (up to a scaling of 
 $t_5$) 
 in terms of $\phi$ and $\Psi$ as 
 \bea
 \nonumber 
 \cases{
 \phi_{t_5} + \phi_{xxxxx} + 10 \phi \phi_{xxx} + 25 \phi_x \phi_{xx} 
	 + 20 \phi^2 \phi_x = 0  \vspace{1.5mm}, \cr
 \Psi_{t_5} = 9 \Psi_{xxxxx} + 30 \phi \Psi_{xxx} + 45 \phi_x \Psi_{xx} 
	  + (35\phi_{xx} + 20 \phi^2) \Psi_x 
 \cr \qquad \;\hspace{1pt} \mbox{}
	 + (10\phi_{xxx} + 20 \phi \phi_x )\Psi.
 \cr}
 \eea
 The first equation is nothing but the Kaup--Kupershmidt equation, 
 while the second equation together with (\ref{KKscat}) constitutes 
 a Lax representation for it. 
 \vspace{12pt}
 \\
 {\it Symmetrization}. 
 In the case where $U$ is 
 scalar, we consider the linear change 
 of variables
 \[
 u = q+r, \quad U = 
	 \sqrt{\a} (q-r).
 \]
 Here, $\a$ is a nonzero constant. 
 Then we can rewrite
 (\ref{sol2}) as a system of two symmetrically coupled 
 equations, 
 \bea
 \nonumber 
 \cases{
 q_t = \hf q_{xxx} -\hf r_{xxx} + (\hf+ \a)(q-r) q_{xx} + (\hf-\a)(q-r) r_{xx}
 \cr \qquad \mbox{}
	 + (1-\hf \a)q_x^2 
	 + \a q_x r_x -(1+\hf\a) r_x^2 - \a (q-r)^2 q_x 
 \cr \qquad \mbox{}
	 + 3\a (q-r)^2 r_x -\a^2 (q-r)^4
     \vspace{1.5mm}, \cr
 r_t = -\hf q_{xxx} + \hf r_{xxx} - (\hf- \a)(q-r) q_{xx} - (\hf+\a)
 (q-r) r_{xx}
 \cr \qquad \mbox{}
	 - (1+\hf \a)q_x^2 
	 + \a q_x r_x +(1 - \hf \a) r_x^2 +3\a (q-r)^2 q_x 
 \cr \qquad \mbox{}
	 - \a (q-r)^2 r_x-\a^2 (q-r)^4.
 \cr}
 \eea
 This system does not belong to the class 
 of nondegenerate 
 systems studied in \cite{Foursov2,Foursov3}. 

 \subsubsection{System (\ref{sol3})}
 \label{Solution-3}

 For system (\ref{sol3}), 
 if we define 
 new variables $w$ and $W$ by 
 \bea
 \label{Miura1}
 \cases{
 w \equiv -u_x -\hf u^2 + \hf \sca{U}{U} \vspace{1.5mm},
 \cr
 W \equiv U_x + uU,
 \cr}
 \eea
 they satisfy the following system:
 \bea
 \label{cDS2}
 \cases{
 w_t = -3\sca{W}{W_x}
 \vspace{1.5mm} ,\cr
 W_t = W_{xxx} + w_x W + 2w W_x.
 \cr}
 \eea
 This system coincides with the 
 multi-component 
 Drinfel'd--Sokolov 
 system (\ref{cDS}), up to a scaling of $W$. 
 The Miura map 
 (\ref{Miura1}) is a 
 multi-component 
generalization 
 of that for the case of 
 scalar $U$ 
 in \cite{Wilson,Drinfeld}. 
 \vspace{12pt}
 \\
 {\it Relation to system $(\ref{sol1})$}. 
 If we 
 introduce a new scalar variable $v$ by 
 \begin{equation}
 v \equiv u_x - u^2 + \sca{U}{U},
 \label{sol3-1}
 \end{equation}
 system (\ref{sol3}) is changed into the following system: 
 \bea
 \label{334}
 \cases{
 v_t = \bigl( 3 v \sca{U}{U} + 3 \sca{U}{U_{xx}} -3 \sca{U}{U}^2 \bigr)_x
 \vspace{1.5mm} ,\cr
 U_t = U_{xxx} + v_x U + v U_x -3\sca{U}{U_x} U.
 \cr}
 \eea
 Then, it 
 is straightforward to obtain (\ref{sol1}) (for $\hat{u}$ and $U$) 
 from (\ref{334}) through 
 potentiation $v=\hat{u}_x$.
 Combining (\ref{sol3-1}) and (\ref{Miura1}), 
 we obtain 
 the relation $v - 2 w = 3 u_x$, and consequently, 
 \[
 \hat{u} -2 \int^x w \,\d x'= 3u.
 \] 
 Using this relation, we can also 
 rewrite (\ref{Miura1}) as a transformation 
 between 
 system (\ref{sol1}) and the multi-component 
 Drinfel'd--Sokolov system 
 (\ref{cDS2}). 
 \vspace{12pt}
 \\
 {\it Symmetrization}. 
 In the case where $U$ is 
 scalar, 
 we consider 
 the linear change of variables
 \[
 u = q+r, \quad U = 
	 \sqrt{\a} (q-r).
 \]
 Here, $\a$ is a nonzero constant. 
 Then we can rewrite
 (\ref{sol3}) as a system of two symmetrically coupled 
 equations,
 \bea
 \label{modDS}
 \cases{
 q_t = \hf \bigl[ q_{xx} -r_{xx} + (1+ \a)(q-r) q_{x} + (1-\a)
	 (q-r) r_{x}
 \cr \qquad \mbox{}
	 - (1- \a)q^3 -(1+\a) q^2 r +(1-\a) qr^2 + (1+\a)r^3 \bigr]_x
     \vspace{1.5mm}, \cr
 r_t = \hf \bigl[-q_{xx} + r_{xx} - (1- \a)(q-r) q_{x} - (1+\a)
	 (q-r) r_{x}
 \cr \qquad \mbox{}
 + (1+\a)q^3+ (1- \a)q^2 r -(1+\a) q r^2 -(1-\a) r^3 \bigr]_x.
 \cr}
 \eea
 This system does not belong to the class of nondegenerate systems studied 
 in \cite{Foursov2,Foursov3}. 
 However, it 
 was found 
 in connection with the Kac--Moody Lie algebras and 
 written in a 
 Hamiltonian form about twenty years ago \cite{Wilson,Drinfeld}. 
 More specifically, 
 system (\ref{modDS}) with $\a=-1$ coincides 
 with (3)--(4) in \cite{Wilson} 
 for 
 the ${\rm D}_3^{(2)}$ case, up to a scaling of variables 
 (see also the generalized mKdV equation in \cite{Drinfeld} 
 for 
 the ${\rm A}_3^{(2)}$ case). 

 \subsubsection{System (\ref{sol9})}
 \label{Solution-9}

 System (\ref{sol9}) is merely 
 a 
 potential form of the multi-component 
 Zakharov--Ito system (\ref{cIto}). 
 \vspace{12pt}
 \\
 {\it Symmetrization}. 
 In the case where $U$ is 
 scalar, 
 we set 
 \[
 u = q+r, \quad U = q-r.
 \]
 Then we can rewrite (\ref{sol9}) as a system of two 
 symmetrically coupled pKdV 
 equations,
 \bea
 \nonumber
 \cases{
 q_t = \hf q_{xxx} + \hf r_{xxx} + 2q_x^2 + r_x^2 
     \vspace{1.5mm}, \cr
 r_t = \hf q_{xxx} + \hf r_{xxx} + q_x^2 + 2r_x^2.
 \cr}
 \eea
 This system is identical to (37) in \cite{Foursov2} or (3.10) 
 in \cite{Foursov3}. 

 \subsubsection{System (\ref{sol10})}
 \label{Solution-10}
 {\it Remark}. 
 If we consider separately the two cases 
\mbox{$a=0$} and \mbox{$a \neq 0$}, 
 we can reduce the number of parameters in system (\ref{sol10}) 
 by scaling variables. In the former case the parameter $b$ can also 
 be fixed at any nonzero value, while in the latter case only the 
 parameter $a$ can be scaled away. 
 However, this case distinction is neither 
 necessary nor essential, as is demonstrated below. 
 \vspace{12pt}
 \\
 If we define new variables $w$ and $W$ by 
 \bea
 \nonumber 
 \cases{
 w \equiv u_x + \frac{a}{2} \sca{U}{U}
 \vspace{1.5mm} ,\cr
 W \equiv \sqrt{\sca{U}{U}} U,
 \cr}
 \eea
 they satisfy the following system:
 \bea
 \label{cIto2}
 \cases{
 w_t = w_{xxx} + 6 ww_x + \bigl( b - \frac{a^2}{4} \bigr) \sca{W}{W}_x
 \vspace{1.5mm} ,\cr
 W_t = 2 (w W)_x.
 \cr}
 \eea
 Thus the essential parameter is 
$b-a^2/4$ 
rather than $a$ or $b$. 
 If $b - a^2/4 \neq 0$, 
 system (\ref{cIto2}) 
 coincides with the 
 multi-component 
 Zakharov--Ito system (\ref{cIto}), up to a scaling of variables. 
 When $b-a^2/4 =0$, 
 (\ref{cIto2}) is 
 a triangular system that consists of 
 the KdV equation and the 
 linear equation for $W$ 
 with 
 KdV-equation-dependent coefficients. This triangular system was 
 studied from a point of view of symmetries in \cite{Kamp} 
 (see also \cite{Foursov0,Fuchs,Gurses}). 
 As we have noted 
 in section~\ref{sub-cIto}, 
 we can 
 relate with $W$ the inverse square of 
 a solution of the 
 KdV linear problem. 
 \vspace{12pt}
 \\
 {\it Symmetrization}. 
 In the case where $U$ is 
 scalar, 
 we set 
 \[
 u = q+r, \quad U = q-r, \quad a = 1+2\a, \quad b = 2\beta.
 \]
 Then we can rewrite 
 (\ref{sol10}) as a system of two 
 symmetrically coupled pKdV 
 equations,
 \bea
 \label{sol1-sym}
 \cases{
 q_t = \hf q_{xxx} +\hf r_{xxx} + (1+ \a)(q-r) q_{xx} -\a(q-r) r_{xx}
	 + (3+ \a)q_x^2 
 \cr \qquad \mbox{}
	 + (2-2\a) q_x r_x +(1+\a) r_x^2 + (2+4\a) (q-r)^2 q_x 
	 +\beta (q-r)^4
     \vspace{1.5mm}, \cr
 r_t = \hf q_{xxx} +\hf r_{xxx} + \a(q-r) q_{xx} - (1+\a)(q-r) r_{xx}
	 + (1+ \a)q_x^2 
 \cr \qquad \mbox{}
	 + (2-2\a) q_x r_x +(3+\a) r_x^2 + (2+4\a) (q-r)^2 r_x 
	 +\beta (q-r)^4.
 \cr}
 \eea
 This system is identical to (12) in \cite{Foursov2}. 
 Foursov claimed in \cite{Foursov2} that 
 this system was 
 either reduced to 
 the 
 representative case $\a=0$ $(a=1)$ or decoupled 
 by a linear change of dependent 
 variables. 
 However, in fact this 
 is not true. 
 As far as we consider a linear change of 
 variables, 
 we need one more 
 representative case, 
$\a = -{1 \over 2}$ 
$(a=0)$, 
 in which (\ref{sol1-sym}) cannot be 
 decoupled. 
 Therefore, we can 
 say that (at least) one system is missing from 
 the final list 
 of Foursov--Olver given in \cite{Foursov3}. 

 \subsubsection{System (\ref{sol14})}
 \label{Solution-14}

 System (\ref{sol14}) 
 is merely a 
 potential form of the Jordan KdV 
 system (\ref{cKdV}). We 
 see in section~\ref{Solution-16} 
 that this system is 
 also 
 connected with 
 system (\ref{sol16}) through 
 a Miura-type transformation. 
 \vspace{12pt}
 \\
 {\it Symmetrization}. 
 In the case where $U$ is 
 scalar, 
 we consider the linear change of variables
 \[
 u=\frac{1}{2} (q+r), \quad U=\frac{\ii}{2} (q-r).
 \]
 Then (\ref{sol14}) is decoupled 
 into two 
 pKdV equations,
 \bea
 \nonumber 
 \cases{
 q_t = q_{xxx} + 3 q_x^2 
     \vspace{1.5mm}, \cr
 r_t = r_{xxx} + 3 r_x^2.
 \cr}
 \eea
 This corresponds to (27) in \cite{Foursov2}. 

 \subsubsection{System (\ref{sol19})}
 \label{Solution-19}

 System (\ref{sol19}) is 
 connected with 
 system (\ref{sol22}) through a Miura-type transformation. 
 We 
 discuss the integrability of these two systems 
 in section~\ref{Solution-22}. 
 \vspace{12pt}
 \\
 {\it Symmetrization}. 
 In the case where $U$ is 
 scalar, 
 we consider the linear change of variables
 \[
 u=q+r, \quad U= \sqrt{3} (q-r).
 \]
 Then we can rewrite (\ref{sol19}) as a system of 
 two symmetrically coupled pKdV equations,
 \bea
 \nonumber 
 \cases{
 q_t = q_{xxx} + 3(q-r) q_{xx}+ 3 (q_x + r_x) q_x +3 (q-r)^2 q_x
     \vspace{1.5mm}, \cr
 r_t = r_{xxx} - 3(q-r) r_{xx}+ 3 (q_x + r_x) r_x +3 (q-r)^2 r_x.
 \cr}
 \eea
 This system coincides with (34) in \cite{Foursov2} or (3.9) 
 in \cite{Foursov3}. 

 \subsubsection{System (\ref{sol18})}
 \label{Solution-18}

 System (\ref{sol18}) is 
 connected with 
 system (\ref{sol23}) through a Miura-type transformation. 
 We 
 discuss the integrability of these two systems 
 in section~\ref{Solution-23}. 
 \vspace{12pt}
 \\
 {\it Symmetrization}. 
 In the case where $U$ is 
 scalar, 
 we consider the linear change of variables
 \[
 u=\hf (q+r), \quad U = \frac{\sqrt{6}}{2} (q-r).
 \]
 Then we can rewrite (\ref{sol18}) as a system of 
 two symmetrically coupled pKdV equations,
 \bea
 \nonumber 
 \cases{
 q_t = q_{xxx} + 3(q-r) q_{xx}+ 3 q_x^2 + 3 (q-r)^2 q_x
     \vspace{1.5mm}, \cr
 r_t = r_{xxx} - 3(q-r) r_{xx}+ 3 r_x^2 + 3 (q-r)^2 r_x.
 \cr}
 \eea
 This system coincides with (28) in \cite{Foursov2} or (3.8) 
 in \cite{Foursov3}. 

 \subsubsection{System (\ref{sol20})}
 \label{Solution-20}

 System (\ref{sol20}) is 
 connected with 
 system (\ref{sol24}) through a Miura-type transformation. 
 We 
 discuss the integrability of these two systems 
 in section~\ref{Solution-24}. 
 \vspace{12pt}
 \\
 {\it Symmetrization}. 
 In the case where $U$ is 
 scalar, 
 we consider the linear change of variables
 \[
 u= q+r, \quad U = {\sqrt{3}} (q-r).
 \]
 Then we can rewrite (\ref{sol20}) as a system of 
 two symmetrically coupled pKdV equations,
 \bea
 \nonumber 
 \cases{
 q_t = -\frac{1}{2} q_{xxx} +\frac{3}{2} r_{xxx} - 6(q-r) r_{xx} + 6 r_x^2 
	 +12 (q-r)^2 r_x 
 \cr \qquad \mbox{}+3 (q-r)^4
     \vspace{1.5mm}, \cr
 r_t = \frac{3}{2} q_{xxx} -\frac{1}{2} r_{xxx} + 6(q-r) q_{xx} + 6 q_x^2 
	 +12 (q-r)^2 q_x 
 \cr \qquad \mbox{}+3 (q-r)^4.
 \cr}
 \eea
 This system is identical to (13) in \cite{Foursov2} with 
 $\a =6$. Thus, 
 it is equivalent to 
 (41) in \cite{Foursov2} or 
 (3.12) in \cite{Foursov3}, up to 
 a linear change of dependent 
 variables. 

 \subsubsection{System (\ref{sol21})}
 \label{Solution-21}

 For system (\ref{sol21}), 
 if we introduce 
 a new variable $w$ by 
 \begin{equation}
 w \equiv u_x + 2\sca{U}{U},
 \label{map1}
 \end{equation}
 it solves the KdV equation,
 \begin{subequations}
 \begin{equation}
 w_t = w_{xxx} + 2ww_x.
 \label{KdV39}
 \end{equation}
 Therefore, 
 system (\ref{sol21}) is reduced to 
 a triangular form. 
 The equation for $U$ 
is 
rewritten in terms of $w$ 
 as 
 \begin{equation}
 U_t = 4U_{xxx} + w_x U + 2wU_x.
 \label{KdV_ei}
 \end{equation} 
 \label{KdV1}
 \end{subequations}
 We note 
 that the vector equation (\ref{KdV_ei}) 
 is of 
 the same form as the time 
 part 
 of the linear problem 
 for the KdV equation (\ref{KdV39}). 
 This relation between system (\ref{sol21}) and the 
 KdV equation resembles the relation 
 between the  $5^{\mbox{\scriptsize th }}$order symmetry 
 of system (\ref{sol2}) and the Kaup--Kupershmidt equation 
 shown in section~\ref{Solution-2}. 
 A recursion operator and a Lax representation for 
 the triangular system (\ref{KdV1}) 
 were given in \cite{Gurses} and \cite{Sakov}, respectively. 
 Once we obtain $w(x,t)$ and $U(x,t)$, 
 we can 
 determine $u(x,t)$ by using (\ref{map1}). 
 \vspace{12pt}
 \\
 {\it Symmetrization}. 
 In the case where $U$ is 
 scalar, 
 we consider the linear change of variables
 \[
 u= 6 (q+r), \quad U = {\sqrt{3}}\hspace{1pt}\ii (q-r).
 \]
 Then we can rewrite (\ref{sol21}) as a system of 
 two symmetrically coupled pKdV equations,
 \bea
 \nonumber 
 \cases{
 q_t = \frac{5}{2} q_{xxx} -\frac{3}{2} r_{xxx} + 6(q-r) q_{xx} + 6 q_x^2 
     +12 q_x r_x - 6r_x^2 
 \cr \qquad \mbox{}
 -12 (q-r)^2 (q_x-r_x) -3 (q-r)^4
     \vspace{1.5mm}, \cr
 r_t = -\frac{3}{2} q_{xxx} +\frac{5}{2} r_{xxx} -6(q-r) r_{xx} - 6 q_x^2 
     +12 q_x r_x + 6r_x^2 
 \cr \qquad \mbox{}
 +12 (q-r)^2 (q_x-r_x) -3 (q-r)^4.
 \cr}
 \eea
 This system coincides with (42) in \cite{Foursov2} or (3.13) 
 in \cite{Foursov3}. 

 \subsubsection{System (\ref{sol4})}
 \label{Solution-4}

 For system (\ref{sol4}), if 
 we introduce 
 a new variable $w$ by 
 \begin{equation}
 w \equiv -u_x + \hf u^2 -\hf \sca{U}{U}, 
 \label{KdV2}
 \end{equation}
 it solves 
 the KdV equation,
\begin{subequations}
 \begin{equation}
 w_t = w_{xxx} -3ww_x.
 \label{w-KdV}
 \end{equation}
 Therefore, 
 system (\ref{sol4}) is 
 reduced to a triangular form. 
 Still, 
 it is a 
 very interesting system. 
 To see this, we rewrite 
 the equation for $u$ in terms of $w$ as 
 \begin{equation}
 u_t = -2 u_x^2 + u^2 u_x -3 wu_x -w_x u -w_{xx}. 
 \label{uweq}
 \end{equation} 
\end{subequations}
 Then, the 
 reduction $w=0$ 
 changes 
 this equation 
 to 
 a nontrivial closed equation for $u$. 
 With a rescaling of variables, it reads 
 \begin{equation}
 u_{t} = u_x (u_x-u^2). 
 \label{shockwave}
 \end{equation} 
 Equation (\ref{shockwave}) 
possesses an infinite set of commuting symmetries
 \[
 u_{\tau_n} = u_x (u_x- u^2)^n, \quad n \in {\mathbb R}.
 \]
 We can easily obtain a travelling-wave solution of 
 (\ref{shockwave}) 
 with two arbitrary constants \cite{FL}, 
 which is called a complete solution in the theory of 
 partial differential equations. 
 However, 
 we 
 do not 
 know any explicit 
 formula for 
 the general solution of (\ref{shockwave}). 
 Using (\ref{KdV2}), we can rewrite 
 the 
 equation for $U$ as a linear equation 
 with a $u,w$-dependent coefficient. 
 \vspace{12pt}
 \\
 {\it Symmetrization}. 
 In the case where $U$ is 
 scalar, 
 we consider the linear change of variables
 \[
 u= q+r, \quad U = q-r.
 \]
 Then we can rewrite (\ref{sol4}) as a system of 
 two symmetrically coupled mKdV equations,
 \bea
 \nonumber 
 \cases{
 q_t = \hf q_{xxx} + \hf r_{xxx} + \hf(q-r) (q_{xx}-r_{xx}) -(q_x -r_x)r_x
 \cr \qquad \mbox{}
	 +(q^2-5qr)q_x -(q^2+qr)r_x 
     \vspace{1.5mm}, \cr
 r_t = \hf q_{xxx} + \hf r_{xxx} + \hf(q-r) (q_{xx}-r_{xx}) +(q_x- r_x)q_x
 \cr \qquad \mbox{}
	 -(qr+r^2)q_x +(-5qr+r^2)r_x.
 \cr}
 \eea
 This system coincides with 
 (57) in \cite{Foursov2} 
 or (3.18) in \cite{Foursov3}, up to a scaling of variables. 

 \subsubsection{System (\ref{sol5})}
 \label{Solution-5}

 In system (\ref{sol5}), 
 $u$ and $\sca{U}{U}$ satisfy 
 the same equations 
 as 
 in system (\ref{sol4}). Thus, 
 if we define 
 $w$ by (\ref{KdV2}), we 
 obtain (\ref{w-KdV}) and (\ref{uweq}) again. 
 The only difference 
 between the two systems 
 (\ref{sol5}) 
 and 
 (\ref{sol4}) 
 lies in 
 the forms of 
 equations for $U$, 
 which can be rewritten as linear equations 
 with $u,w$-dependent coefficients. 
 \vspace{12pt}
 \\
 {\it Symmetrization}. 
 In the case where $U$ is 
 scalar, 
 (\ref{sol5}) is identical to 
 (\ref{sol4}). Thus, through symmetrization, 
 we just 
 obtain the same result as in 
 section~\ref{Solution-4}.

 \subsubsection{System (\ref{sol6})}
 \label{Solution-6}

 System (\ref{sol6}) 
 has already been obtained in \cite{Razb} 
 as a reduction of 
 a bi-Hamiltonian system 
 (see also \cite{Kuper2}). 
 If we introduce 
 a new variable $w$ by 
 \[
 w \equiv -u_x + \hf u^2 -\hf \sca{U}{U}, 
 \]
 it solves the KdV equation, 
 \begin{equation}
 w_t = w_{xxx} -3ww_x.
 \label{KdVeq}
 \end{equation}
 Therefore, 
 system (\ref{sol6}) is 
 reduced to a triangular form. 
 Substituting 
 $\hf \sca{U}{U} = -u_x + \hf u^2 -w$ into 
 the equations for $u$ and $U$ respectively, 
 we obtain 
 \bea
 \nonumber 
 \cases{
 u_t = -(wu+w_x)_x
 \vspace{1.5mm} ,\cr
 U_t = - (wU)_x.
 \cr}
 \nonumber 
 \eea
 Thus the reduction $w=0$ is trivial in this system. 
 We mention again (cf.\ section~\ref{Solution-10}) 
 that the above equation for $U$ coupled to 
 the KdV equation 
 (\ref{KdVeq}) was studied in \cite{Kamp}. 
\vspace{12pt}
\\
{\it Remark}. 
%System (\ref{sol6}) admits an interesting one-parameter deformation, 
%System (\ref{sol6}) admits an integrable one-parameter deformation, 
Actually, 
a one-parameter deformation of system 
(\ref{sol6}), 
 \bea
 \label{sol6_def}
 \cases{
 u_t = 3a u_x 
	+ u_{xxx} - \frac{3}{2} u^2 u_x + \frac{1}{2} u_x \sca{U}{U}
    +u \sca{U}{U_x} + \sca{U}{U_{xx}}  + \sca{U_x}{U_x}
 \vspace{1.5mm}, \cr
 U_t = a U_x + u_{xx} U + u_x U_x -u u_x U - \hf u^2 U_x
       + \hf \sca{U}{U}U_x + \sca{U}{U_x} U,
 \cr}
 \eea
is 
still 
integrable. Indeed, 
if we introduce 
%a new scalar variable 
$w$ 
in this 
%the deformed 
case 
by 
 \[
 w \equiv -a -u_x + \hf u^2 -\hf \sca{U}{U},
\]
system (\ref{sol6_def}) 
is changed into the following system: 
 \bea
% \label{cIto3}
\nonumber
 \cases{
 w_t = w_{xxx} -3ww_x + 2a \sca{U}{U_x}
 \vspace{1.5mm} ,\cr
 U_t = - (w U)_x.
 \cr}
 \eea
%If 
When $a \neq 0$, 
this system 
%(\ref{cIto3}) 
coincides with the 
multi-component Zakharov--Ito system (\ref{cIto}), 
up to a scaling of variables. 
 \vspace{12pt}
 \\
 {\it Symmetrization}. 
 In the case where $U$ is 
 scalar, 
 we consider the linear change of variables
 \[
 u= q+r, \quad U = q-r.
 \]
 Then we can rewrite (\ref{sol6}) as a system of 
 two symmetrically coupled mKdV equations,
 \bea
 \nonumber 
 \cases{
 q_t = \hf q_{xxx} + \hf r_{xxx} + (q-r) q_{xx} +(q_x -r_x)q_x
	 -4 qrq_x -2 q^2 r_x
     \vspace{1.5mm}, \cr
 r_t = \hf q_{xxx} + \hf r_{xxx} - (q-r) r_{xx} -(q_x -r_x)r_x
	 -2 r^2 q_x-4 qrr_x.
 \cr}
 \eea
 This system coincides with 
 (58) in \cite{Foursov2} 
 or (3.19) in \cite{Foursov3}.

 \subsubsection{System (\ref{sol7})}
 \label{Solution-7}

 For system (\ref{sol7}), 
 if we introduce 
 a new variable $w$ by 
 \[
 w\equiv -u_x + \hf u^2 -\hf \sca{U}{U}, 
 \]
 it satisfies 
 the KdV equation, 
 \begin{subequations}
 \begin{equation}
 w_t = w_{xxx} -3ww_x.
 \label{KdVeq2}
 \end{equation}
 Therefore, 
 system (\ref{sol7}) is 
 reduced to 
 a triangular form. 
The equation for $u$ 
is 
rewritten in terms of $w$ as 
 \begin{equation}
 u_t = -u^2 u_x + \hf u^4 -w_{xx} +  w u_x -w_x u-2wu^2 + 2w^2.
 \label{uweq2}
 \end{equation}
 \label{counter}
 \end{subequations}
 The triangular system (\ref{counter}) 
possesses (at least) two 
 higher 
 symmetries. 
 The first higher symmetry 
 is given by 
 \bea
 \nonumber 
 \cases{
 w_{t_5} = w_{xxxxx} - 5 ww_{xxx} -10 w_x w_{xx} +\frac{15}{2} w^2 w_x
 \vspace{1.5mm} ,\cr
 u_{t_5} = -\hf u^4 w +2u^2 w^2 - 2w^3 +u^2 wu_x - \hf w^2 u_x -u^3 w_x 
	 +5uww_x 
 \cr \qquad \; \mbox{}
 +2u u_x w_x +3w_x^2 -u^2w_{xx} +5ww_{xx} + u_x w_{xx} 
	 -uw_{xxx} -w_{xxxx},
 \cr}
 \eea
 which obviously vanishes under the reduction $w=0$. 
 Similarly, the 
 second one 
 vanishes under the same 
 reduction. 
 On the other hand, the reduction $w=0$ 
 changes (\ref{counter}) 
 to a nontrivial 
 closed equation for $u$, 
 \[
 u_t = -u^2 u_x + \hf u^4. 
 \]
 As far as we could check with the help of a computer, 
 this equation seems to have no 
 polynomial 
 higher symmetry. 
 We can construct its general solution in implicit form 
 using the method of characteristic curves. 
 \vspace{12pt}
 \\
 {\it Symmetrization}. 
 In the case where $U$ is 
 scalar, 
 we consider the linear change of variables
 \[
 u= q+r, \quad U = q-r.
 \]
 Then we can rewrite (\ref{sol7}) as a system of 
 two 
 symmetrically coupled mKdV equations,
 \bea
 \nonumber 
 \cases{
 q_t = \hf q_{xxx} + \hf r_{xxx} + \hf (q-r) (q_{xx}-r_{xx}) -(q_x -r_x)r_x
 \cr \qquad \mbox{}
	 -(3qr+r^2) q_x +(-3qr +r^2) r_x +\hf (q-r)^3 q
     \vspace{1.5mm}, \cr
 r_t = \hf q_{xxx} + \hf r_{xxx} + \hf (q-r) (q_{xx}-r_{xx}) +(q_x -r_x)q_x
 \cr \qquad \mbox{}
	 +(q^2-3qr) q_x -(q^2 +3qr) r_x -\hf (q-r)^3 r.
 \cr}
 \eea
 This system coincides with 
 (61) in \cite{Foursov2} 
 or (3.20) in \cite{Foursov3}, up to a scaling of variables. 

 \subsubsection{System (\ref{sol8})}
 \label{Solution-8}

 For system (\ref{sol8}), 
 if we introduce 
 a new variable $w$ by 
 \begin{equation}
 w\equiv -u_x + \hf u^2 -\hf \sca{U}{U}, 
 \label{Ricc}
 \end{equation}
 it satisfies 
 the KdV equation, 
 \[
 w_t = w_{xxx} -3ww_x.
 \]
 Therefore, 
 system (\ref{sol8}) is rewritten 
 in a triangular form. 
 Substituting $\hf \sca{U}{U} = -u_x + \hf u^2 -w$ into 
 the equations for $u$ and $U$ respectively, we obtain 
 \bea
 \label{uweq3}
 \cases{
 u_t = -w_x u -\hf wu^2 -w_{xx} +w^2
 \vspace{1.5mm} ,\cr
 U_t = -\hf (w_x + wu)U.
 \cr}
 \eea
 Thus the reduction $w=0$ is trivial in this system. 
 We note that 
 in (\ref{uweq3}) 
 no term involves $x$-derivatives of 
\mbox{$u,\hspace{1pt} U$} such as 
\mbox{$u_x, \hspace{1pt}U_x, \hspace{1pt}u_{xx}, \hspace{1pt}U_{xx}$}. 
 Then, for a given solution of the KdV equation 
 $w(x,t)$, the equation for $u(x,t)$ 
 can be 
 regarded as a Riccati equation with $x$ fixed. 
 Once we obtain $w(x,t)$ and $u(x,t)$, 
 we can integrate 
 the equation for $U$ 
 as 
 \[
 U (x,t) = \e^{-\hf \int^t_0 (w_x + wu) \d t'} U (x,0).
 \]
 \\
 {\it Remark}. 
 Actually, 
 system (\ref{sol8}) is the $3^{\mbox{\scriptsize rd }}$order 
 symmetry of a 
 nontrivial $1^{\mbox{\scriptsize st }}$order system,
 \bea
 \label{sol8-1}
 \cases{
 u_{t_1} = u_{x} +\hf \sca{U}{U} 
     \vspace{1.5mm}, \cr
 U_{t_1} = \hf u U.
 \cr}
 \eea
 For this system, 
 $w$ defined by (\ref{Ricc}) 
 obeys 
 the linear 
 equation $w_{t_1} = w_x$ and the equation for $u$ 
 is rewritten as $u_{t_1} = \hf u^2 -w$. 
 System (\ref{sol8-1}) in the $N=1$ case 
 as well as its higher symmetries 
 was studied in \cite{Bogo,Tu,Boiti2} 
 (see also \cite{Hu} for its soliton-like 
 solutions). 
 \vspace{12pt}
 \\
 {\it Symmetrization}. 
 In the case where $U$ is 
 scalar, 
 we consider the linear change of variables
 \[
 u= q+r, \quad U = q-r.
 \]
 Then we can rewrite (\ref{sol8}) as a system of 
 two symmetrically coupled mKdV equations, 
 \bea
 \nonumber 
 \cases{
 q_t = \hf q_{xxx} + \hf r_{xxx} + \frac{1}{4} (q-r) 
	 (3q_{xx} - r_{xx}) +\hf (q_x -r_x)^2
 \cr \qquad \mbox{}
	 + \hf (q^2-6qr-r^2) q_x -(q^2+2qr) r_x -q^2r (q-r)
     \vspace{1.5mm}, \cr
 r_t = \hf q_{xxx} + \hf r_{xxx} + \frac{1}{4} (q-r) 
	 (q_{xx} - 3r_{xx}) +\hf (q_x -r_x)^2
 \cr \qquad \mbox{}
	 - (2qr +r^2) q_x -\hf (q^2+6qr -r^2) r_x + qr^2 (q-r).
 \cr}
 \eea
This system 
coincides with 
(23) in \cite{Foursov2} 
 with 
 $\a=1$, up to a 
 scaling of dependent variables. Thus, it 
is equivalent 
to 
(62) in \cite{Foursov2} or 
 (3.21) in \cite{Foursov3}, up to 
 a linear change of 
 variables. 

 \subsubsection{System (\ref{sol15})}
 \label{Solution-15}

 System (\ref{sol15}) is just a disguised form of a single 
 vector equation. Indeed, if 
 we introduce 
 an $(N+1)$-component vector $W$ by $W\equiv (u, U) =(u, U_1, \ldots, U_N)$, 
 system (\ref{sol15}) 
 can be rewritten in the form 
 \bea
 W_t = W_{xxx} + \sca{W}{W} W_x.
 \nonumber 
 \eea
 This is a well-known vector mKdV equation and its integrability has 
 been established in the literature \cite{Svi0,Svi,Adler,TW1}. 
 \vspace{12pt}
 \\
 {\it Symmetrization}. 
 In the case where $U$ is 
 scalar, 
 we consider the linear change of variables
 \[
 u= \hf (q+r), \quad U = \frac{\ii}{2} (q-r).
 \]
 Then we can rewrite (\ref{sol15}) as a system of 
 two symmetrically coupled mKdV equations,
 \bea
 \nonumber 
 \cases{
 q_t = q_{xxx} + qrq_x
     \vspace{1.5mm}, \cr
 r_t = r_{xxx} +qrr_x.
 \cr}
 \eea
 This system is known as (the non-reduced form of) the complex mKdV 
 equation \cite{AKNS}. 
 It is identical to (43) in \cite{Foursov2} or, 
 (3.14) in \cite{Foursov3} with a correction of misprints. 

 \subsubsection{System (\ref{sol17})}
 \label{Solution-17}

 System (\ref{sol17}) is also 
 a disguised form of a single 
 vector equation. Indeed, if 
 we introduce an $(N+1)$-component vector 
 $W$ by $W\equiv (u, U) =(u, U_1, \ldots, U_N)$, 
 system (\ref{sol17}) can be rewritten in the form 
 \bea
 W_t = W_{xxx} + \sca{W}{W} W_x + \sca{W}{W_x}W.
 \nonumber 
 \eea
 This is another 
 well-known vector mKdV equation, 
 for which 
 a Lax representation was given in \cite{YO} for the $N=1$ case 
 and in \cite{Konope} for the general $N$ case. 
 \vspace{12pt}
 \\
 {\it Symmetrization}. 
 In the case where $U$ is 
 scalar, 
 we consider the linear change of variables
 \[
 u= q+r, \quad U = \ii (q-r).
 \]
 Then we can rewrite (\ref{sol17}) as a system of 
 two symmetrically coupled mKdV equations, 
 \bea
 \label{sol17-sym}
 \cases{
 q_t = q_{xxx} + 6qrq_x + 2 q^2 r_x
     \vspace{1.5mm}, \cr
 r_t = r_{xxx} +2r^2 q_x + 6qrr_x.
 \cr}
 \eea
 This system coincides with 
 (48) in \cite{Foursov2} or 
 (3.15) in \cite{Foursov3}, up to a scaling of variables. 

 \subsubsection{System (\ref{sol16})}
 \label{Solution-16}

 System (\ref{sol16}) 
 is known as a Jordan mKdV system \cite{Svi0}. 
 Let us briefly summarize its integrability. 
 It is well-known 
 that the matrix mKdV equation,
 \begin{equation}
 Q_t = Q_{xxx} -3 (Q_x Q^2 + Q^2 Q_x),
 \label{matrix-mKdV}
 \end{equation}
 admits a Lax representation \cite{Zakh,TW1,Konope,Linden2}. 
 Then, 
 system (\ref{sol16}) is also integrable, because it is 
 obtained from 
 (\ref{matrix-mKdV}) through 
 the following reduction:
 \[
 Q = u {\mathbf 1} + \sum_{j=1}^{N} U_j {\mathbf e}_j, 
 \hspace{5mm} \{ {\mathbf e}_i, {\mathbf e}_j \}_{+} = -2 \delta_{i j} 
	 {\mathbf 1}.
 \]
 We mention that 
 (\ref{sol16}) admits a 
 generalization to 
 a system for 
 two vector unknowns 
 preserving the integrability \cite{Sak}. 
 \vspace{12pt}
 \\
 {\it Relation to systems $(\ref{cKdV})$ and $(\ref{sol14})$}. 
 If we define new variables $w$ and $W$ by \cite{Svi0}
 \bea
 \nonumber 
 \cases{
 w \equiv \pm u_x -u^2 + \sca{U}{U}
 \vspace{1.5mm} ,\cr
 W \equiv U_x  \mp 2u U,
 \cr} 
 \eea
 they satisfy the following system: 
 \bea
 \nonumber
 \cases{
 w_t = w_{xxx} + 3 \bigl( w^2 -\sca{W}{W} \bigr)_x
 \vspace{1.5mm} ,\cr
 W_t = W_{xxx} + 6(wW)_x.
 \cr}
 \eea
 This system coincides with 
 the Jordan KdV system (\ref{cKdV}) and, 
through 
 potentiation of it, 
 we 
 obtain 
 system (\ref{sol14}).
 \vspace{12pt}
 \\
 {\it Symmetrization}. 
 In the case where $U$ is 
 scalar, 
 we consider the linear change of variables
 \[
 u= \hf (q+r), \quad U = \frac{\ii}{2}(q-r).
 \]
 Then 
 (\ref{sol16}) is decoupled into two 
 mKdV equations,
 \bea
 \nonumber 
 \cases{
 q_t = q_{xxx} - 6q^2 q_x 
     \vspace{1.5mm}, \cr
 r_t = r_{xxx} - 6r^2 r_x.
 \cr}
 \eea
 This corresponds to (50) in \cite{Foursov2}.

 \subsubsection{System (\ref{sol22})}
 \label{Solution-22}

 We note that through introduction of 
 a new scalar 
 variable $w$ by 
 \[
 w\equiv u_x -u^2,
 \]
 system (\ref{sol22}) is 
 transformed to 
 a system of 
 coupled KdV-mKdV type, 
 \bea
 \label{compact}
 \cases{
 w_t = w_{xxx} + 6ww_x +w_x \sca{U}{U} + 2w \sca{U}{U}_x
 + \hf \sca{U}{U}_{xxx}
 \vspace{1.5mm} ,\cr
 U_t = U_{xxx} + 3(wU)_x + \sca{U}{U}U_x + \frac{3}{2} \sca{U}{U}_x U.
 \cr}
 \eea
 Let us demonstrate that 
 system (\ref{sol22}) admits a Lax representation. 
 We consider a pair of linear equations for a column-vector function 
 $\vt{\psi}$, 
 \[
 \vt{\psi}_x = \hat{U} \vt{\psi}, 
\hspace{5mm} \vt{\psi}_t = \hat{V} \vt{\psi}, 
 \]
with the matrices $\hat{U}$ and $\hat{V}$ of the following form:
\begin{subequations}
\begin{equation}
\hat{U}= \left(
\begin{array}{cc}
-\ii \z I_l  & Q \\
 R & \ii \z I_m + P\\
\end{array}
\right),
\nonumber 
\end{equation}
\begin{equation}
\nonumber 
\hat{V} = 
\left(
\begin{array}{c|c}
\begin{array}{l}
-4\ii \z^3 I_l -2\ii \z QR 
\\
 \mbox{} + Q_x R -QR_x + 2QPR
\end{array}
&
\begin{array}{l}
4\z^2 Q + 2\ii \z (Q_x + QP) -Q_{xx} 
\\
 \mbox{} -2Q_x P -QP_x + 2QRQ -QP^2
\end{array}
\\
\hline 
\begin{array}{l}
4\z^2 R +2\ii \z (-R_x + PR) 
\\
 \mbox{} 
-R_{xx} + P_x R + 2PR_x 
\\ 
\mbox{} + 2RQR -P^2 R
\end{array}
&
\begin{array}{l}
4\ii \z^3 I_m + 2\ii \z RQ -P_{xx} + R_x Q 
\\
\mbox{} 
-RQ_x + P_x P -PP_x + 2PRQ 
\\ \mbox{}
+ 2RQP -P^3 -3 g_x P
\end{array}
\end{array}
\right).
\end{equation}
\label{UV-sol22}
\end{subequations}
Here, 
$\z$ is the spectral parameter, 
$I_l$ and $I_m$ are 
the $l \times l$ and $m \times m$ unit matrices respectively, 
$Q$, $R$ and $P$ are $l \times m$, $m \times l$ and $m \times m$ 
matrices respectively, and 
$g$ is a 
scalar function. 
The compatibility condition \mbox{$\vt{\psi}_{xt} = \vt{\psi}_{tx}$} 
implies the so-called 
zero-curvature condition, 
\[
\hat{U}_t - \hat{V}_x + \hat{U}\hat{V} - \hat{V}\hat{U} =O.
\]
Then, 
substituting 
(\ref{UV-sol22}) into this 
condition, 
we obtain a 
system of three matrix 
equations,
\bea
\label{threemm}
\cases{
Q_t + Q_{xxx} + 3(Q_x P)_x -3Q_x RQ-3QRQ_x + 3Q_x P^2 
\cr \quad \mbox{}+ 3QP_x P -3g_x QP =O
\vspace{1.5mm} ,\cr
R_t + R_{xxx} -3(PR_x)_x -3R_x QR -3RQR_x + 3P^2 R_x 
\cr \quad \mbox{}+ 3PP_x R + 3g_x PR=O
\vspace{1.5mm} ,\cr
P_t + P_{xxx} + 3(g_x P)_x -3 (PRQ+RQP)_x +3PP_x P 
\cr \quad \mbox{} +3P^2 RQ -3RQP^2 =O.
\cr}
\eea
We note that 
this system admits the 
reduction $R = {}^t \hspace{-0.25mm} Q$, ${}^t \hspace{-0.5mm} P = -P$, 
where the superscript ${}^t$ 
denotes the 
transposition. 
In particular, if we choose
\bea
\nonumber 
\cases{
Q = (u, W_1, \ldots, W_N) \equiv (u,W)
\vspace{1.5mm} ,\cr
R 
= \left(
\begin{array}{c}
u \\
W_1 \\
\vdots \\
W_N
\end{array}
\right) 
=
\left(
\begin{array}{c}
 u \\
 {}^t \hspace{-0.15mm} W \\
\end{array}
\right)
\vspace{1.5mm} ,\cr
P = \left(
\begin{array}{cc}
0 & 
\begin{array}{ccc}
V_1 & \cdots & V_N
\end{array}
\\
\begin{array}{c}
-V_1 \\ \vdots \\ -V_N
\end{array}
& \mbox{\LARGE $O$} 
\end{array}
\right)
\equiv \left(
\begin{array}{cc}
0 & V \\
-{}^t \hspace{-0.20mm} V & O\\
\end{array}
\right),
\cr}
\eea
system (\ref{threemm}) is reduced to the 
system
\bea
\label{threem}
\cases{
u_t + u_{xxx} -6u^2 u_x -3u_x (\sca{W}{W} +\sca{V}{V})
-3u (\sca{W}{W_x} + \sca{V}{V_x}) 
\cr \quad \mbox{}
+3g_x \sca{W}{V} - 3 \sca{W_x}{V}_x = 0
\vspace{1.5mm} ,\cr
W_t + W_{xxx} + 3 (u_x V)_x -3u g_x V -3uu_x W -3u^2 W_x 
-3\sca{W}{W}W_x
\cr \quad \mbox{}
-3\sca{W}{W_x} W -3\sca{W}{V}_x V ={\mathbf 0}
\vspace{1.5mm} ,\cr
V_t + V_{xxx} + 3 (g_x V)_x -3(u^2 V)_x -3 (\sca{W}{V}W)_x 
-3 \sca{V}{V_x}V
\cr \quad \mbox{}
-3u \sca{V}{V} W + 3u \sca{W}{V}V={\mathbf 0},
\cr}
\eea
together with one constraint,
\[
(u \, {}^t \hspace{-0.20mm} V W -u \, {}^t \hspace{-0.20mm} WV)_x 
- \sca{W}{V} ({}^t \hspace{-0.20mm} V W -{}^t \hspace{-0.20mm} WV)=O.
\]
When $W$ and $V$ are scalar, i.e.\ $N=1$, this 
constraint is 
satisfied 
automatically 
and
we obtain a three-component mKdV system. 
There is another 
case, the case 
$W=V$, 
for which 
the constraint is 
satisfied. 
Then, if we 
set 
\[
g=u, \hspace{5mm} W=V = \frac{\ii}{\sqrt{3}} U,
\]
and change the sign of time $t$ ($t \to -t$), 
system (\ref{threem}) collapses 
to system (\ref{sol22}). 
\vspace{12pt}
\\
{\it Relation to system $(\ref{sol19})$}. 
If we 
introduce a new scalar variable $v$ by 
\[
v\equiv u_x - u^2 + \frac{1}{3} \sca{U}{U},
\]
system (\ref{sol22}) is changed into the following system: 
\bea
\label{vUeq}
\cases{
v_t = \bigl( v_{xx} +3v^2 +  v \sca{U}{U} + \sca{U}{U_{xx}} \bigr)_x
\vspace{1.5mm} ,\cr
U_t = U_{xxx} + 3 v_x U + 3v U_x +\sca{U}{U_x} U.
\cr}
\eea
Then, 
it is straightforward to obtain (\ref{sol19}) (for $\hat{u}$ and $U$) 
from (\ref{vUeq}) through 
potentiation $v=\hat{u}_x$. 
It should be mentioned here that the 
authors encountered two papers \cite{Kersten,KSY} on 
the integrability of (\ref{vUeq}) 
in the case of 
scalar $U$, 
after they had obtained 
all the presented 
results 
independently. 
\vspace{12pt}
\\
{\it Symmetrization}. 
In the case where $U$ is 
scalar, 
we consider the linear change of variables 
 \[
 u= q+r, \quad U = \sqrt{3}(q-r).
 \]
 Then we can rewrite (\ref{sol22}) as a system of 
two symmetrically coupled mKdV equations,
 \bea
 \cases{
 q_t = \bigl[ q_{xx} + 3 (q-r) q_x + q^3 -12 q^2 r + 3qr^2 \bigr]_x
     \vspace{1.5mm}, \cr
 r_t = \bigl[ r_{xx} -3 (q-r)r_x + 3q^2 r -12 qr^2 + r^3 \bigr]_x.
 \cr}
\nonumber
 \eea
This system coincides with 
(55) in \cite{Foursov2} 
or (3.17) in \cite{Foursov3}. 
Moreover, if we introduce 
new variables $\hat{q}$ and $\hat{r}$ by 
 \[
 \hat{q} \equiv \sqrt{3} \hspace{1pt} \ii \hspace{1pt}q 
	\e^{\int^x (q-r) \d x'}, \hspace{5mm}
 \hat{r} \equiv \sqrt{3} \hspace{1pt}\ii \hspace{1pt} 
	r \e^{-\int^x (q-r) \d x'}, 
 \]
they satisfy the 
system of 
coupled mKdV 
equations (\ref{sol17-sym}). 

\subsubsection{System (\ref{sol23})}
\label{Solution-23}

Through introduction of 
a new scalar 
variable $w$ by 
 \[
w \equiv u_x -u^2,
 \]
system (\ref{sol23}) is 
transformed to 
a system that looks very 
similar to (\ref{compact}),
 \bea
 \label{compact2}
 \cases{
 w_t = w_{xxx} + 6ww_x +w_x \sca{U}{U} + 2w \sca{U}{U}_x
 + \hf \sca{U}{U}_{xxx}
 \vspace{1.5mm} ,\cr
 U_t = U_{xxx} + 6(wU)_x + \sca{U}{U}U_x + 2 \sca{U}{U}_x U.
 \cr}
 \eea
System (\ref{compact2}) is a multi-component 
generalization 
of a flow of 
the Jaulent--Miodek hierarchy \cite{JM}. Let us 
demonstrate that (\ref{compact2}) admits a 
Lax representation. We consider a 
pair of 
linear equations for a column-vector function $\vt{\psi}$,
\bea
\label{JMLax}
\cases{
\vt{\psi}_{xx} + (Q+ \z R )\vt{\psi} = \z^2 \vt{\psi}
\vspace{1.5mm} ,\cr
\vt{\psi}_t = \bigl( 4\z^2 I + 2\z R + 2Q + \frac{3}{2} R^2 \bigr) \vt{\psi}_x 
        - \bigl[ \z R_x + Q_x + \frac{3}{4} (R^2)_x \bigr] \vt{\psi}.
\cr}
\eea
Here, $\z$ is the spectral 
parameter, and 
$Q$ and $R$ are 
square matrices with 
the same dimension. 
The compatibility condition $\vt{\psi}_{xxt} = \vt{\psi}_{txx}$ 
for (\ref{JMLax}) 
implies 
a system of two matrix equations,
\bea
\label{matrixJM}
\cases{
Q_t = Q_{xxx} + 3(Q^2)_x + \frac{3}{4} (R^2)_{xxx} + \frac{3}{2} R^2 Q_x 
        + \frac{3}{4} \bigl[ Q (R^2)_x + 3(R^2)_x Q \bigr]
\vspace{1.5mm} ,\cr
R_t = R_{xxx} + 3(QR+RQ)_x + \frac{3}{4} 
\bigl[ 3(R^2)_x R + R (R^2)_x + 2R^2 R_x \bigr],
\cr}
\eea
together with one constraint,
\[
\bigl[ Q,R^2 \bigr] = O. 
\]
If we consider the 
reduction,
\[
Q = w {\mathbf 1}, \hspace{5mm}
R = \frac{\sqrt{6}}{3} \ii
\sum_{j=1}^{N} U_j {\mathbf e}_j, 
\hspace{5mm} \{ {\mathbf e}_i, {\mathbf e}_j \}_{+} = -2 \delta_{i j} 
        {\mathbf 1}, 
\]
the constraint 
is automatically 
satisfied 
and 
system (\ref{matrixJM}) 
collapses to 
system (\ref{compact2}). 
This Lax representation 
for (\ref{compact2}) can be rewritten as 
that for (\ref{sol23}) \cite{Alonso}. 
\vspace{12pt}
\\
{\it Relation to system $(\ref{sol18})$}. 
If we 
introduce a new scalar variable $v$ by 
\begin{equation}
v\equiv u_x - u^2 + \frac{1}{6} \sca{U}{U},
\label{sol23-18}
\end{equation}
system (\ref{sol23}) is 
changed 
into the following system: 
\bea
\label{380}
\cases{
v_t = \bigl( v_{xx} +3v^2 +  2v \sca{U}{U} + \sca{U}{U_{xx}} 
        + \hf \sca{U_x}{U_x} \bigr)_x
\vspace{1.5mm} ,\cr
U_t = U_{xxx} + 6 v_x U + 6 v U_x +2\sca{U}{U_x} U.
\cr}
\eea
Then, it is 
straightforward to 
obtain (\ref{sol18}) (for $\hat{u}$ and $U$) 
from (\ref{380}) through 
potentiation $v=\hat{u}_x$. 
\vspace{12pt}
\\
{\it Symmetrization}. 
In the case where $U$ is 
scalar, 
we consider the linear change of variables
\[
u= \hf (q+r), \quad U = \frac{\sqrt{6}}{2}(q-r).
\]
Then we can rewrite (\ref{sol23}) as a system of 
two symmetrically coupled mKdV equations,
\bea
\cases{
q_t = \bigl[ q_{xx} + 3 (q-r) q_x + q^3 -6 q^2 r + 3qr^2 \bigr]_x
    \vspace{1.5mm}, \cr
r_t = \bigl[ r_{xx} -3 (q-r)r_x + 3q^2 r -6 qr^2 + r^3 \bigr]_x.
\cr}
\nonumber
\eea
This system coincides with 
(51) in \cite{Foursov2} 
or (3.16) in \cite{Foursov3}. 
It is known as a flow of the modified Jaulent--Miodek 
hierarchy \cite{Alonso} (see also 
(7.37) in \cite{Nijhoff}). 
While 
elaborating on this paper, the authors encountered 
one paper \cite{Das} 
on the three-component generalization of this 
flow 
[(\ref{sol23}) with $N=2$]. 

 \subsubsection{System (\ref{sol24})}
 \label{Solution-24}

For system (\ref{sol24}), 
if we define new variables $w$ and $W$ by 
 \bea
 \label{Miura3}
 \cases{
 w \equiv u_x + u^2 + \frac{1}{6} \sca{U}{U} \vspace{1.5mm},
 \cr
 W \equiv U_x + 2uU,
 \cr}
 \eea
 they satisfy the following system:
 \bea
 \label{cHS2}
 \cases{
 w_t = w_{xxx} -6ww_x + 2\sca{W}{W_x}
 \vspace{1.5mm} ,\cr
 W_t = -2 W_{xxx} + 6 w W_x.
 \cr}
 \eea
 This system coincides with the multi-component 
Hirota--Satsuma system (\ref{cHS}), up to a scaling of variables. 
The Miura map 
(\ref{Miura3}) is a 
multi-component 
generalization 
of that for the case of 
scalar $U$ in \cite{Wilson,Drinfeld} 
and 
that 
for the case of 
two-component vector $U$ in \cite{Wu}. 
 \vspace{12pt}
 \\
 {\it Relation to system $(\ref{sol20})$}. 
If we 
introduce a new scalar variable $v$ by 
 \begin{equation}
 v\equiv u_x - u^2 -\frac{1}{6} \sca{U}{U},
 \label{sol24-20}
 \end{equation}
system (\ref{sol24}) is 
changed into the following system (cf.\ (4.3) 
in \cite{Leble}): 
 \bea
 \label{385}
 \cases{
 v_t = \bigl( v_{xx} + 3v^2 
	+ 4 v \sca{U}{U} + 2 \sca{U}{U_{xx}} +\sca{U_x}{U_x} 
	 + \frac{2}{3} \sca{U}{U}^2 \bigr)_x
 \vspace{1.5mm} ,\cr
 U_t = -2 U_{xxx} -6 v_x U -6 v U_x -4\sca{U}{U_x} U.
 \cr}
 \eea
Then, it is straightforward to 
obtain (\ref{sol20}) (for $\hat{u}$ and $U$) 
from (\ref{385}) through 
potentiation $v=\hat{u}_x$. 
Combining 
(\ref{sol24-20}) 
and 
(\ref{Miura3}), 
we obtain the relation $v + w = 2 u_x$, and consequently, 
 \[
 \hat{u} + \int^x w 
 \hspace{1pt}\d x' = 2u.
 \] 
Using this relation, we can also rewrite (\ref{Miura3}) as 
a transformation between 
system (\ref{sol20}) and the multi-component 
Hirota--Satsuma system (\ref{cHS2}). 
\vspace{12pt}
 \\
 {\it Symmetrization}. 
In the case where $U$ is 
scalar, 
we consider the linear change of variables
 \[
 u= -\hf (q+r), \quad U = \frac{\sqrt{6}}{2}\ii (q-r).
 \]
Then we can rewrite (\ref{sol24}) as a system of 
two symmetrically coupled mKdV equations, 
 \bea
 \cases{
 q_t = \bigl[ -\hf q_{xx} + \frac{3}{2} r_{xx} 
	 +3 (q-r) q_x -2 r^3 \bigr]_x
     \vspace{1.5mm}, \cr
 r_t = \bigl[ \frac{3}{2} q_{xx}  - \hf r_{xx} - 3(q-r)r_x -2 q^3
	 \bigr]_x.
 \cr}
\nonumber
 \eea
 This system is identical to (63) in \cite{Foursov2} 
 or (3.22) in \cite{Foursov3}. 
 It was found 
 in connection with the Kac--Moody Lie algebras 
 and written in a 
Hamiltonian form about twenty years ago 
 (cf.\ the ${\rm C}_2^{(1)}$ case in \cite{Wilson} 
 or the ${\rm B}_2^{(1)}$ case in \cite{Drinfeld}). 

 \newpage
 \noindent
 \section{The case ${\lambda_1 = \lambda_2= {1 \over 2} }$
 $\;\,$-- coupled Ibragimov--Shabat 
 equations --}  
 \label{lambda1-2}
 \setcounter{equation}{0}

 In this section, we 
 classify $2^{\mbox{\scriptsize nd }}$order and 
$3^{\mbox{\scriptsize rd }}$order systems 
 in the $\lambda_1=\lambda_2 = {1 \over 2}$ 
 (Ibragimov--Shabat weighting \cite{Ibragimov}) case. 
 In the first part (section~\ref{com-lis3}), we present 
 a complete list of such systems with a specific order symmetry. 
 In the second part (section~\ref{int-all3}), we 
 prove that the listed 
 systems 
 are linearizable. 

 \subsection{List of systems with a higher symmetry}
 \label{com-lis3}

 The general ansatz for a 
 $\lambda_1=\lambda_2 = \frac{1}{2}$ 
 homogeneous evolutionary 
 system of $2^{\mbox{\scriptsize nd }}$order for a scalar function 
 $u$ and a vector function $U$ takes the form
 \bea
 \label{l12o2}
 \cases{
 u_{t_2} = a_1 u_{xx} + a_2 u^2 u_x + a_3 u^5 
	 + a_4 u_x \sca{U}{U} + a_5 u \sca{U}{U_x} 
 \cr \qquad \; \mbox{}
	 + a_6 u^3 \sca{U}{U} + a_7 u \sca{U}{U}^2 
     \vspace{1.5mm}, \cr
 U_{t_2} = a_{8} U_{xx} + a_{9} u u_x U + a_{10} u^2 U_x 
	 + a_{11} u^4 U + a_{12} \sca{U}{U} U_x
 \cr \qquad \; \mbox{}
	 + a_{13} \sca{U}{U_x} U
	 + a_{14} u^2 \sca{U}{U} U 
	 + a_{15} \sca{U}{U}^2 U. 
 \cr}
 \eea
 The following constraints guarantee the order to be 2 and
 the system not to be triangular:
 \[
 (a_1, a_8) \neq (0,0), \hspace{5mm} (a_4,a_5,a_6,a_7) 
 \neq (0,0,0,0), \hspace{5mm} 
 (a_9, a_{10}, a_{11},a_{14}) \neq (0,0,0,0).
 \]
 Similarly, the general ansatz for a 
 $3^{\mbox{\scriptsize rd }}$order system 
 takes the form
 \bea
 \label{l12o3}
 \cases{
 u_{t_3} = b_1 u_{xxx} + b_2 u^2 u_{xx} + b_3 u u_x^2 
	 + b_4 u^4 u_x + b_5 u^7 + b_6 u_{xx} \sca{U}{U} 
 \cr \qquad \; \mbox{}
	 + b_7 u_x \sca{U}{U_x} + b_8 u \sca{U_x}{U_x} 
	 + b_9 u \sca{U}{U_{xx}} + b_{10} u^2 u_x \sca{U}{U}
 \cr \qquad \; \mbox{}
	 + b_{11} u^3 \sca{U}{U_x} + b_{12} u^5 \sca{U}{U} 
	 + b_{13} u_x \sca{U}{U}^2 
 \cr \qquad \; \mbox{}
	 + b_{14} u \sca{U}{U} \sca{U}{U_x}
	 + b_{15} u^3 \sca{U}{U}^2 + b_{16} u \sca{U}{U}^3 
     \vspace{1.5mm}, \cr
 U_{t_3} = b_{17} U_{xxx} + b_{18} u u_{xx} U + b_{19} u_x^2 U 
	 + b_{20} u u_x U_x + b_{21} u^2 U_{xx} 
 \cr \qquad \; \mbox{}
	 + b_{22} u^3 u_x U + b_{23} u^4 U_x 
	 + b_{24} u^6 U + b_{25} \sca{U}{U}U_{xx}
	 + b_{26} \sca{U}{U_x} U_x 
 \cr \qquad \; \mbox{}
	 + b_{27} \sca{U_x}{U_x}U
	 + b_{28} \sca{U}{U_{xx}} U + b_{29} uu_x \sca{U}{U}U
 \cr \qquad \; \mbox{}
	 + b_{30} u^2 \sca{U}{U}U_x + b_{31} u^2 \sca{U}{U_x} U
	 + b_{32} u^4 \sca{U}{U} U 
 \cr \qquad \; \mbox{}
	 + b_{33} \sca{U}{U}^2 U_x 
	 + b_{34} \sca{U}{U} \sca{U}{U_x} U + b_{35} u^2 \sca{U}{U}^2 U
 \cr \qquad \; \mbox{}
	 + b_{36} \sca{U}{U}^3 U,
 \cr}
 \eea
 for which the following constraints guarantee the order to be 3 and
 the system not to be triangular: $(b_1, b_{17}) \neq (0,0)$ and
 at least one of $b_6, \ldots, b_{16}$ 
 and one of $b_{18}, \ldots, b_{24}, b_{29}, \ldots, b_{32},b_{35}$ 
 must not vanish. 
 However, when we consider a $3^{\mbox{\scriptsize rd }}$order symmetry 
for a $2^{\mbox{\scriptsize nd }}$order system, we 
 relax these constraints as follows (cf.\ section~\ref{Comp_as}): 
 $(b_1, b_{17}) \neq (0,0)$ and at least one of $b_1, \ldots, b_{16}$ 
 and one of $b_{17}, \ldots, b_{36}$ must not vanish. 
 \vspace{6pt}
 \begin{proposition}
 {\it
 No $2^{\mbox{\scriptsize nd }}$order 
 system of the form $(\ref{l12o2})$ with 
 a $3^{\mbox{\scriptsize rd }}$order symmetry 
 of the form $(\ref{l12o3})$ 
 or a $4^{\mbox{\scriptsize th }}$order symmetry 
 exists. 
 }
 \end{proposition}
 \vspace{6pt}
 \begin{theorem}
 {\it
 Any $3^{\mbox{\scriptsize rd }}$order 
 system of the form $(\ref{l12o3})$ 
 with a $5^{\mbox{\scriptsize th }}$order 
 symmetry has to coincide with either of the following 
two systems up to 
a scaling of $t_3,x,u,U \,($we omit the subscript of $t_3)$:
 \bea
 \label{solut1}
 \cases{
 u_t = (a+1) (u_{xxx} + 3 u^2 u_{xx} + 9 u u_x^2 +3 u^4 u_x 
	  +3 u_{xx} \sca{U}{U} 
  \cr \qquad \mbox{}
	  +6u_x \sca{U}{U_x} + 3u_x \sca{U}{U}^2) +2a u \sca{U}{U_{xx}} 
	  \cr \qquad \mbox{}     
  + (2a+3) u \sca{U_x}{U_x}
  + (10a+6 ) u_x u^2 \sca{U}{U} + 2a u^3 \sca{U}{U_x} 
  \cr \qquad \mbox{} +6a u \sca{U}{U} \sca{U}{U_x} 
  + a u^5 \sca{U}{U}+ 2a u^3 \sca{U}{U}^2 + au \sca{U}{U}^3 
      \vspace{2.5mm}, \cr
 U_t = U_{xxx} +3\sca{U}{U} U_{xx} + 6\sca{U}{U_x}U_x + 3\sca{U_x}{U_x}U
	  + 3 \sca{U}{U}^2 U_x 
  \cr \qquad \mbox{} 
  -2a u_{xx} uU + (a+3) u_x^2 U + 6 uu_x U_x +3u^2 U_{xx} 
   -6 a u_x u^3 U 
  \cr \qquad \mbox{}
  + 3u^4 U_x 
  -2a u_xu \sca{U}{U} U -4a u^2 \sca{U}{U_x}U + 6u^2 \sca{U}{U} U_x 
  \cr \qquad \mbox{}
  -a u^6 U -2a u^4 \sca{U}{U}U -a u^2 \sca{U}{U}^2 U,
  \hspace{13mm} a :  {\rm arbitrary},
  \cr}
 \eea
 \vspace{-2.5mm}
 \bea
  \label{solut2}
  \cases{
  u_t = u_{xxx} + 3 u^2 u_{xx} + 9 u u_x^2 +3 u^4 u_x 
	  +3 u_{xx} \sca{U}{U} +6u_x \sca{U}{U_x} 
  \cr \qquad \mbox{}
	  +2u \sca{U}{U_{xx}} + 2u \sca{U_x}{U_x} 
	  + 10 u_x u^2 \sca{U}{U} + 2u^3 \sca{U}{U_x} 
  \cr \qquad \mbox{}
	  + 3u_x \sca{U}{U}^2 
	  + 6u \sca{U}{U} \sca{U}{U_x} 
	  +u^5 \sca{U}{U} +2u^3 \sca{U}{U}^2 
  \cr \qquad \mbox{}
	  + u\sca{U}{U}^3 
      \vspace{2.5mm}, \cr
  U_t = -2 u_{xx} u U + u_x^2 U -6u_x u^3 U -2 u_x u \sca{U}{U} U 
	  -4u^2 \sca{U}{U_x} U 
  \cr \qquad \mbox{} 
	  -u^6 U -2u^4 \sca{U}{U}U -u^2 \sca{U}{U}^2 U.
  \cr}
 \eea
 }
 \end{theorem}
 \vspace{6pt}
 \noindent
Both system (\ref{solut1}) and system (\ref{solut2}) admit the reduction 
$U = \vt{0}$. From this viewpoint, they are considered as generalizations 
of the 
Ibragimov--Shabat equation \cite{Ibragimov}. 
In addition, 
system (\ref{solut1}) admits the reduction 
$u=0$ which 
changes it to a vector 
analogue of the Ibragimov--Shabat equation \cite{SWo0,SWo}, 
 \begin{equation}
 U_t = U_{xxx} +3\sca{U}{U} U_{xx} + 6\sca{U}{U_x}U_x + 3\sca{U_x}{U_x}U
	  + 3 \sca{U}{U}^2 U_x.
 \label{vIS}
 \end{equation}
We can linearize 
(\ref{solut1}) and (\ref{solut2}) 
through the same 
change of dependent variables. 
In fact, both of them 
are 
$3^{\mbox{\scriptsize rd }}$order symmetries of a 
nontrivial $1^{\mbox{\scriptsize st }}$order system, 
 \bea
 \label{1stIS}
 \cases{
  u_{t_1} = u_{x} + u \sca{U}{U} 
      \vspace{1.5mm}, \cr
  U_{t_1} = -u^2 U, 
  \cr}
 \eea
which is naturally linearizable in the same way. 

 \subsection{Integrability of systems (\ref{solut1}) and (\ref{solut2})}
 \label{int-all3}

 \subsubsection{System (\ref{solut1})}
 \label{Solut-1}
 We note that 
 system (\ref{solut1}) possesses 
 the following conservation law:
  \bea
  && \bigl( u^2 + \sca{U}{U} \bigr)_t 
  \cr &= \hspace{-2mm}& 
	  \bigl[ (a+1) (2uu_{xx} -u_x^2 + 6 u^3 u_{x} + u^6)
	  +2\sca{U}{U_{xx}} -\sca{U_x}{U_x} 
  \cr && \mbox{}         
	  +(4a+6)u^2 \sca{U}{U_x} 
	  + (2a+6) u_x u \sca{U}{U} +(2a+3) u^4 \sca{U}{U} 
  \cr && \mbox{} 
	  + (a+3) u^2 \sca{U}{U}^2
	  +6 \sca{U}{U} \sca{U}{U_x} + \sca{U}{U}^3 \bigr]_x.
  \label{cons1}
  \eea
Then, if we introduce 
new variables $w$ and $W$ by 
  \bea
  \label{ISlin}
  \cases{
  w \equiv u \e^{\int^x (u^2 + \sca{U}{U}) \d x'}\vspace{1.5mm}\hspace{-1pt},
  \cr
  W \equiv U \e^{\int^x (u^2 + \sca{U}{U}) \d x'}\hspace{-1pt},
  \cr}
  \eea
 they satisfy a pair of linear equations,
  \bea
  \cases{
  w_t = (a+1) w_{xxx}
  \vspace{1.5mm}, \cr
  W_t = W_{xxx}.
  \cr }
  \nonumber
  \eea
 If we set $U=\vt{0}$ or $u=0$, (\ref{ISlin}) is reduced to 
 the linearizing transformation for the 
 Ibragimov--Shabat equation \cite{SoSh,Calog} and 
 that for its 
vector 
analogue (\ref{vIS}), respectively. 

  \subsubsection{System (\ref{solut2})}
  \label{Solut-2}
System (\ref{solut2}) is obtained from (\ref{solut1}) by 
rescaling $t$ appropriately 
and taking the limit $a \to \infty$. 
As this fact implies in combination with 
(\ref{cons1}), 
system (\ref{solut2}) possesses 
 the following conservation 
  law:
  \bea
  \bigl( u^2 + \sca{U}{U} \bigr)_t 
  & \hspace{-3mm}=\hspace{-3mm}& 
	  \bigl( 2uu_{xx} -u_x^2 + 6 u^3 u_{x} + u^6
	  + 2u_x u \sca{U}{U} 
  \cr && \mbox{} 
	  +4u^2 \sca{U}{U_x} +2 u^4 \sca{U}{U} 
	  + u^2 \sca{U}{U}^2 \bigr)_x.
  \nonumber
  \eea
Then, by the same change of variables
as in section~\ref{Solut-1}, 
  \bea
  \cases{
  w= u \e^{\int^x (u^2 + \sca{U}{U}) \d x'}\vspace{1.5mm}\hspace{-1pt},\cr
  W= U \e^{\int^x (u^2 + \sca{U}{U}) \d x'}\hspace{-1pt},
  \cr}
  \nonumber
  \eea
 system (\ref{solut2}) is decoupled into one linear equation 
 and one trivial equation,
  \bea
  \cases{
  w_t = w_{xxx}
  \vspace{1.5mm}, \cr
  W_t = \mathbf{0}.
  \cr} 
  \nonumber
  \eea

 \newpage
 \noindent
 \section{The case ${\lambda_1=\frac{1}{3}}$, ${\lambda_2= \frac{2}{3}}$
 $\;\,$-- negative results --}
 \label{lambda1-3}
 \setcounter{equation}{0}

 In this section, we 
 search for $2^{\mbox{\scriptsize nd }}$order 
 and $3^{\mbox{\scriptsize rd }}$order systems with 
 a specific order 
 symmetry 
 in the case of $\lambda_1 =\frac{1}{3}$, $\lambda_2 =\frac{2}{3}$. 
 However, 
 the 
 results 
 turn out to be negative, 
 as is shown 
below. 

 The general ansatz for a $\lambda_1 =\frac{1}{3}$, $\lambda_2= \frac{2}{3}$ 
 homogeneous evolutionary 
 system of $2^{\mbox{\scriptsize nd }}$order for a scalar function 
 $u$ and a vector function $U$ takes the form
 \bea
 \label{l1323o2}
 \cases{
 u_{t_2} = a_1 u_{xx} + a_2 u^3 u_{x} + a_3 u^7 
	 + a_4 \sca{U}{U_x}+ a_5 u^3 \sca{U}{U} 
     \vspace{1.5mm}, \cr
 U_{t_2} = a_{6} U_{xx} + a_{7} u^2 u_x U + a_{8} u^3 U_x + a_9 u^6 U
	 + a_{10} u^2 \sca{U}{U} U.
 \cr}
 \eea
 The following constraints guarantee the order to be 2 and
 the system not to be triangular:
 \[
 (a_1, a_6) \neq (0,0), \hspace{5mm} (a_4,a_5) \neq (0,0), 
	 \hspace{5mm} (a_7,a_8,a_9,a_{10}) \neq (0,0,0,0).
 \]
 Similarly, 
 the general ansatz for a 
 $3^{\mbox{\scriptsize rd }}$order system 
 takes the form
 \bea
 \label{l1323o3}
 \cases{
 u_{t_3} = b_1 u_{xxx} + b_2 u^3 u_{xx} + b_3 u^2 u_x^2 + b_4 u^6 u_x 
	 + b_5 u^{10} + b_6 \sca{U}{U_{xx}} 
 \cr \qquad \; \mbox{}
	 + b_7 \sca{U_x}{U_x} 
	 + b_8 u^2 u_x \sca{U}{U} + b_9 u^3 \sca{U}{U_x} 
	 + b_{10} u^6 \sca{U}{U} 
 \cr \qquad \; \mbox{}
	 + b_{11} u^2 \sca{U}{U}^2
     \vspace{1.5mm}, \cr
 U_{t_3} = b_{12} U_{xxx} + b_{13} u^2 u_{xx} U + b_{14} u u_x^2 U
	 + b_{15} u^2 u_x U_x + b_{16} u^3 U_{xx} 
 \cr \qquad \; \mbox{}
	 + b_{17} u^5 u_x U + b_{18} u^6 U_x + b_{19} u^9 U
	 + b_{20} u u_x \sca{U}{U} U 
 \cr \qquad \; \mbox{}
	 + b_{21} u^2 \sca{U}{U} U_x 
	 + b_{22} u^2 \sca{U}{U_x}U + b_{23} u^5 \sca{U}{U}U 
 \cr \qquad \; \mbox{}
	 +b_{24} u \sca{U}{U}^2 U, 
 \cr}
 \eea
 for which the following constraints guarantee the order to be 3 and 
 the system not to be triangular: $(b_1, b_{12}) \neq (0,0)$ and
 at least one of $b_6, \ldots, b_{11}$ 
 and one of $b_{13}, \ldots, b_{24}$ must not vanish. 
 However, when we consider a $3^{\mbox{\scriptsize rd }}$order symmetry 
for a $2^{\mbox{\scriptsize nd }}$order system, we 
 relax these constraints as follows (cf.\ section~\ref{Comp_as}): 
 $(b_1, b_{12}) \neq (0,0)$ and at least one of $b_1, \ldots, b_{11}$ 
 and one of $b_{12}, \ldots, b_{24}$ must not vanish. 
 \vspace{6pt}
 \begin{proposition}
 {\it
No $2^{\mbox{\scriptsize nd }}$order system 
 of the form $(\ref{l1323o2})$ with a 
$3^{\mbox{\scriptsize rd }}$order symmetry of the form 
$(\ref{l1323o3})$ 
or a $4^{\mbox{\scriptsize th }}$order symmetry 
exists. 
}
 \end{proposition}
 \vspace{6pt}
 \begin{proposition}
 {\it
No $3^{\mbox{\scriptsize rd }}$order system 
 of the form $(\ref{l1323o3})$ with 
 a $5^{\mbox{\scriptsize th }}$order symmetry 
 exists. 
 }
 \end{proposition}

 \newpage
 \noindent
 \section{The case ${\lambda_1=\frac{2}{3}}$, ${\lambda_2= \frac{1}{3}}$}
 \setcounter{equation}{0}
 \label{lambda2-3}

 In this section, we 
 classify 
 $2^{\mbox{\scriptsize nd }}$order and 
$3^{\mbox{\scriptsize rd }}$order systems in the case of $\lambda_1=\frac{2}{3}$, 
 $\lambda_2 =\frac{1}{3}$. 
 In the first part (section~\ref{com-lis4}), we present 
 complete lists of such systems with a specific order 
 symmetry. 
 In the second part (section~\ref{int-all4}), we 
 prove that 
 the listed systems are linearizable. 

 \subsection{Lists of systems with a higher symmetry}
 \label{com-lis4}
 The general ansatz for a $\lambda_1 =\frac{2}{3}$, $\lambda_2= \frac{1}{3}$ 
 homogeneous evolutionary 
 system of $2^{\mbox{\scriptsize nd }}$order for a scalar function 
 $u$ and a vector function $U$ takes the form
 \bea
 \label{l2313o2}
 \cases{
 u_{t_2} = a_1 u_{xx} + a_2 u^4 + a_3 \sca{U}{U_{xx}} + a_4 \sca{U_x}{U_x} 
	 + a_5 u^3 \sca{U}{U}
 \cr \qquad \; \mbox{}
	 + a_6 u^2 \sca{U}{U}^2
	 + a_7 u \sca{U}{U}^3 + a_8 \sca{U}{U}^4
     \vspace{1.5mm}, \cr
 U_{t_2} = a_{9} U_{xx} + a_{10} u^3 U + a_{11} u^2 \sca{U}{U} U 
	 + a_{12} u \sca{U}{U}^2 U + a_{13} \sca{U}{U}^3 U.
 \cr}
 \eea
 The following constraints guarantee the order to be 2 and
 the system not to be triangular:
 \bea
 && (a_1, a_3, a_9) \neq (0,0,0), \hspace{5mm} 
 (a_3,a_4,a_5,a_6,a_7,a_8) \neq (0,0,0,0,0,0), 
 \nonumber \\
 &&	
 (a_{10},a_{11},a_{12}) \neq (0,0,0).
 \nonumber
 \eea
 Similarly, the general ansatz for a 
 $3^{\mbox{\scriptsize rd }}$order system 
 takes the form
 \bea
 \label{l2313o3}
 \cases{
 u_{t_3} = b_1 u_{xxx} + b_2 u^3 u_{x} + b_3 \sca{U}{U_{xxx}} 
	 + b_4 \sca{U_x}{U_{xx}} + b_5 u^2 u_x \sca{U}{U} 
 \cr \qquad \; \mbox{}
	 + b_6 u^3 \sca{U}{U_x} + b_7 u u_x \sca{U}{U}^2 
	 + b_8 u^2 \sca{U}{U} \sca{U}{U_x} 
 \cr \qquad \; \mbox{}
	 + b_9 u_x \sca{U}{U}^3
	 + b_{10} u \sca{U}{U}^2 \sca{U}{U_x} 
	 + b_{11} \sca{U}{U}^3 \sca{U}{U_x}
     \vspace{1.5mm}, \cr
 U_{t_3} = b_{12} U_{xxx} + b_{13} u^2 u_{x} U + b_{14} u^3 U_x 
	 + b_{15} u u_x \sca{U}{U} U 
 \cr \qquad \; \mbox{}
	 + b_{16} u^2 \sca{U}{U} U_{x} 
	 + b_{17} u^2 \sca{U}{U_x} U + b_{18} u_x \sca{U}{U}^2 U 
 \cr \qquad \; \mbox{}
	 + b_{19} u \sca{U}{U}^2 U_x 
	 + b_{20} u \sca{U}{U} \sca{U}{U_x} U 
	 + b_{21} \sca{U}{U}^3 U_x 
 \cr \qquad \; \mbox{}
	 + b_{22} \sca{U}{U}^2 \sca{U}{U_x} U,
 \cr}
 \eea
 for which 
 the following constraints guarantee the order to be 3 and
 the system not to be triangular: $(b_1, b_3, b_{12}) \neq (0,0,0)$ and
 at least one of $b_3, \ldots, b_{11}$ 
 and one of $b_{13}, \ldots, b_{20}$ must not vanish. 
 However, when we consider a $3^{\mbox{\scriptsize rd }}$order symmetry 
for a $2^{\mbox{\scriptsize nd }}$order system, we 
 relax these constraints as follows (cf.\ section~\ref{Comp_as}): 
 $(b_1, b_3, b_{12}) \neq (0,0,0)$ and at least one of $b_1, \ldots, b_{11}$ 
 and one of $b_{12}, \ldots, b_{22}$ must not vanish. 
 \vspace{6pt}
 \begin{theorem}
 {\it
 Any $2^{\mbox{\scriptsize nd }}$order 
 system of the form $(\ref{l2313o2})$ 
 with a $3^{\mbox{\scriptsize rd }}$order symmetry 
 of the form $(\ref{l2313o3})$ 
 has to coincide with the following system up to a scaling of 
 $t_2,x,u,U \,($we omit the subscript of $t_2)$:
 \bea
 \label{l2313-1}
 \cases{
 u_t = u_{xx} + 2 \sca{U}{U_{xx}} + 2 \sca{U_x}{U_x} + 2u \sca{U}{U}^3 
	 + 2 \sca{U}{U}^4 
     \vspace{1.5mm}, \cr
 U_t = - u \sca{U}{U}^2 U - \sca{U}{U}^3 U.
 \cr}
 \eea
 }
 \end{theorem}
 \vspace{6pt}
 \begin{theorem}
 {\it
 Any $2^{\mbox{\scriptsize nd }}$order 
 system of the form $(\ref{l2313o2})$ with a $4^{\mbox{\scriptsize th }}$order 
 symmetry has to coincide with either $(\ref{l2313-1})$ or 
 the following system up to a scaling of 
 $t_2,x,u,U$:
 \bea
 \label{l2313-2}
 \cases{
 u_t = -2 \sca{U}{U_{xx}} - 2 u^3 \sca{U}{U} 
	 -6 u^2 \sca{U}{U}^2 -6 u \sca{U}{U}^3 - 2 \sca{U}{U}^4 
     \vspace{1.5mm}, \cr
 U_t = U_{xx} + u^3 U + 3 u^2 \sca{U}{U} U + 3 u \sca{U}{U}^2 U 
	 + \sca{U}{U}^3 U.
 \cr}
 \eea
 }
 \end{theorem}
 \vspace{6pt}
 \begin{theorem}
 {\it
 Any $3^{\mbox{\scriptsize rd }}$order 
 system of the form 
 $(\ref{l2313o3})$ with a $5^{\mbox{\scriptsize th }}$order 
 symmetry has to coincide with either of the following 
two systems 
 up to a scaling of $t_3,x,u,U \,($we omit the subscript of $t_3)$:
 \bea
 \label{l2313-3}
 \cases{
 u_t = u_{xxx} + 2 \sca{U}{U_{xxx}} + 6 \sca{U_x}{U_{xx}} + 2u_x \sca{U}{U}^3 
	 + 4 \sca{U}{U}^3 \sca{U}{U_x} 
     \vspace{1.5mm}, \cr
 U_t = - u_x \sca{U}{U}^2 U - 2 \sca{U}{U}^2 \sca{U}{U_x} U,
 \cr}
 \eea
 \vspace{-2.5mm}
 \bea
 \label{l2313-4}
 \cases{
 u_t = u_{xxx} + 2 \sca{U}{U_{xxx}} + 6 \sca{U_x}{U_{xx}} + 2u_x \sca{U}{U}^3 
	 + 4 \sca{U}{U}^3 \sca{U}{U_x} 
     \vspace{1.5mm}, \cr
 U_t = - u_x \sca{U}{U}^2 U -4u \sca{U}{U} \sca{U}{U_x} U 
	 +4u \sca{U}{U}^2 U_x 
 \cr \qquad \mbox{}
	 - 6 \sca{U}{U}^2 \sca{U}{U_x} U+ 4 \sca{U}{U}^3 U_x.
 \cr}
 \eea
 }
 \end{theorem}
 \vspace{6pt}
 \noindent
 We note 
that (\ref{l2313-3}) is the 
 $3^{\mbox{\scriptsize rd }}$order symmetry of 
 the $2^{\mbox{\scriptsize nd }}$order system (\ref{l2313-1}).

 \subsection{Integrability of systems (\ref{l2313-1})--(\ref{l2313-4})}
 \label{int-all4}
 \subsubsection{Systems (\ref{l2313-1}) and (\ref{l2313-3})}

We 
present a procedure for 
solving system (\ref{l2313-1}) only, 
because 
its $3^{\mbox{\scriptsize rd }}$order symmetry 
(\ref{l2313-3}) can be solved in the same 
way. 
For system (\ref{l2313-1}), if we introduce 
a new variable $w$ by 
\begin{equation}
w \equiv u + \sca{U}{U}, 
\label{wuUU}
\end{equation}
it solves the linear equation, 
 \[
 w_t = w_{xx}. 
 \]
Once we know 
$w(x,t)$ by solving this equation, 
we obtain from the relation 
$\left( \sca{U}{U}^{-2} \right)_t = 4 w$ that 
 \[
 \frac{1}{\sca{U(x,t)}{U(x,t)}^2} = 
	 4\int_0^t w(x, t') \d t' + \frac{1}{\sca{U(x,0)}{U(x,0)}^2}. 
 \]
Then we can determine $u(x,t)$ by using (\ref{wuUU}). 
Finally, noting the relation $\bigl( \sca{U}{U}^{-\hf} U \bigr)_t = \vt{0}$, 
we obtain the following expression for $U(x,t)$: 
 \[
 U (x,t) = \frac{1}{\left[ 1+ 4 \sca{U(x,0)}{U(x,0)}^2 \int_0^t w(x, t') \d t' 
	 \right]^{\frac{1}{4}}} U(x,0). 
 \]

 \subsubsection{System (\ref{l2313-2})}

For 
system (\ref{l2313-2}), 
we have the relation 
$(u+ \sca{U}{U})_t=0$. 
Thus, we can set 
 \[
 u+ \sca{U}{U} \equiv \phi (x),
 \]
where the function $\phi (x)$ does not depend on $t$. 
Then the equation for $U$ is rewritten in terms of $\phi(x)$ as 
 \begin{equation}
 U_t = U_{xx} + \phi^3 U. 
 \label{timedep-S}
 \end{equation}
The solutions of (\ref{timedep-S}) are 
given by 
 \[
 U(x,t) = \int \d \lambda \, \e^{\lambda t} \Psi(x; \lambda),
 \]
 where $\Psi(x; \lambda)$ is a solution of the 
 ordinary differential equation
 \[
 \Psi_{xx} + \phi^3 \Psi = \lambda \Psi. 
 \]
The following commutation 
relation indicates 
that system (\ref{l2313-2}) possesses a polynomial 
higher symmetry 
of every even order (cf.\ \cite{MikShYam,MiShSok}): 
 \[
 \left[ \partial_x^2 + \phi^3 , \, (\partial_x^2 + \phi^3)^n \right] = 0, 
 \hspace{5mm} n =1, 2, \ldots \,\hspace{1pt}.
 \]

 \subsubsection{System (\ref{l2313-4})}

For system (\ref{l2313-4}), if we introduce 
a new variable $w$ by 
\[
w \equiv u+\sca{U}{U}, 
\]
it solves the linear equation, 
 \[
 w_t = w_{xxx}. 
 \]
Once we 
know $w(x,t)$, 
we obtain from the 
relation $\left( \sca{U}{U}^{-2} \right)_t = 4 w_x$ that
 \begin{equation}
 \frac{1}{\sca{U(x,t)}{U(x,t)}^2} = 
	 4\int_0^t w_x (x, t') \d t' + \frac{1}{\sca{U(x,0)}{U(x,0)}^2}. 
 \label{U^2}
 \end{equation}
Then the equation for $\sca{U}{U}^{-\hf} U$ 
can be rewritten as
 \begin{eqnarray}
 \left(
 \frac{1}{\sqrt{\sca{U}{U}}} U
 \right)_t & \hspace{-2mm} = \hspace{-2mm} & 
 4 (u+ \sca{U}{U}) \sca{U}{U}^2 
 \left(
 \frac{1}{\sqrt{\sca{U}{U}}} U
 \right)_x 
 \nonumber \\
 & \hspace{-2mm} = \hspace{-2mm} & 
 \frac{4 w(x,t)}
 {4 \int^t_0 w_x (x,t') \d t' +\frac{1}{\sca{U(x,0)}{U(x,0)}^2} }
 \left(
 \frac{1}{\sqrt{\sca{U}{U}}} U
 \right)_x. 
 \nonumber 
 \end{eqnarray}
The general solution of this equation is given by 
 \bea
 && \frac{1}{\sqrt{\sca{U}{U}}} U_j = f_j \left(
	 4 \int^t_0 w (x,t') \d t'  
	 + \int^x \frac{1}{\sca{U(x',0)}{U(x',0)}^2} \hspace{1pt}\d x' \right), 
 \nonumber \\ 
 && \hspace{80mm} j=1, 2, \ldots, N, 
 \label{UUU}
 \eea
where $f_1(z), \ldots, f_N(z)$ 
are arbitrary functions of $z$, 
except that they must 
satisfy 
one constraint, 
$\sum_{j=1}^N \left[ f_j(z) \right]^2 =1$. 
Combining (\ref{UUU}) with 
(\ref{U^2}), 
we arrive at 
the following formula: 
 \[
 U_j (x,t) = \frac{1}{(\xi_x)^{\frac{1}{4}}}  f_j(\xi),
 \hspace{5mm} j=1, 2, \ldots, N,
 \]
 where $\xi(x,t) \equiv 4 \int^t_0 w (x,t') \d t' 
	 + \int^x \sca{U(x',0)}{U(x',0)}^{-2}\d x' $.

 \newpage
 \noindent
 \section{Concluding remarks}
In this paper, we have 
presented 
a classification of 
integrable evolutionary systems 
in $1+1$ dimensions 
for 
one scalar unknown $u(x,t)$ and one vector unknown 
$U(x,t)$. 
We focused on 
polynomial systems that are homogeneous under 
a suitable weighting of 
$\partial_x$, $\partial_t$, $u(x,t)$, $U(x,t)$, 
and 
considered 
five distinct 
weightings 
for
\mbox{$u,\hspace{1pt}U$} 
relative to 
a fixed weight of $\partial_x$. 
Then, 
with the help of a computer algebra program, 
we 
obtained 
the complete lists, 
up to a scaling of variables, 
of 
$2^{\mbox{\scriptsize nd }}$order systems with a 
$3^{\mbox{\scriptsize rd }}$order or a 
$4^{\mbox{\scriptsize th }}$order symmetry 
and $3^{\mbox{\scriptsize rd }}$order 
systems with a $5^{\mbox{\scriptsize th }}$order symmetry. 
We demonstrated 
the integrability of 
nearly all listed systems 
by constructing a 
Lax representation or a linearizing transformation, 
or, 
in some cases, 
by 
identifying 
an integrable closed subsystem 
contained in 
the 
system under 
investigation. 

The following tables give a quick overview 
of the systems found. 
Note we use ``MT'' 
as an abbreviation for ``Miura-type transformation'', 
including Miura map plus potentiation. 
In these tables, 
we 
set 
the weight of $\partial_x$ at 
unity, 
without any loss of generality. 
For 
full 
details 
regarding Lax representations, 
transformations, 
references, etc., 
the reader is referred to 
the corresponding 
part 
of the paper 
identified through the equation number. 
Here, 
we would like to make 
a few remarks on 
our classification results: 
\begin{itemize}
\item
The most interesting 
classification 
results 
are 
obtained 
for the case 
$\lambda_1 = \lambda_2 =1$, namely, 
the Burgers/pKdV/mKdV weighting. 
The lists in this case 
consist of 
a large 
number of 
systems, which 
are shown to have a 
very wide variety of 
underlying structures. 
We 
compared these lists thoroughly 
with 
the lists of 
two-component systems by 
Foursov--Olver \cite{Foursov1,Foursov2,Foursov3}, 
refined 
and generalized 
their work, 
as described 
in the introduction. 
\item
We found 
a 
number of 
pairs/triplets of 
scalar-vector systems 
connected 
through 
transformations of dependent variables. 
Besides 
standard Miura 
transformations 
that 
map 
both 
the 
scalar 
and 
vector variables 
to new ones 
(see, {\it e.g.\ }\/(\ref{Miura1})), 
we 
also 
found 
Miura-type 
transformations 
that 
act only on 
the scalar variable 
and 
do not change 
the vector 
variable 
(see, {\it e.g.\ }\/(\ref{sol3-1}) 
combined with 
potentiation $v=\hat{u}_x$). 
For some other 
systems, 
we showed that 
a new scalar variable 
defined 
in terms of the 
old scalar and vector variables satisfies 
a 
closed 
integrable 
equation, 
such as the KdV equation or a linear equation. 
\item
The search for such transformations
in our case of 
scalar-vector systems 
is simple 
in comparison to 
scalar-scalar systems.
For instance, 
the ansatz that a new scalar variable 
depends on the original 
vector variable $U$ only through scalar products 
$\sca{\partial_x^m U}{\partial_x^n U}$ 
narrows down the candidates for 
such transformations considerably. 
This 
leads us to 
the counter-intuitive observation 
that 
scalar-vector systems are, in a sense, 
more tractable than 
scalar-scalar systems. 
This is probably one 
reason why, unlike our 
work, 
Foursov and Olver 
proved 
integrability\footnote{They 
discussed 
the existence of 
a recursion operator or a bi-Hamiltonian structure.} 
for 
only a small proportion 
of their two-component 
systems \cite{Foursov1,Foursov2,Foursov3}. 
 \end{itemize}

Finally, we 
mention 
some 
problems not solved 
in this paper: 
 \begin{itemize}
 \item
How can 
the integrability of the three systems 
(\ref{sol12'}), (\ref{sol12}) and (\ref{sol11}) 
be established 
along the lines of this paper? 
The 
main 
obstacle 
is that we 
know neither a linearizing transformation 
nor a proper 
Lax representation 
for 
the two-component Burgers system (\ref{cBurgers}). 
The 
dependence of the functional form of 
travelling-wave solutions on the boundary conditions 
and the velocity implies that 
(\ref{cBurgers}) 
is a 
highly nontrivial system 
and not linearizable by a naive 
extension of the Hopf--Cole transformation. 
\item 
Some scalar-vector systems are 
converted to 
a triangular form, i.e.\ 
a closed subsystem plus remaining equations coupled to it. 
When the remaining equations contain 
a nonlinear PDE 
in 
its 
own 
variable 
(cf.\ 
(\ref{uweq}) or (\ref{uweq2})), 
it 
seems to be 
especially 
difficult 
to solve 
them 
explicitly 
for a 
given solution of the 
subsystem. 
Is there any 
method, 
like an extension of the 
inverse scattering method, 
for dealing with 
such triangular systems analytically?
  \item
Can one 
construct an explicit formula for the general solution of 
(\ref{shockwave})? 
This 
equation is obtained 
from system (\ref{sol4}) or (\ref{sol5}) 
by converting 
it to a triangular form and 
then 
considering the 
special case in which 
the solution 
of the 
subsystem, KdV equation in this case, 
is identically zero. 
 \end{itemize}

Although we concentrated our attention on 
the five distinct 
weightings 
for 
\mbox{$u, \hspace{1pt}U$} 
in this paper, 
we 
also found 
integrable 
systems 
that are 
homogeneous under 
a 
different weighting 
of variables. 
Namely, we obtained 
systems of 
coupled KdV-mKdV type, {\it e.g.\ }\/(\ref{334}), 
(\ref{compact}), 
(\ref{compact2}) and 
(\ref{385})\footnote{We note that, 
in addition to a 
scaling of variables, 
these systems 
admit another 
equivalence 
transformation 
$\tilde{u} = u + k \sca{U}{U}$, $\hspace{1pt}\tilde{U}=U$.}, 
together with the proof of their integrability.
We are planning 
to complete 
a classification of 
integrable 
systems of this type, i.e.\ 
scalar-vector 
systems with weights 
$\lambda_1 = 2, \; \lambda_2 = 1$, 
in a subsequent paper. 
Preliminary 
results can be 
viewed on the web page 
\\
\hspace*{22mm}
{\tt http://lie.math.brocku.ca/twolf/htdocs/sv/over.html}~.

 \begin{table}
 \scriptsize
 \centering
 \caption{An overview of the 
considered classes 
with unit weighting of $\partial_x$}
 \vspace*{6pt}\hspace*{-1.0cm}\begin{tabular}{|c|c|l|l|} \hline
 $\!\!$weights$\!\!$ & $\!\!$weights of$\!\!$ &          &       \\
 $\!\!(\lambda_1,\lambda_2)\!\!$ & $\!\! \partial_t$, $\partial_\tau$
 in$\!\!$ & system & comments \\
 $\!\!$of $u, U\!\!$ & $\!\!$sys., sym.$\!\!$&      &
 \\ \hline \hline 
      & 2, 3 & \mbox{none} &
 \\ \cline{2-4} 
 (2,\,2)& 2, 4 & \mbox{none} &
 \\ \cline{2-4} 
      & 3, 5 & \,\,(\ref{cDS}) $\cases{
 u_t = \sca{U}{U_x}
     \vspace{1.0mm}, \cr
 U_t = U_{xxx} + u_x U + 2u U_x.
\cr}_{\vphantom {\displaystyle \sum}}^{\vphantom {\displaystyle \sum}}
$ 
  & \parbox{5cm}{
 multi-component generalization of 
a Drinfel'd--Sokolov system \cite{Wilson,Drinfeld}, 
see \cite{Melnikov,Strampp1,Strampp2}}
 \\ \cline{3-4} 
      &       & \,\,(\ref{cKdV}) 
$\cases{
 u_t = u_{xxx} + 6 uu_x - 6 \sca{U}{U_x}
     \vspace{1.0mm}, \cr
 U_t = U_{xxx} + 6 u_x U + 6 u U_x.
\cr}_{\vphantom {\displaystyle \sum}}^{\vphantom {\displaystyle \sum}}
$
 & \parbox{5cm}{
 known as a Jordan KdV system \cite{SWo,Svi0,Svi,Ad}}
 \\ \cline{3-4} 
      &       & \,\,(\ref{cIto}) $\cases{
 u_t = u_{xxx} + 3 uu_x +3 \sca{U}{U_x}
     \vspace{1.0mm}, \cr
 U_t = u_x U + u U_x.
\cr}_{\vphantom {\displaystyle \sum}}^{\vphantom {\displaystyle \sum}}
$
  & \parbox{5cm}{multi-component 
generalization of Zakharov--Ito 
system \cite{Zakh,Ito}, 
see \cite{Kuper}}
 \\ \cline{3-4} 
      &       & \,\,(\ref{cHS}) $\cases{
 u_t = u_{xxx} + 6 uu_x -12 \sca{U}{U_x}
     \vspace{1.0mm}, \cr
 U_t = -2 U_{xxx} - 6u U_x.
\cr
}_{\vphantom {\displaystyle \sum}}^{\vphantom {\displaystyle \sum}}
$
  & \parbox{5cm}{ 
multi-component generalization of 
Hirota--Satsuma system \cite{HiSa}, see \cite{HiroOh}}
 \\ \hline 
 (1,\,1)& 
$\begin{array}{c} 
\\ \mbox{2, 3} \\ \mbox{or}\\ 
\mbox{2, 4} \end{array}$
& \,\,(\ref{sol25'}) $\cases{
 u_{t} = \frac{1}{3} (1+2a) (u_{xx} + 2uu_x) + \frac{4}{3} \sca{U}{U_x}
 \vspace{1.0mm}, \cr
 U_{t} = U_{xx} + \frac{1}{3} (1-a) u_x U + uU_x + \frac{1}{12} (1-4a) u^2 U 
 \cr \qquad\, \mbox{} 
- \frac{1}{3} \sca{U}{U}U,
 \hspace{25mm} a :  {\rm arbitrary}.
\cr
}_{\vphantom {\displaystyle \sum}}^{\vphantom {\displaystyle \sum}}
$
  & \parbox{5cm}{linearized 
by an extended Hopf--Cole transformation}
 \\ \cline{3-4} 
      &       & \,\,(\ref{sol13'}) $\cases{
 u_{t} = u_{xx} + 2uu_x +2 \sca{U}{U_x}
 \vspace{1.0mm}, \cr
 U_{t} = - \hf u_x U - \hf u^2 U - \hf \sca{U}{U}U.
\cr}_{\vphantom {\displaystyle \sum}}^{\vphantom {\displaystyle \sum}}
$
  & \parbox{5cm}{is scaling limit of (\ref{sol25'}), 
linearized 
by the same transformation}
 \\ \cline{3-4} 
      &       & \,\,(\ref{sol12'}) $\cases{
 u_{t} = u_{xx} + 2uu_x + \sca{U}{U_x}
 \vspace{1.0mm} ,\cr
 U_{t} = \hf u_x U + u U_x.
\cr}_{\vphantom {\displaystyle \sum}}^{\vphantom {\displaystyle \sum}}
$
  & \parbox{5cm}{
contains 
two-component Burgers system 
(\ref{cBurgers}) as closed subsystem, 
{\it integrability 
unproven}
} 
 \\ \cline{2-4} 
      & 3, 5      & \,\,(\ref{sol25}) $\cases{
 u_t = a (u_{xxx} + 3 uu_{xx} + 3 u_x^2 +3 u^2 u_x) + u_x \sca{U}{U} 
    +2u \sca{U}{U_x} 
 \cr \qquad\, \mbox{}
	 + 2\sca{U}{U_{xx}} + 2\sca{U_x}{U_x} 
     \vspace{1.0mm}, \cr
 U_t = U_{xxx} + \hf (1-a) u_{xx} U + \frac{3}{2} u_x U_x 
	+ \frac{3}{2}u U_{xx} +\frac{3}{4} (1-2a) uu_x U 
 \cr \qquad\, \mbox{} + \frac{3}{4} u^2 U_x 
	  - \sca{U}{U_x}U +\frac{1}{8}(1-4a) u^3 U - \hf u \sca{U}{U} U,
 \cr \hspace{68mm} a :  {\rm arbitrary}.
\cr}_{\vphantom {\displaystyle \sum}}^{\vphantom {\displaystyle \sum}}
$
  & \parbox{5cm}{is symmetry of (\ref{sol25'})}
 \\ \cline{3-4} 
      &       & \,\,(\ref{sol13}) $\cases{
 u_t = u_{xxx} + 3uu_{xx} + 3u_x^2 +3 u^2 u_x + u_x \sca{U}{U} 
    +2u \sca{U}{U_x} 
 \cr \qquad\, \mbox{}
	 + 2\sca{U}{U_{xx}} + 2\sca{U_x}{U_x} 
     \vspace{1.0mm}, \cr
 U_t = -\hf u_{xx} U -\frac{3}{2} uu_x U - \sca{U}{U_x}U -\hf u^3 U 
       - \hf u \sca{U}{U} U.
\cr}_{\vphantom {\displaystyle \sum}}^{\vphantom {\displaystyle \sum}}
$
  & \parbox{5cm}{is symmetry of (\ref{sol13'}), 
scaling limit of (\ref{sol25})}
 \\ \cline{3-4} 
      &       & \,\,(\ref{sol12}) $\cases{
 u_t = u_{xxx} + 3uu_{xx} + 3u_x^2 +3 u^2 u_x + u_x \sca{U}{U} 
    +2u \sca{U}{U_x} 
 \cr \qquad\, \mbox{}
	 + \sca{U}{U_{xx}} + \sca{U_x}{U_x} 
     \vspace{1.0mm}, \cr
 U_t = \hf u_{xx} U + u_x U_x + uu_x U +u^2 U_x + \hf \sca{U}{U} U_x 
	 + \hf \sca{U}{U_x}U.
\cr}_{\vphantom {\displaystyle \sum}}^{\vphantom {\displaystyle \sum}}
$
  & \parbox{5cm}{is symmetry of (\ref{sol12'}), see there}
 \\ \cline{3-4} 
      &       & \,\,(\ref{sol11}) $\cases{
 u_t = u_{xxx} + 3uu_{xx} + 3u_x^2 +3 u^2 u_x + u_x \sca{U}{U} 
    +2u \sca{U}{U_x} 
 \cr \qquad\, \mbox{}
	 + \sca{U}{U_{xx}} + \sca{U_x}{U_x} 
     \vspace{1.0mm}, \cr
 U_t = \hf u_{xx} U + u_x U_x + uu_x U +u^2 U_x + \sca{U}{U} U_x.
\cr}_{\vphantom {\displaystyle \sum}}^{\vphantom {\displaystyle \sum}}
$
  & \parbox{5cm}{contains 
$3^{\mbox{\tiny rd }}$order symmetry of 
two-component Burgers 
system (\ref{cBurgers}), 
{\it integrability unproven}}
 \\ \cline{3-4} 
      &       & (\ref{sol1}) $\cases{
 u_t = 3u_x \sca{U}{U} +3 \sca{U}{U_{xx}} 
       -3 \sca{U}{U}^2  \vspace{1.0mm}, \cr
 U_t = U_{xxx} + u_{xx} U + u_x U_x -3\sca{U}{U_x}U.
\cr}_{\vphantom {\displaystyle \sum}}^{\vphantom {\displaystyle \sum}}
$
  & \parbox{5cm}{obtained from (\ref{sol3}) by MT}
 \\ \cline{3-4} 
      &       & (\ref{sol2}) $\cases{
 u_t = 2u_x \sca{U}{U} +2 \sca{U}{U_{xx}}- \sca{U_x}{U_x}
       -2 \sca{U}{U}^2  \vspace{1.0mm}, \cr
 U_t = U_{xxx} + u_{xx} U + 2u_x U_x -2\sca{U}{U}U_x-2\sca{U}{U_x}U.
\cr}_{\vphantom {\displaystyle \sum}}^{\vphantom {\displaystyle \sum}}
$
  & \parbox{5cm}{linearizable by change of variable, 
related to 
Kaup--Kupershmidt equation in a certain manner}
 \\ \cline{3-4} 
      &       & (\ref{sol3}) $\cases{
 u_t = u_x \sca{U}{U} +2u \sca{U}{U_x} 
       + \sca{U}{U_{xx}} +  \sca{U_x}{U_x}  \vspace{1.0mm}, \cr
 U_t = U_{xxx} + u_{xx} U + u_x U_x -2 uu_x U -u^2 U_x 
       + \sca{U}{U}U_x 
	- \sca{U}{U_x}U.
\cr}_{\vphantom {\displaystyle \sum}}^{\vphantom {\displaystyle \sum}}
$
  & \parbox{5cm}{connected with 
(\ref{cDS}) and (\ref{sol1}) by MT}
 \\ \cline{3-4} 
      &       & (\ref{sol9}) $\cases{
 u_t = u_{xxx} + \frac{3}{2} u_x^2 + \frac{3}{2}\sca{U_x}{U_x} 
     \vspace{1.0mm}, \cr
 U_t = u_x U_x.
\cr}_{\vphantom {\displaystyle \sum}}^{\vphantom {\displaystyle \sum}}
$
  & \parbox{5cm}{is potential form of 
(\ref{cIto})}
 \\ \cline{3-4} 
      &       & (\ref{sol10}) $\cases{
 u_t = u_{xxx} + 3 u_x^2 + 2a u_x \sca{U}{U} 
    + a \sca{U}{U_{xx}}+a \sca{U_x}{U_x} +b \sca{U}{U}^2  
     \vspace{1.0mm}, \cr
 U_t = u_{xx} U + 2 u_x U_x + a \sca{U}{U}U_x + a \sca{U}{U_x} U, 
 \quad (a,b) \neq (0,0).
\cr}_{\vphantom {\displaystyle \sum}}^{\vphantom {\displaystyle \sum}}
$
  & \parbox{5cm}
{for $b \neq a^2/4$, 
transformed to 
(\ref{cIto}); 
for $b = a^2/4$, 
to 
KdV equation + linear vector 
equation coupled to it}
 \\ \hline 
 \end{tabular}
 \normalsize
 \end{table}

 \begin{table}
 \scriptsize
 \centering
 \caption{Continuation}
 \vspace*{6pt}\hspace*{-1cm}\begin{tabular}{|c|c|l|l|} \hline
 $\!\!$weights$\!\!$ & $\!\!$weights of$\!\!$ &          &       \\
 $\!\!(\lambda_1,\lambda_2)\!\!$ & $\!\! \partial_t$, $\partial_\tau$
 in$\!\!$ & system & comments \\
 $\!\!$of $u, U\!\!$ & $\!\!$sys., sym.$\!\!$&      &             
 \\ \hline \hline 
 (1,\,1)& 3, 5 & (\ref{sol14}) $\cases{
 u_t = u_{xxx} + 3 u_x^2 - 3\sca{U_x}{U_x} 
     \vspace{1.0mm}, \cr
 U_t = U_{xxx} + 6 u_x U_x.
\cr}_{\vphantom {\displaystyle \sum}}^{\vphantom {\displaystyle \sum}}$
  & \parbox{5cm}{is potential form of 
(\ref{cKdV}),
 MT connects to (\ref{sol16})}
 \\ \cline{3-4} 
      &       & (\ref{sol19}) $\cases{
 u_t = u_{xxx} + 3 u_x^2 + u_x \sca{U}{U} + \sca{U}{U_{xx}}
     \vspace{1.0mm}, \cr
 U_t = U_{xxx} + 3 u_{xx} U + 3 u_x U_x + \sca{U}{U_x}U.
\cr}_{\vphantom {\displaystyle \sum}}^{\vphantom {\displaystyle \sum}}$
  & \parbox{5cm}{obtained from (\ref{sol22}) by MT}
 \\ \cline{3-4} 
      &       & (\ref{sol18}) $\cases{
 u_t = u_{xxx} + 3 u_x^2 + 2 u_x \sca{U}{U} + \sca{U}{U_{xx}}
 + \hf \sca{U_x}{U_x} 
     \vspace{1.0mm}, \cr
 U_t = U_{xxx} + 6 u_{xx} U + 6u_x U_x + 2 \sca{U}{U_x}U.
\cr}_{\vphantom {\displaystyle \sum}}^{\vphantom {\displaystyle \sum}}$
  & \parbox{5cm}{obtained 
 from (\ref{sol23}) by MT}
  \\ 
 \cline{3-4} 
       &       & (\ref{sol20}) $\cases{
  u_t = u_{xxx} +3 u_x^2 + 4 u_x \sca{U}{U} 
	+ 2 \sca{U}{U_{xx}} + \sca{U_x}{U_x} + \frac{2}{3} \sca{U}{U} ^2
      \vspace{1.0mm}, \cr
  U_t = -2 U_{xxx} - 6 u_{xx} U - 6u_x U_x -4 \sca{U}{U_x} U.
 \cr}_{\vphantom {\displaystyle \sum}}^{\vphantom {\displaystyle \sum}}$
   & \parbox{5cm}{obtained 
 from (\ref{sol24}) by MT}
  \\ 
 \cline{3-4} 
       &       & (\ref{sol21}) $\cases{
  u_t = u_{xxx} + u_x^2 
	-12 \sca{U}{U_{xx}} + 12 \sca{U_x}{U_x} -4 \sca{U}{U} ^2
      \vspace{1.0mm}, \cr
  U_t = 4 U_{xxx} + u_{xx} U +2 u_x U_x +4 \sca{U}{U} U_x +4 \sca{U}{U_x}U.
 \cr}_{\vphantom {\displaystyle \sum}}^{\vphantom {\displaystyle \sum}}$
   & \parbox{5cm}{reduced to 
 triangular system: KdV equation + linear vector equation 
coupled to it}
 \\ 
 \cline{3-4} 
       &       & (\ref{sol4}) $\cases{
  u_t = u_{xxx} - \frac{3}{2} u^2 u_x + \frac{3}{2} u_x \sca{U}{U} 
     +u \sca{U}{U_x} + \sca{U}{U_{xx}}  + \sca{U_x}{U_x} 
  \vspace{1.0mm}, \cr
  U_t = -u_x U_x - \hf u^2 U_x + \frac{3}{2} \sca{U}{U}U_x.
 \cr}_{\vphantom {\displaystyle \sum}}^{\vphantom {\displaystyle \sum}}$
   & \parbox{5cm}
{converted to triangular system: 
KdV eq.\ 
+ nonlinear eq.\ 
with interesting reduction 
(\ref{shockwave}) + linear vector eq.}
  \\ \cline{3-4} 
       &       & (\ref{sol5}) $\cases{
  u_t = u_{xxx} - \frac{3}{2} u^2 u_x + \frac{3}{2} u_x \sca{U}{U} 
     +u \sca{U}{U_x} + \sca{U}{U_{xx}}  + \sca{U_x}{U_x} 
  \vspace{1.0mm}, \cr
  U_t = -u_x U_x - \hf u^2 U_x + \hf \sca{U}{U}U_x + \sca{U}{U_x}U.
 \cr}_{\vphantom {\displaystyle \sum}}^{\vphantom {\displaystyle \sum}}$
  & \parbox{5cm}{exactly like for (\ref{sol4}) only different 
linear vector eq.}
  \\ \cline{3-4} 
       &       & (\ref{sol6}) $\cases{
  u_t = u_{xxx} - \frac{3}{2} u^2 u_x + \frac{1}{2} u_x \sca{U}{U} 
     +u \sca{U}{U_x} + \sca{U}{U_{xx}}  + \sca{U_x}{U_x} 
  \vspace{1.0mm}, \cr
  U_t = u_{xx} U + u_x U_x -u u_x U - \hf u^2 U_x 
	+ \hf \sca{U}{U}U_x 
 \cr \qquad\, \mbox{} 
	+ \sca{U}{U_x} U.
 \cr}_{\vphantom {\displaystyle \sum}}^{\vphantom {\displaystyle \sum}}$
  & \parbox{5cm}{obtained in \cite{Razb}, converted to 
triangular system: KdV equation + 
linear equations coupled to it, 
admits deformation connected with (\ref{cIto}) by MT} 
  \\ \cline{3-4} 
       &       & (\ref{sol7}) $\cases{
  u_t = u_{xxx} - \frac{3}{2} u^2 u_x + \frac{3}{2} u_x \sca{U}{U} 
     +u \sca{U}{U_x} + \sca{U}{U_{xx}} + \sca{U_x}{U_x} 
  \cr \qquad\, \mbox{}+ \hf \sca{U}{U}^2 
      \vspace{1.0mm}, \cr
  U_t = - u_x U_x - \hf u^2 U_x 
	- \hf \sca{U}{U}U_x + \hf u \sca{U}{U} U.
 \cr}_{\vphantom {\displaystyle \sum}}^{\vphantom {\displaystyle \sum}}$
  & \parbox{5cm}{contains interesting 
triangular 
system (\ref{counter}) 
as closed subsystem: 
KdV equation + 
nonlinear 
equation coupled to it}
  \\ \cline{3-4} 
       &       & (\ref{sol8}) $\cases{
  u_t = u_{xxx} - \frac{3}{2} u^2 u_x + u_x \sca{U}{U} 
     +u \sca{U}{U_x} + \sca{U}{U_{xx}} + \sca{U_x}{U_x} 
  \cr \qquad\, \mbox{}-\frac{1}{4} u^2 \sca{U}{U} + \frac{1}{4} \sca{U}{U}^2 
      \vspace{1.0mm}, \cr
  U_t = \hf u_{xx} U + \hf \sca{U}{U_x}U -\frac{1}{4} u^3 U 
	+ \frac{1}{4}u \sca{U}{U} U.
 \cr}_{\vphantom {\displaystyle \sum}}^{\vphantom {\displaystyle \sum}}$
   & \parbox{5cm}{is 
symmetry of 
$1^{\mbox{\tiny st }}$order system (\ref{sol8-1}), 
converted to triangular system: KdV eq.\ + 
Riccati eq.\ coupled to it 
+ linear vector 
eq.\ coupled to them}
  \\ \cline{3-4} 
       &       & (\ref{sol15}) $\cases{
  u_t = u_{xxx} + u^2 u_x + u_x \sca{U}{U}
      \vspace{1.0mm}, \cr
  U_t = U_{xxx} + u^2 U_x + \sca{U}{U} U_x.
 \cr}_{\vphantom {\displaystyle \sum}}^{\vphantom {\displaystyle \sum}}$
   & \parbox{5cm}{equivalent to a single vector mKdV equation 
 for vector $(u,U)$}
  \\ \cline{3-4} 
       &       & (\ref{sol17}) $\cases{
  u_t = u_{xxx} + 2 u^2 u_x + u_x \sca{U}{U} + u \sca{U}{U_x}
      \vspace{1.0mm}, \cr
  U_t = U_{xxx} + uu_x U + u^2 U_x + \sca{U}{U} U_x + \sca{U}{U_x}U.
 \cr}_{\vphantom {\displaystyle \sum}}^{\vphantom {\displaystyle \sum}}$
   & \parbox{5cm}{equivalent to a single vector mKdV equation 
 for vector $(u,U)$}
  \\ \cline{3-4} 
       &       & (\ref{sol16}) $\cases{
  u_t = u_{xxx} -6 u^2 u_x + 6u_x \sca{U}{U} +12u \sca{U}{U_x} 
      \vspace{1.0mm}, \cr
  U_t = U_{xxx} - 12 uu_x U - 6u^2 U_x + 6 \sca{U}{U}U_x.
 \cr}_{\vphantom {\displaystyle \sum}}^{\vphantom {\displaystyle \sum}}$
   & \parbox{5cm}{known as a Jordan mKdV system \cite{Svi0}, 
connected with (\ref{cKdV}) and 
(\ref{sol14}) by MT}
  \\ \cline{3-4} 
       &       & (\ref{sol22}) $\cases{
  u_t = u_{xxx} -6 u^2 u_x + u_x \sca{U}{U} + 2u \sca{U}{U_x} 
	  + \sca{U}{U_{xx}}+ \sca{U_x}{U_x} 
      \vspace{1.0mm}, \cr
   U_t = U_{xxx} + 3 u_{xx} U + 3 u_x U_x - 6 uu_x U -3 u^2 U_x 
	   + \sca{U}{U} U_x 
  \cr \qquad\, \mbox{} + 3 \sca{U}{U_x}U.
  \cr}_{\vphantom {\displaystyle \sum}}^{\vphantom {\displaystyle \sum}}$
    & \parbox{5cm}{admits 
 Lax representation, 
 connected with (\ref{sol19}) by MT}
  \\ \cline{3-4} 
       &       & (\ref{sol23}) $\cases{
  u_t = u_{xxx} -6 u^2 u_x + u_x \sca{U}{U} + 2u \sca{U}{U_x} 
	  + \sca{U}{U_{xx}}+ \sca{U_x}{U_x} 
      \vspace{1.0mm}, \cr
  U_t = U_{xxx} + 6 u_{xx} U + 6u_x U_x -12 uu_x U -6 u^2 U_x 
	  + \sca{U}{U} U_x 
  \cr \qquad\, \mbox{} + 4 \sca{U}{U_x}U.
 \cr}_{\vphantom {\displaystyle \sum}}^{\vphantom {\displaystyle \sum}}$
   & \parbox{5cm}{multi-component generalization of 
 a modified Jaulent--Miodek flow \cite{Alonso}, 
 admits Lax representation, 
 connected with (\ref{sol18}) by MT}
  \\ \cline{3-4} 
       &       & (\ref{sol24}) $\cases{
  u_t = u_{xxx} -6 u^2 u_x + u_x \sca{U}{U} + 2u \sca{U}{U_x} 
	+ \sca{U}{U_{xx}} + \sca{U_x}{U_x} 
      \vspace{1.0mm}, \cr
  U_t = -2 U_{xxx} - 6 u_{xx} U - 6u_x U_x + 12uu_x U + 6u^2 U_x 
	  + \sca{U}{U} U_x 
  \cr \qquad\, \mbox{} -2 \sca{U}{U_x}U.
 \cr}_{\vphantom {\displaystyle \sum}}^{\vphantom {\displaystyle \sum}}$
   & \parbox{5cm}{connected with 
 (\ref{cHS}) and (\ref{sol20}) by MT}
  \\ \hline 
  \end{tabular}
  \normalsize
  \end{table}

  \begin{table}
  \scriptsize
  \centering
  \caption{Continuation}
  \vspace*{6pt}\hspace*{-1cm}\begin{tabular}{|c|c|l|l|} \hline
  $\!\!$weights$\!\!$ & $\!\!$weights of$\!\!$ &          &       \\
  $\!\!(\lambda_1,\lambda_2)\!\!$ & $\!\! \partial_t$, $\partial_\tau$
  in$\!\!$ & system & comments \\
  $\!\!$of $u, U\!\!$ & $\!\!$sys., sym.$\!\!$&      &             
  \\ \hline \hline 
   & 2, 3 & \mbox{none} &
  \\ \cline{2-4} 
   & 2, 4 & \mbox{none} &
  \\ \cline{2-4} 
  $\bigl(\frac{1}{2},\frac{1}{2}\bigr)$ 
   & 3, 5 & (\ref{solut1}) $\cases{
  u_t = (a+1) (u_{xxx} + 3 u^2 u_{xx} + 9 u u_x^2 +3 u^4 u_x 
	   +3 u_{xx} \sca{U}{U} 
   \cr \qquad\, \mbox{}
	   +6u_x \sca{U}{U_x} + 3u_x \sca{U}{U}^2) +2a u \sca{U}{U_{xx}} 
	   \cr \qquad\, \mbox{}     
   + (2a+3) u \sca{U_x}{U_x}
   + (10a+6 ) u_x u^2 \sca{U}{U} + 2a u^3 \sca{U}{U_x} 
   \cr \qquad\, \mbox{} +6a u \sca{U}{U} \sca{U}{U_x} 
   + a u^5 \sca{U}{U}+ 2a u^3 \sca{U}{U}^2 + au \sca{U}{U}^3 
       \vspace{1.5mm}, \cr
  U_t = U_{xxx} +3\sca{U}{U} U_{xx} + 6\sca{U}{U_x}U_x + 3\sca{U_x}{U_x}U
	   + 3 \sca{U}{U}^2 U_x 
   \cr \qquad\, \mbox{} 
   -2a u_{xx} uU + (a+3) u_x^2 U + 6 uu_x U_x +3u^2 U_{xx} 
    -6 a u_x u^3 U 
   \cr \qquad\, \mbox{}
   + 3u^4 U_x 
   -2a u_xu \sca{U}{U} U -4a u^2 \sca{U}{U_x}U + 6u^2 \sca{U}{U} U_x 
   \cr \qquad\, \mbox{}
   -a u^6 U -2a u^4 \sca{U}{U}U -a u^2 \sca{U}{U}^2 U,
   \hspace{11mm} a : {\rm arbitrary}.
 \cr}_{\vphantom {\displaystyle \sum}}^{\vphantom {\displaystyle \sum}}$
  & \parbox{5cm}{is 
symmetry of 
$1^{\mbox{\tiny st }}$order system (\ref{1stIS}), 
extension of 
vector Ibragimov--Shabat equation, 
linearizable by change of variables}
 \\ \cline{3-4} 
      &       & (\ref{solut2}) $\cases{
  u_t = u_{xxx} + 3 u^2 u_{xx} + 9 u u_x^2 +3 u^4 u_x 
	  +3 u_{xx} \sca{U}{U} +6u_x \sca{U}{U_x} 
  \cr \qquad\, \mbox{}
	  +2u \sca{U}{U_{xx}} + 2u \sca{U_x}{U_x} 
	  + 10 u_x u^2 \sca{U}{U} + 2u^3 \sca{U}{U_x} 
  \cr \qquad\, \mbox{}
	  + 3u_x \sca{U}{U}^2 
	  + 6u \sca{U}{U} \sca{U}{U_x} 
	  +u^5 \sca{U}{U} +2u^3 \sca{U}{U}^2 
  \cr \qquad\, \mbox{}
	  + u\sca{U}{U}^3 
      \vspace{1.5mm}, \cr
  U_t = -2 u_{xx} u U + u_x^2 U -6u_x u^3 U -2 u_x u \sca{U}{U} U 
	  -4u^2 \sca{U}{U_x} U 
  \cr \qquad\, \mbox{} 
	  -u^6 U -2u^4 \sca{U}{U}U -u^2 \sca{U}{U}^2 U.
\cr}_{\vphantom {\displaystyle \sum}}^{\vphantom {\displaystyle \sum}}$
  & \parbox{5cm}
{is 
symmetry of $1^{\mbox{\tiny st }}$order system (\ref{1stIS}), 
scaling limit of (\ref{solut1}), 
linearized by 
the same 
change of variables}
 \\ \hline 
    & 2, 3 & \mbox{none} &
 \\ \cline{2-4} 
 $\bigl(\frac{1}{3},\frac{2}{3}\bigr)$ 
     & 2, 4 & \mbox{none} &
 \\ \cline{2-4} 
      & 3, 5 & \mbox{none} &
 \\ \hline 
 $\bigl(\frac{2}{3},\frac{1}{3}\bigr)$    
      & 2, 3 & (\ref{l2313-1}) $\cases{
 u_t = u_{xx} + 2 \sca{U}{U_{xx}} + 2 \sca{U_x}{U_x} + 2u \sca{U}{U}^3 
	 + 2 \sca{U}{U}^4 
     \vspace{1.0mm}, \cr
 U_t = - u \sca{U}{U}^2 U - \sca{U}{U}^3 U.
\cr}_{\vphantom {\displaystyle \sum}}^{\vphantom {\displaystyle \sum}}$
  & \parbox{5cm}{ultralocal change of variables gives linear equations}
 \\ \cline{2-4} 
      &  &  $(\ref{l2313-1})_{\vphantom {\displaystyle A}}^{
	\vphantom {\displaystyle A}}$  & 
 \\ \cline{3-4} 
      & \raisebox{3mm}{2, 4}  & (\ref{l2313-2}) $\cases{
 u_t = -2 \sca{U}{U_{xx}} - 2 u^3 \sca{U}{U} 
	 -6 u^2 \sca{U}{U}^2 -6 u \sca{U}{U}^3 - 2 \sca{U}{U}^4 
     \vspace{1.0mm}, \cr
 U_t = U_{xx} + u^3 U + 3 u^2 \sca{U}{U} U + 3 u \sca{U}{U}^2 U 
	 + \sca{U}{U}^3 U.
\cr}_{\vphantom {\displaystyle \sum}}^{\vphantom {\displaystyle \sum}}$
  & \parbox{5cm}{ultralocal change of variables gives linear equations}
%$\hspace{0.6cm}$\vrule width 3.8cm height 0.4pt depth 0pt}
 \\ \cline{2-4} 
      & 3, 5 & (\ref{l2313-3}) $\cases{
 u_t = u_{xxx} + 2 \sca{U}{U_{xxx}} + 6 \sca{U_x}{U_{xx}} + 2u_x \sca{U}{U}^3 
	 + 4 \sca{U}{U}^3 \sca{U}{U_x} 
     \vspace{1.0mm}, \cr
 U_t = - u_x \sca{U}{U}^2 U - 2 \sca{U}{U}^2 \sca{U}{U_x} U.
\cr}_{\vphantom {\displaystyle \sum}}^{\vphantom {\displaystyle \sum}}$
  & \parbox{5cm}{is 
symmetry of (\ref{l2313-1})
%,$\hspace{2pt}$\raisebox{2pt}{\vrule width 1.50cm height 0.4pt depth 0pt}
}
 \\ \cline{3-4} 
      &       & (\ref{l2313-4}) $\cases{
 u_t = u_{xxx} + 2 \sca{U}{U_{xxx}} + 6 \sca{U_x}{U_{xx}} + 2u_x \sca{U}{U}^3 
	 + 4 \sca{U}{U}^3 \sca{U}{U_x} 
     \vspace{1.0mm}, \cr
 U_t = - u_x \sca{U}{U}^2 U -4u \sca{U}{U} \sca{U}{U_x} U 
	 +4u \sca{U}{U}^2 U_x 
 \cr \qquad\, \mbox{}
	 - 6 \sca{U}{U}^2 \sca{U}{U_x} U+ 4 \sca{U}{U}^3 U_x.
\cr}_{\vphantom {\displaystyle \sum}}^{\vphantom {\displaystyle \sum}}$
  & \parbox{5cm}
{ultralocal change of variables gives linear equations}
%$\hspace{0.6cm}$\vrule width 3.8cm height 0.4pt depth 0pt}
 \\ \hline      
 \end{tabular}
 \normalsize
 \end{table}

 \newpage
 \noindent
 \section*{Acknowledgments} 
 \addcontentsline{toc}{section}{Acknowledgments}
 One of the authors (T.T.) would like to thank 
 Prof.\ Junkichi 
 Satsuma, Prof.\ Miki Wadati, 
 Prof.\ Tetsuji Tokihiro 
 and Prof.\ Yasuhiro Ohta for 
their useful comments. 
T.T.\ is also indebted to 
Prof.\ Hideshi Yamane for his kind help. 
 This research was supported in part by a JSPS 
 Fellowship for Young Scientists.


\begin{thebibliography}{99}
 \addcontentsline{toc}{section}{\refname}

 \bibitem{IbSha}
 Ibragimov N H and Shabat A B 1980
 Evolutionary equations with nontrivial Lie--B\"{a}cklund group 
 \FAA {\bf 14} 19--28

 \bibitem{Fokas2}
 Fokas A S 1980 
 A symmetry approach to exactly solvable evolution equations 
 \JMP {\bf 21} 1318--1325

  \bibitem{SoSh}
  Sokolov V V and Shabat A B 1984
  Classification of integrable evolution equations 
  {\it Soviet Sci.\ Rev.\ Sect.} \/C {\bf 4} 221--280

 \bibitem{MS1}
 Mikhailov A V and Shabat A B 1985 
 Integrability conditions for systems of two equations 
 of the form $\vt{u}\sb t=A(\vt{u})\vt{u}\sb {xx}+F(\vt{u},\vt{u}\sb x)$.\ I 
 \TMP {\bf 62} 107--122

 \bibitem{MS2}
 Mikhailov A V and Shabat A B 1986 
 Integrability conditions for systems of two equations 
 of the form $\vt{u}\sb t=A(\vt{u})\vt{u}\sb {xx}+F(\vt{u},\vt{u}\sb x)$.\ II 
 \TMP {\bf 66} 31--44

  \bibitem{Fokas} Fokas A S 1987 Symmetries and integrability 
  \SAM {\bf 77} 253--299

  \bibitem{MikShYam} Mikhailov A V, Shabat A B, and Yamilov R I 1987
  The symmetry approach to the classification of non-linear
  equations.\ Complete lists of integrable systems {\em
  Russian Math.\ Surveys} {\bf 42}(4) 1--63

  \bibitem{MSY}
  Mikhailov A V, Shabat A B and Yamilov R I 1988 
  Extension of the module of invertible transformations.\ Classification of 
  integrable systems \CMP {\bf 115} 1--19

 \bibitem{FujiWata}
 Fujimoto A and Watanabe Y 1989 
 Polynomial evolution equations of not normal type 
 admitting nontrivial symmetries 
 \PL A {\bf 136} 294--299

  \bibitem{MiShSok} Mikhailov A V, Shabat A B, Sokolov V V 1991
  The symmetry approach to classification of integrable equations 
 {\em What is integrability?} \/edited by Zakharov V E 
 (Springer Series in Nonlinear Dynamics, Springer, Berlin) 115--184

 \bibitem{ASY}
 Adler V E, Shabat A B and Yamilov R I 2000 
 Symmetry approach to the integrability problem 
 \TMP {\bf 125} 1603--1661

 \bibitem{Calo1}
 Calogero F and Eckhaus W 1987 
 Nonlinear evolution equations, rescalings, model PDEs and 
 their integrability\hspace{1pt}:\hspace{1pt}I \IP {\bf 3} 229--262

 \bibitem{Calo2}
 Calogero F 1991 
 Why are certain nonlinear PDEs both widely applicable and integrable? 
 {\em What is integrability?} \/edited by Zakharov V E 
 (Springer Series in Nonlinear Dynamics, Springer, Berlin) 1--62 

 \bibitem{Sanders}
 Sanders J A and Wang J P 2004 
On the integrability of systems of second order evolution equations 
with two components 
 \JDE {\bf 203} 1--27

 \bibitem{SWo0}
 Sokolov V V and Wolf T 1999 
 A symmetry test for quasilinear coupled systems 
 \IP {\bf 15} L5--L11

  \bibitem{Bakirov}
  Bakirov I M and Popkov V Yu 1989 
  Completely integrable systems of Brusselator type 
  \PL A {\bf 141} 275--277

 \bibitem{Svi2}
 Svinolupov S I 1989
 On the analogues of the Burgers equation 
 \PL A {\bf 135} 32--36

 \bibitem{Foursov1}
 Foursov M V 2000 
 On integrable coupled Burgers-type equations \PL A {\bf 272} 57--64

 \bibitem{Zharkov1}
 Gerdt V P and Zharkov A Yu 1990 
 Computer classification of integrable coupled KdV-like systems 
 \JSC {\bf 10} 203--207

 \bibitem{Zharkov2}
 Zharkov A Yu 1993 
 Computer classification of the integrable coupled KdV-like 
 systems with unit main matrix 
 \JSC {\bf 15} 85--90

 \bibitem{Meshkov}
 Kulemin I V and Meshkov A G 1997 
 To the classification of integrable systems in $1+1$ dimensions 
 {\em 
 Symmetry in nonlinear mathematical physics} \/vol.\ 1 (Kyiv, 1997) 
 115--121 

 \bibitem{Foursov0}
 Foursov M V 2000 
 On integrable coupled KdV-type systems \IP {\bf 16} 259--274

 \bibitem{Karasu}
 Karasu(Kalkanli) A 1997 
 Painlev\'{e} classification of coupled Korteweg--de Vries systems 
 \JMP {\bf 38} 3616--3622

  \bibitem{Sakov}
  Sakovich S Yu 1999 
  Coupled KdV equations of Hirota--Satsuma type 
  \JNMP {\bf 6} 255--262 (Addendum:\ 2001 ibid.\ {\bf 8} 311--312)

  \bibitem{Foursov2}
  Foursov M V 2000 
  Classification of certain integrable coupled potential KdV and modified 
  KdV-type equations \JMP {\bf 41} 6173--6185

  \bibitem{Foursov3}
  Foursov M V and Olver P J 2000 
  On the classification of symmetrically-coupled integrable 
  evolution equations {\em Symmetries and Differential Equations} \/edited 
by Andreev V K and Shanko Yu V 
  (Institute of Computational Modelling, Krasnoyarsk, Russia, 2000) 
  244--248

 \bibitem{Svi0}
 Svinolupov S I 1993 
 Jordan algebras and integrable systems 
 \FAA {\bf 27} 257--265

  \bibitem{Svi}
  Svinolupov S I and Sokolov V V 1994 
  Vector-matrix generalization of classical integrable equations 
  \TMP {\bf 100} 959--962

 \bibitem{Sak}
 Sakovich S Yu and Tsuchida T 2000 
 Symmetrically coupled higher-order nonlinear Schr\"{o}dinger 
 equations:\ singularity analysis and integrability
 \JPA {\bf 33} 7217--7226

 \bibitem{Olver1}
 Olver P J and Sokolov V V 1998 
 Integrable evolution equations on associative algebras 
 \CMP {\bf 193} 245--268

 \bibitem{Linden1}
 van der Linden J, Capel H W and Nijhoff F W 1989 
 Linear integral equations and multicomponent nonlinear integrable 
 systems II {\em Physica} \/A {\bf 160} 235--273

 \bibitem{Olver2}
 Olver P J and Sokolov V V 1998 
 Non-abelian integrable systems of the derivative nonlinear Schr\"{o}dinger 
 type \IP {\bf 14} L5--L8

 \bibitem{TW3}
 Tsuchida T and Wadati M 1999 
 Complete integrability of derivative nonlinear Schr\"{o}dinger-type equations 
 \IP {\bf 15} 1363--1373

 \bibitem{SWo}
 Sokolov V V and Wolf T 2001 
 Classification of integrable polynomial vector evolution equations 
 \JPA {\bf 34} 11139--11148

 \bibitem{Wang}
 Sanders J A and Wang J P 1998
 On the integrability of homogeneous scalar evolution equations 
 \JDE {\bf 147} 410--434

 \bibitem{Ibragimov}
 Ibragimov N H and Shabat A B 1980
 Infinite Lie--B\"{a}cklund algebras 
 \FAA {\bf 14} 313--315

 \bibitem{Beukers}
 Beukers F, Sanders J A and Wang J P 1998 
 One symmetry does not imply integrability 
 \JDE {\bf 146} 251--260

 \bibitem{Beukers2}
 Beukers F, Sanders J A and Wang J P 2001 
 On integrability of systems of evolution equations 
 \JDE {\bf 172} 396--408

 \bibitem{Kamp2}
 van der Kamp P H and Sanders J A 2002 
 Almost integrable evolution equations 
 {\em Selecta Math.\ New Ser.} {\bf 8} 705--719

 \bibitem{Wolf} 
 Wolf T 2002 
 Applications of {\sc Crack} in the classification of 
 integrable systems, to appear in the CRM Proceedings 
({\it e-print arXiv} \/nlin.SI/0301032)

 \bibitem{TWshort}
 Wolf T 2002 
 Size reduction and partial decoupling of systems of 
 equations {\it J.\ Symb.\ Comput.} {\bf 3} 367--383

  \bibitem{Hereman}
  G\"{o}kta\c{s} \"{U} and Hereman W 1999 
  Algorithmic computation of higher-order symmetries for nonlinear 
  evolution and lattice equations 
  {\em Adv.\ Comput.\ Math.} {\bf 11} 55--80

  \bibitem{Wilson}
  Wilson G 1982 The affine Lie algebra ${\mbox C}_2^{(1)}$ 
  and an equation of Hirota 
  and Satsuma \PL A {\bf 89} 332--334

  \bibitem{Drinfeld}
  Drinfel'd V G and Sokolov V V 1985 
  Lie algebras and equations of Korteweg--de Vries type 
  {\em J.\ Sov.\ Math.} {\bf 30} 1975--2036

 \bibitem{Melnikov}
 Mel'nikov V K 1983 
 On equations for wave interactions 
 \LMP {\bf 7} 129--136

 \bibitem{Strampp1}
 Konopelchenko B and Strampp W 1992 
 New reductions of the Kadomtsev--Petviashvili and two--dimensional 
 Toda lattice hierarchies via symmetry constraints 
 \JMP {\bf 33} 3676--3686

 \bibitem{Strampp2}
 Sidorenko J and Strampp W 1993 
 Multicomponent integrable reductions in 
 the Kadomtsev--Petviashvili hierarchy 
 \JMP {\bf 34} 1429--1446

 \bibitem{Ad}
 Adler V E 2000 
 On the relation between multifield and multidimensional integrable 
 equations {\it e-print arXiv} \/nlin.SI/0011039

 \bibitem{Lax}
 Lax P D 1968 
 Integrals of nonlinear equations of evolution and solitary waves 
 \CPAM {\bf 21} 467--490

 \bibitem{Kamijo}
 Wadati M and Kamijo T 1974 
 On the extension of inverse scattering method 
 \PTP {\bf 52} 397--414

\bibitem{CaloDega}
Calogero F and Degasperis A 1977 
Nonlinear evolution equations solvable by the inverse 
spectral transform.\ II \NC B {\bf 39} 1--54
%(11) 
%no. 1, 

\bibitem{Zakh}
Zakharov V E 1980 
The inverse scattering method 
{\em Solitons} \/edited by Bullough R K and Caudrey P J 
(Topics in Current Physics
17, 
Springer, Berlin) 243--285

 \bibitem{Ito}
 Ito M 1982
 Symmetries and conservation laws of a coupled nonlinear wave equation 
 \PL A {\bf 91} 335--338

 \bibitem{Kuper}
 Kupershmidt B A 1985 
 A coupled Korteweg--de Vries equation with dispersion 
 \JPA {\bf 18} L571--L573

 \bibitem{Boiti1}
 Boiti M, Laddomada C, Pempinelli F and Tu G Z 1983 
 On a new hierarchy of Hamiltonian soliton equations 
 \JMP {\bf 24} 2035--2041

 \bibitem{Bogo}
 Bogolyubov N N and Prikarpatskii A K 1986 
 Complete integrability of the nonlinear Ito and Benney--Kaup systems:\ 
 Gradient algorithm and Lax representation 
 \TMP {\bf 67} 586--596

 \bibitem{HiSa}
 Hirota R and Satsuma J 1981
 Soliton solutions of a coupled Korteweg--de Vries equation 
 \PL A {\bf 85} 407--408

 \bibitem{HiroOh}
 Hirota R and Ohta Y 1991 
 Hierarchies of coupled soliton equations.\ I 
 \JPSJ {\bf 60} 798--809

  \bibitem{Dodd}
  Dodd R and Fordy A 1982 
  On the integrability of a system of coupled KdV equations 
  \PL A {\bf 89} 168--170

 \bibitem{DS2}
 Drinfel'd V G and Sokolov V V 1981 
 New evolution equations having an $(L,\,A)$ pair 
 {\em Trudy Sem.\ S.\ L.\ Soboleva} 
 {\bf 2} 5--9 (in Russian)

  \bibitem{Wu} 
  Wu Y T, Geng X G, Hu X B and Zhu S M 1999 
  A generalized Hirota--Satsuma coupled Korteweg--de Vries equation 
  and Miura transformations 
  \PL A {\bf 255} 259--264

\bibitem{Ma}
Ma W X 1993 A hierarchy of coupled Burgers systems possessing a hereditary 
structure \JPA {\bf 26} L1169--L1174

\bibitem{KK}
Kaup D J 1980 
On the inverse scattering problem for cubic eigenvalue problems of the 
class $\psi_{xxx} + 6Q \psi_x + 6R \psi = \lambda \psi$ 
\SAM {\bf 62} 189--216

\bibitem{Gibbons}
Fordy A P and Gibbons J 1980 
Factorization of operators I.\ Miura transformations 
\JMP {\bf 21} 2508--2510

  \bibitem{Kamp} 
  van der Kamp P H 2002 
  On proving integrability 
  \IP {\bf 18} 405--414

 \bibitem{Fuchs}
 Fuchssteiner B 1982 
 The Lie algebra structure of degenerate Hamiltonian and 
 bi-Hamiltonian systems \PTP {\bf 68} 1082--1104

 \bibitem{Gurses}
 G\"{u}rses M and Karasu A 1998 
 Integrable coupled KdV systems
 \JMP {\bf 39} 2103--2111

 \bibitem{FL}
 Fokas A S and Liu Q M 1994 
 Generalized conditional symmetries and exact solutions 
 of non-integrable equations 
 \TMP {\bf 99} 571--582

 \bibitem{Razb}
 Razboinik S I 1986 
 Vector extensions of modified water wave equations 
 \PL A {\bf 119} 283--286

 \bibitem{Kuper2}
 Kupershmidt B A 1989 
 Modified Korteweg--de Vries equations on Euclidean Lie algebras 
 {\em Int.\ J.\ Mod.\ Phys.} \/B {\bf 3} 853--861

 \bibitem{Tu}
 Tu G Z 1983 
 A new hierarchy of coupled degenerate Hamiltonian equations 
 \PL A {\bf 94} 340--342 

 \bibitem{Boiti2}
 Boiti M, Leon J JP and Pempinelli F 1984 
 A recursive generation of local higher-order sine-Gordon equations 
 and their B\"{a}cklund transformation 
 \JMP {\bf 25} 1725--1734

 \bibitem{Hu}
 Hu X B 1993 
 Generalized Hirota's bilinear equations and their soliton solutions
 \JPA {\bf 26} L465--L471

 \bibitem{Adler}
 Adler V E 1994 
 Nonlinear superposition principle for the Jordan NLS equation 
 \PL A {\bf 190} 53--58

  \bibitem{TW1}
  Tsuchida T and Wadati M 1998 
  The coupled modified Korteweg--de Vries equations 
  \JPSJ {\bf 67} 1175--1187

\bibitem{AKNS}
Ablowitz M J, Kaup D J, Newell A C and Segur H 1973 
Nonlinear-evolution equations of physical significance 
\PRL {\bf 31} 125--127

 \bibitem{YO}
 Yajima N and Oikawa M 1975 
 A class of exactly solvable nonlinear evolution equations 
 \PTP {\bf 54} 1576--1577

  \bibitem{Konope}
  Konopelchenko B G 1983 
  Nonlinear transformations and integrable evolution equations 
  {\em Fortschr.\ Phys.} {\bf 31} 253--296

 \bibitem{Linden2}
 van der Linden J, Nijhoff F W, Capel H W and Quispel G R W 1986
 Linear integral equations and multicomponent nonlinear 
 integrable systems I 
 {\em Physica} \/A {\bf 137} 44--80

 \bibitem{Kersten}
 Kersten P and Krasil'shchik J 2002 
 Complete integrability of the coupled KdV-mKdV system 
 {\em Adv.\ Stud.\ Pure Math.} {\bf 37} 
 {\it Lie Groups, Geometric Structures and Differential 
 Equations---One Hundred Years after Sophus Lie---} \/edited 
by Morimoto T, Sato H and Yamaguchi K 
 151--171 

 \bibitem{KSY}
 Karasu(Kalkanli) A, Sakovich S Yu and Yurdu\c{s}en \'{I} 2003
 Integrability of Kersten--Krasil'shchik 
 coupled KdV-mKdV equations:\ singularity analysis and Lax pair 
 \JMP {\bf 44} 1703--1708

 \bibitem{JM}
 Jaulent M and Miodek I 1976
 Nonlinear evolution equations associated with `energy-dependent 
 Schr\"{o}dinger potentials' 
 \LMP {\bf 1} 243--250

\bibitem{Alonso}
Mart\'{i}nez Alonso L and Guil Guerrero F 1981 
Modified Hamiltonian systems and canonical transformations 
arising from the relationship between generalized Zakharov--Shabat and
energy-dependent Schr\"{o}dinger operators 
\JMP {\bf 22} 2497--2503

\bibitem{Nijhoff}
Nijhoff F W, Quispel G R W, van der Linden J and Capel H W 1983 
On some linear integral equations generating solutions of 
nonlinear partial differential equations {\em Physica} \/A 
{\bf 119} 101--142

\bibitem{Das}
Das A and Popowicz Z 2004 
Bosonic reduction of 
susy generalized Harry Dym equation
\JPA {\bf 37} 8031--8044

\bibitem{Leble}
Leble S B and Ustinov N V 1993 
Darboux transforms, deep reductions and solitons 
\JPA {\bf 26} 5007--5016

\bibitem{Calog}
Calogero F 1987 
The evolution partial differential equation $u\sb t=u\sb {xxx}
+3(u\sb {xx}u\sp 2+3u\sp 2\sb xu)+3u\sb xu\sp 4$ 
\JMP {\bf 28} 538--555

\end{thebibliography}
\end{document}